\documentclass[journal]{IEEEtran} % should be 10pt in the real journal % {IEEEtran2e}

\usepackage{amsmath}
\usepackage{amsfonts}
\usepackage{epsfig}
\usepackage{bar}
\usepackage{color}

    \newcommand{\vect}[1]{{\lowercase{\boldsymbol{#1}}}}
    \newcommand{\mat}[1]{{\uppercase{\boldsymbol{#1}}}}

\newcommand{\nn}{\nonumber}
\newcommand{\limzero}[1]{\lim_{#1\rightarrow 0}}
\newcommand{\liminfty}[1]{\lim_{#1\rightarrow \infty}}
\newcommand{ \divergence }[2]{ \mathsf{D} \left( #1 \| #2 \right) }
\newcommand{ \cnddiv }[3]{ \mathsf{D} \left( #1 \| #2 | #3 \right) }
\newcommand{\Prob}{{\mathsf P}}

\newcommand{\expb}[1]{ \exp \left[ #1 \right] } % exp[.]
\newcommand{\trace}[1]{\mathsf{tr}\left\{#1\right\}}
\newcommand{\Exp}{\mathsf E}
\newcommand{\expect}[1]{{\Exp}\left\{#1\right\}}
\newcommand{\expcnd}[2]{{\Exp}\left\{\left. #1 \,\right|\, #2\right\}}
\newcommand{\var}[1]{\mathsf{var}\left\{#1\right\}}
\newcommand{\inv}[1]{#1^{\scriptscriptstyle -\!1}}

\newcommand{\tran}[1]{#1^{\!\top\!}}
\newcommand{\diag}{\text{diag}}
\newcommand{\etal}{{\em et al.}}
\newcommand{\ith}[1]{\mbox{$#1^{\text{th}}$}}
\newcommand{\eref}[1]{(\ref{#1})}
\newcommand{\half}{\frac{1}{2}} % dongning, 12/12/2001
\newcommand{\oneon}[1]{\frac{1}{#1}}

\renewcommand{\b}{\vect{b}} % _ accent
\newcommand{\e}{\vect{e}}
\newcommand{\f}{\vect{f}}

\newcommand{\n}{\vect{n}}

\newcommand{\x}{\vect{x}}
\newcommand{\y}{\vect{y}}
\newcommand{\z}{\vect{z}}

\newcommand{\xT}{\tran{\vect{x}}}
\newcommand{\yT}{\tran{\vect{y}}}

\renewcommand{\H}{\mat{H}} % ''accent
\newcommand{\I}{\mat{I}}

\newcommand{\N}{\mat{N}}

\newcommand{\W}{\mat{W}}
\newcommand{\X}{\mat{X}}
\newcommand{\Y}{\mat{Y}}
\newcommand{\Z}{\mat{Z}}

\newcommand{\HT}{\tran{\mat{H}}}

\newcommand{\XT}{\tran{\mat{X}}}

\usepackage{verbatim}

\definecolor{gray}{rgb}{.8,.8,.8}%
\definecolor{dgray}{rgb}{.6,.6,.6}%

\def\myfigwidth{3.5in}

\newtheorem{theorem}{Theorem}

\newtheorem{corollary}{Corollary}

\newtheorem{lemma}{Lemma}
\newcommand{\nsp}[1]{\hspace{-#1ex}}
\newcommand{\nind}{\hspace{-6ex}}

\newcommand{\rp}[1]{ \left\{ #1_t \right\} } % {X_t}
\newcommand{\pd}[1]{ {\frac{ \intd }{ \intd {#1} }} }
\newcommand{\ppd}[1]{ {\frac{ \partial }{ \partial {#1} }} }
\newcommand{\pyi}[1]{ q_{#1}(y;\snr) }
\newcommand{\cov}[1]{ \mathsf{Cov}\left\{ #1 \right\} } % Covariance
\newcommand{\exph}[1]{ \expb{ -\half #1 } }

\newcommand{\snr}{ {\mathsf{snr}} }% {\Gamma}%
\newcommand{\Snr}{ \boldsymbol{\Gamma} } % zSnr} }
\newcommand{\zSnr}{ \snr }  % \newcommand{\zSnr}{ \Gamma }

\newcommand{\intd}{{\,\normalfont{\text d}}} % d for integral
\newcommand{\mmse}{ {\mathsf{mmse}} } % MMSE
\newcommand{\cmmse}{ {\mathsf{cmmse}} } % causal filtering MMSE
\newcommand{\pmmse}{ {\mathsf{pmmse}} } % prediction MMSE
\newcommand{\mSigma}{ \boldsymbol{\Sigma} } % matrix
\newcommand{\hX}{ \widehat{X} }
\newcommand{\hvX}{ \widehat{\mat{X}} }

\newcommand{\pyx}{p_{Y|X;\snr}(y|x;\snr)}
\newcommand{\pyX}{p_{Y|X;\snr}(y|X;\snr)}
\newcommand{\pYX}{p_{Y|X;\snr}(Y|X;\snr)}

\newcommand{\vpyx}{p_{\Y|\X;\snr}(\y|\x;\snr)}
\newcommand{\vpYX}{p_{\Y|\X;\snr}(\Y|\X;\snr)}
\newcommand{\vpyX}{p_{\Y|\X;\snr}(\y|\X;\snr)}
\newcommand{\vpy}{p_{\Y;\snr}(\y;\snr)}
\newcommand{\vpY}{p_{\Y;\snr}(\Y;\snr)}

\begin{document}
\bstctlcite{BSTcontrol} % must happen before all affected bib entries

\title{Mutual Information and Minimum Mean-square Error in Gaussian
  Channels}

\author{Dongning Guo,
        Shlomo Shamai (Shitz),
        and Sergio Verd\'u
\def\thefootnote{}% the \thanks{} mark type is empty
\thanks{Dongning Guo was with the Department of Electrical Engineering
  at Princeton University.  He is now with the Department of
  Electrical and Computer Engineering at Northwestern University,
  Evanston, IL, 60208, USA.  Email: dGuo@Northwestern.EDU.  Shlomo
  Shamai (Shitz) is with the Department of Electrical Engineering,
  Technion-Israel Institute of Technology, 32000 Haifa, Israel.
  Email: sshlomo@ee.technion.ac.il.  Sergio Verd\'u is with the
  Department of Electrical Engineering, Princeton University,
  Princeton, NJ 08544, USA.  Email: Verdu@Princeton.EDU.  } }

\maketitle%

%\markboth{IEEE Transactions on Information Theory,~Vol.~?,
%  No.~?,~month~2005}{Guo \MakeLowercase{\textit{et al.}}: Mutual
%  Information and Minimum Mean-square Error in Gaussian Channels}

\begin{abstract}%
  This paper deals with arbitrarily distributed finite-power input
  signals observed through an additive Gaussian noise channel.  It
  shows a new formula that connects the input-output mutual
  information and the minimum mean-square error (MMSE) achievable by
  optimal estimation of the input given the output.  That is, the
  derivative of the mutual information (nats) with respect to the
  signal-to-noise ratio (SNR) is equal to half the MMSE, regardless of
  the input statistics.  This relationship holds for both scalar and
  vector signals, as well as for discrete-time and continuous-time
  noncausal MMSE estimation. % (smoothing).
  
  This fundamental information-theoretic result has an unexpected
  consequence in continuous-time nonlinear estimation: For any input
  signal with finite power, the causal filtering MMSE achieved at SNR
  is equal to the average value of the noncausal smoothing MMSE
  achieved with a channel whose signal-to-noise ratio is chosen
  uniformly distributed between 0 and SNR.
\end{abstract}

\begin{keywords}
  Gaussian channel, minimum mean-square error (MMSE), mutual
  information, nonlinear filtering, optimal estimation, smoothing,
  Wiener process.
\end{keywords}

% \IEEEpeerreviewmaketitle

\section{Introduction}
\label{s:int}

% DG041018:  change in the first 2 sentences
% This paper is centered around two basic quantities that
% measure the noisiness of channels.
% It is well-known that the MMSE is achieved by conditional mean estimation.  

This paper is centered around two basic quantities in information
theory and estimation theory, namely, the {\em mutual information}
between the input and the output of a channel, and the {\em minimum
  mean-square error} (MMSE) in estimating the input given the output.
The key discovery is a relationship between the mutual information and
MMSE that holds regardless of the input distribution, as long as the
input-output pair are related through additive Gaussian noise.

Take for example the simplest scalar real-valued Gaussian channel with
an arbitrary and fixed input distribution.  Let the signal-to-noise
ratio (SNR) of the channel be denoted by $\snr$.  Both the
input-output mutual information and the MMSE are monotone functions of
the SNR, denoted by $I(\snr)$ and $\mmse(\snr)$ respectively.  This
paper finds that the mutual information in nats and the MMSE satisfy
the following relationship regardless of the input statistics:
\begin{equation}  \label{e:pie}
  \pd{\snr} I(\snr) = \half \mmse(\snr).
\end{equation}

Simple as it is, the identity \eref{e:pie} was unknown before this
work.  It is trivial that one can compute the value of one monotone
function given the value of another (e.g., by simply composing the
inverse of the latter function with the former); what is quite
surprising here is that the overall transformation \eref{e:pie} not
only is strikingly simple but is also independent of the input
distribution.  In fact, this relationship and its variations hold
under arbitrary input signaling and the broadest settings of Gaussian
channels, including discrete-time and continuous-time channels, either
in scalar or vector versions.

In a wider context, the mutual information and mean-square error are
at the core of information theory and estimation theory respectively.
The input-output mutual information is an indicator of how much coded
information can be pumped through a channel reliably given a certain
input signaling, whereas the MMSE measures how accurately each
individual input sample can be recovered using the channel output.
Interestingly, \eref{e:pie} shows the strong relevance of mutual
information to estimation and filtering and provides a non-coding
operational characterization for mutual information.  Thus not only is
the significance of an identity like \eref{e:pie} self-evident, but
the relationship is intriguing and deserves thorough exposition.

At zero SNR, the right hand side of \eref{e:pie} is equal to one half
of the input variance.  In that special case the formula, and in
particular, the fact that at low-SNR mutual information is insensitive
to the input distribution has been remarked before \cite{Verdu90IT,
  LapSha02IT, Verdu02IT}.  Relationships between the local behavior of
mutual information at vanishing SNR and the MMSE of the estimation of
the output given the input are given in \cite{PreVer04IT}.

Formula \eref{e:pie} can be proved using the new ``incremental
channel'' approach which gauges the decrease in mutual information due
to an infinitesimally small additional Gaussian noise.  The change in
mutual information can be obtained as the input-output mutual
information of a derived Gaussian channel whose SNR is infinitesimally
small, a channel for which the mutual information is essentially
linear in the estimation error, and hence relates the rate of mutual
information increase to the MMSE.

Another rationale for the relationship \eref{e:pie} traces to the
geometry of Gaussian channels, or, more tangibly, the geometric
properties of the likelihood ratio associated with signal detection in
Gaussian noise.  Basic information-theoretic notions are firmly
associated with the likelihood ratio, and foremost the mutual
information is expressed as the expectation of the log-likelihood
ratio of conditional and unconditional measures.  The likelihood ratio
also plays a fundamental role in detection and estimation, e.g., in
hypothesis testing it is compared to a threshold to decide which
hypothesis to take.  Moreover, the likelihood ratio is central in the
connection of detection and estimation, in either continuous-time
\cite{Kailat68IC, Kailat69IT, Kailat70IT} or discrete-time setting
\cite{JafGup72IC}.  In fact, Esposito \cite{Esposi68IC} and Hatsell
and Nolte \cite{HatNol71IT} noted simple relationships between
conditional mean estimation and the gradient and Laplacian of the
log-likelihood ratio respectively, although they did not import mutual
information into the picture.  Indeed, the likelihood ratio bridges
information measures and basic quantities in detection and estimation,
and in particular, the estimation errors (e.g., \cite{MazBag95IT}).

% and on the other to find achievable bounds for
%mutual information based on estimation errors associated with linear
%estimators (e.g., \cite{GasVet02Infocom}).

In continuous-time signal processing, both the causal (filtering) MMSE
and noncausal (smoothing) MMSE are important performance measures.
Suppose for now that the input is a stationary process with arbitrary
but fixed statistics.  Let $\cmmse(\snr)$ and $\mmse(\snr)$ denote the
causal and noncausal MMSEs respectively as a function of the SNR.
This paper finds that formula \eref{e:pie} holds literally in this
continuous-time setting, i.e., the derivative of the mutual
information rate is equal to half the noncausal MMSE.  Furthermore, by
using this new information-theoretic identity, an unexpected
fundamental result in nonlinear filtering is unveiled.  That is, the
filtering MMSE is equal to the mean value of the smoothing MMSE:
\begin{equation}  \label{e:iee}
  \cmmse(\snr) = \expect{ \mmse(\Gamma) }
\end{equation}
where $\Gamma$ is chosen uniformly distributed between 0 and $\snr$.
In fact, stationarity of the input is not required if the MMSEs are
defined as time averages.

Relationships between the causal and noncausal estimation errors have
been studied for the particular case of linear estimation (or Gaussian
inputs) in \cite{AndChi71EL}, where a bound on the loss due to the
causality constraint is quantified.  Capitalizing on earlier research
on the ``estimator-correlator'' principle by Kailath and others (see
\cite{KaiPoo98IT}), Duncan \cite{Duncan68IC, Duncan70JAM},
Zakai\footnote{Duncan's Theorem was independently obtained by Zakai in
  the more general setting of inputs that may depend causally on the
  noisy output in a 1969 unpublished Bell Labs Memorandum (see
  \cite[{ref.~[53]}]{Kailat70PI}).}  and Kadota \etal\ 
\cite{KadZak71IT} pioneered the investigation of relations between the
mutual information and causal filtering of continuous-time signals
observed in white Gaussian noise.  In particular, Duncan showed that
the input-output mutual information can be expressed as a
time-integral of the causal MMSE \cite{Duncan70JAM}.  Duncan's
relationship has proven to be useful in many applications in
information theory and statistics \cite{KadZak71IT, KadZak71ITa,
  BarSha88IT, ChaSha99IT}.  There are also a number of other works in
this area, most notably those of Liptser \cite{Liptse74PPI} and
Mayer-Wolf and Zakai \cite{MayZak83}, where the rate of increase in
the mutual information between the sample of the input process at the
current time and the entire past of the output process is expressed in
the causal estimation error and certain Fisher informations.  Similar
results were also obtained for discrete-time models by Bucy
\cite{Bucy79IS}.  In \cite{Shmele85RIE} Shmelev devised a general,
albeit complicated, procedure to obtain the optimal smoother from the
optimal filter.

\begin{comment}
The relationship between the causal and noncausal estimation errors
has been studied for the particular case of linear estimation (or
Gaussian inputs) in \cite{AndChi71EL}, where a bound on the loss due
to causality constraint is quantified.  Duncan \cite{Duncan68IC,
Duncan70JAM}, Zakai \cite[{ref.~[53]}]{Kailat70PI} and Kadota \etal\
\cite{KadZak71IT} pioneered the investigation of relations between the
mutual information and conditional mean filtering~\cite{Duncan68IC,
Duncan70JAM}, which capitalized on earlier research on the
``estimator-correlator'' principle by Price~\cite{Price56IRE},
Kailath~\cite{Kailat63}, and others (see \cite{KaiPoo98IT}).  There
are also a number of other works in this area, most notably those of
Liptser \cite{Liptse74PPI} and Mayer-Wolf and Zakai~\cite{MayZak83},
where the rate of increase in the mutual information between the
sample of the input process at the current time and the entire past of
the output process is expressed in the causal estimation error and
some Fisher informations.  Similar results were also obtained for
discrete-time models by Bucy~\cite{Bucy79IS}.  In~\cite{Shmele85RIE}
Shmelev devised a general, albeit complicated, procedure to obtain the
optimal smoother from the optimal filter.  As one shall see, also
relevant to this work is a relationship between differential entropy
and Fisher's information known as de Bruijn's identity
\cite{Stam59IC}.
%\cite{Costa85IT, CovTho91}.
\end{comment}

The new relationship \eref{e:pie} in continuous-time and Duncan's
Theorem are proved in this paper using the incremental channel
approach with increments in additional noise and additional
observation time respectively.  Formula \eref{e:iee} connecting
filtering and smoothing MMSEs is then proved by comparing \eref{e:pie}
to Duncan's theorem.  A non-information-theoretic proof is not yet
known for \eref{e:iee}.

In the discrete-time setting, identity \eref{e:pie} still holds, while
the relationship between the mutual information and the causal MMSEs
takes a different form: We show that the mutual information is
sandwiched between the filtering error and the prediction error.

\begin{comment}
  The new findings in this chapter are related to many previous works.
  The fact that likelihood ratios connect detection and estimation has
  been used to find various useful identities and bounds (e.g.
  \cite{ZivZak69IT, ChaZak75IT} and more recently \cite{BelSte97IT}).
\end{comment}

The remainder of this paper is organized as follows.  Section
\ref{s:dt} gives the central result \eref{e:pie} for both scalar and
vector channels along with four different proofs and discussion of
applications.  Section \ref{s:ct} gives the continuous-time channel
counterpart along with the fundamental nonlinear filtering-smoothing
relationship \eref{e:iee}, and a fifth proof of \eref{e:pie}.
Discrete-time channels are briefly dealt with in Section \ref{s:dc}.
Section \ref{s:g} studies general random transformations observed in
additive Gaussian noise, and offers a glimpse at feedback channels.
Section \ref{s:im} gives new representations for entropy, differential
entropy, and mutual information for arbitrary distributions.

%%%%%%%%%%%%%%%%%%%%%%%%%%%%%%%%%%%%%%%%%%%%%%%%%%%%%%%%%%%%
%%%%%%%%%%%%%%%%%%%%%%%%%%%%%%%%%%%%%%%%%%%%%%%%%%%%%%%%%%%%
% DG041018: new section title
\section{Scalar and Vector Gaussian Channels}
% \section{Discrete-time Gaussian-noise Channels}
\label{s:dt}

\subsection{The Scalar Channel}
\label{s:sc}

Consider a pair or real-valued random variables $(X,Y)$ related
by\footnote{In this paper, random objects are denoted by upper-case
  letters and their values denoted by lower-class letters.  The
  expectation $\expect{\cdot}$ is taken over the joint distribution of
  the random variables within the brackets.}
\begin{equation}
  Y = \sqrt{\snr} \, X + N
  \label{e:ch}
\end{equation}
where $\snr\geq0$ and the $N\sim \mathcal{N} (0,1)$ is a standard
Gaussian random variable independent of $X$.  Then $X$ and $Y$ can be
regarded as the input and output respectively of a single use of a
scalar Gaussian channel with a signal-to-noise ratio of
$\snr$.\footnote{If $\Exp X^2=1$ then $\snr$ complies with the usual
  notion of signal-to-noise power ratio; otherwise $\snr$ can be
  regarded as the gain in the output SNR due to the channel.  Results
  in this paper do not require $\Exp X^2=1$.}  The input-output
conditional probability density is described by
\begin{equation} \label{e:pyx}
  \pyx = \oneon{\sqrt{2\pi}} \exph{ \left( y-\sqrt{\snr}\, x \right)^2 }.
\end{equation}
Upon the observation of the output $Y$, one would like to infer the
information bearing input $X$.  The {\em mutual information} between
$X$ and $Y$ is:
\begin{equation}  \label{e:ixy}
  I(X;Y) = \expect{ \log \frac{ \pYX }{ p_{Y;\snr}(Y;\snr) } }.
\end{equation}
where $p_{Y;\snr}$ denotes the well-defined marginal probability
density function of the output:
\begin{equation}
  p_{Y;\snr}(y;\snr) = \expect{ \pyX }. %_{Y|X}(y|X) }. % , \quad \forall y.
\end{equation}
The mutual information is clearly a function of $\snr$, which we
denote by
\begin{equation}
  I(\snr) = I\left( X;\sqrt{\snr}\, X + N \right).
  \label{e:is}
\end{equation}
The error of an estimate, $f(Y)$, of the input $X$ based on the
observation $Y$ can be measured in mean-square sense:
\begin{equation}
%  \mathsf{mse}\left(\hX\right) =
  \expect{ \left( X - f(Y) \right)^2 }.
  \label{e:mse}
\end{equation}
It is well-known that the minimum value of \eref{e:mse}, referred to
as the {\em minimum mean-square error} or MMSE, is achieved by the
conditional mean estimator:
\begin{equation}
  \hX(Y;\snr) = \expcnd{ X }{ Y;\snr }.
  \label{e:cm}
\end{equation}
The MMSE is also a function of $\snr$, which is denoted by
\begin{equation}
  \mmse(\snr) = \mmse\left( X\,|\,\sqrt{\snr}\, X+N \right).
  % \expect{ \left( X - \expcnd{X}{Y;\snr} \right)^2 }.
  \label{e:es}
\end{equation}

\begin{comment}
Let the input distribution (probability measure) be $P_X$ and the
probability density function of $Y$ conditioned on $X$ be $p_{Y|X}$.
The {\em mutual information} between $X$ and $Y$ is:
\begin{equation}
  I(X;Y) = \expect{ \log \frac{ p_{Y|X}(Y|X) }{ p_Y(Y) } }
  \label{e:ixy}
\end{equation}
where $\expect{\cdot}$ takes the expectation over the joint
distribution of the random variables within the brackets.  Here, $p_Y$
denotes the marginal probability density function of the output, which
is well-defined:
\begin{equation}
  p_Y(y) = \expect{ p_{Y|X}(y|X) }. % , \quad \forall y.
\end{equation}

Let the distribution of the input be $P_X$, which does not depend on
$\snr$.  The marginal probability density function of the output
exists:
\begin{equation}
  p_{Y;\snr}(y;\snr) = \expect{ \pyX }.
\end{equation}
\end{comment}

To start with, consider the special case when the input distribution
$P_X$ is standard Gaussian.  The input-output mutual information is
then the well-known channel capacity under input power constraint
\cite{Shanno48BSTJ}:
\begin{equation}
  I(\snr) = \half \log(1+\snr).
  % I(\snr) = \capacity(\snr) = \half \log(1+\snr).
  \label{e:ig}
\end{equation}
Meanwhile, the conditional mean estimate of the Gaussian input is
merely a scaling of the output:
\begin{equation}
  \hX(Y;\snr) = \frac{ \sqrt{\snr} }{ 1+\snr } \, Y,
  \label{e:hxg}
\end{equation}
and hence the MMSE is:
\begin{equation}
  \mmse(\snr) = \oneon{1+\snr}.
  \label{e:eg}
\end{equation}
An immediate observation is
\begin{equation} \label{e:piee}
  \pd{\snr} I(\snr) = \half \mmse(\snr) \, \log e,
\end{equation}
where the base of logarithm is consistent with the mutual information
unit.  To avoid numerous $\log e$ factors, henceforth we adopt natural
logarithms and use nats as the unit of all information measures.  It
turns out that the relationship \eref{e:piee} holds not only for
Gaussian inputs, but for any finite-power input.
\begin{theorem}  \label{th:di}         % discrete-time, iid
  Let $N$ be standard Gaussian, independent of $X$.  For every input
  distribution\, $P_X$ that satisfies\, $\Exp X^2 <\infty$,
  \begin{equation}
    \pd{\snr} I\left( X; \sqrt{\snr}\,X+N \right)
    = \half \mmse\left( X\,|\,\sqrt{\snr}\,X+N \right).
    % \pd{\snr} I(\snr) = \half \mmse(\snr).
    % I(X;\stsnr\cdot X+N) = \int_0^\snr \mmse(\gamma) \, \intd \gamma.
    \label{e:ie}
  \end{equation}
\end{theorem}
\begin{proof}
  See Section \ref{s:ic}.
  % See Section \ref{s:sic}.
  % and Appendix \ref{a:di} for two different proofs.
\end{proof}
The identity \eref{e:ie} reveals an intimate and intriguing connection
between Shannon's mutual information and optimal estimation in the
Gaussian channel \eref{e:ch}, namely, the rate of the mutual
information increase as the SNR increases is equal to half the MMSE
achieved by the optimal (in general nonlinear) estimator.

In addition to the special case of Gaussian inputs, Theorem
\ref{th:di} can also be verified for another simple and important
input signaling: $\pm1$ with equal probability.  The conditional mean
estimate for such an input is given by
\begin{equation}
  \hX(Y;\snr) = \tanh\left( \sqrt{\snr}\, Y \right).
\end{equation}
The MMSE and the mutual information are obtained as:
\begin{equation}
  \mmse(\snr) = 1 - \int^\infty_{-\infty}
  % \frac{ e^{-\half y^2} }{ \sqrt{2\pi} }
  \frac{ e^{-\frac{y^2}{2}} }{ \sqrt{2\pi} }
  \tanh( \snr - \sqrt{\snr}\, y ) \,\intd y,
  \label{e:eb}
\end{equation}
and (e.g., \cite[p.~274]{Blahut87} and \cite[Problem 4.22]{Gallag68})
\begin{equation}
  I(\snr) = \snr - \int^\infty_{-\infty}
  % \frac{ e^{-\half y^2} }{ \sqrt{2\pi} }
  \frac{ e^{-\frac{y^2}{2}} }{ \sqrt{2\pi} }
  \log \cosh( \snr - \sqrt{\snr}\, y ) \,\intd y
  \label{e:ib}
\end{equation}
respectively.  Appendix \ref{a:dim} verifies that \eref{e:eb} and
\eref{e:ib} satisfy \eref{e:ie}.

\begin{figure}
  % Figure generated using /math/I-mmse/scalar-ch.nb
  \begin{center}
    \includegraphics[width=\myfigwidth]{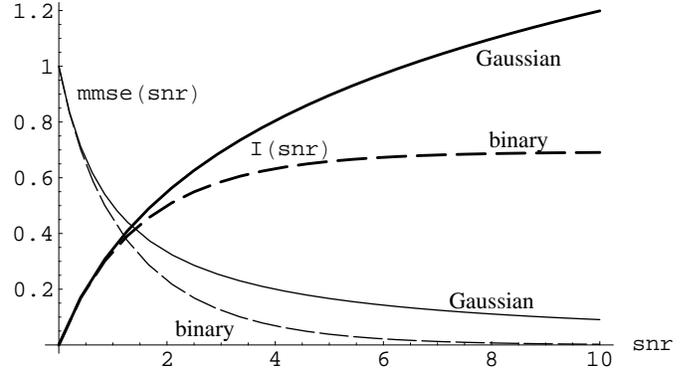}
    \caption{The mutual information (in nats) and MMSE of scalar
      Gaussian channel with Gaussian and equiprobable binary inputs,
      respectively.}
  \label{f:sc}
  \end{center}
\end{figure}

For illustration purposes, the MMSE and the mutual information are
plotted against the SNR in Figure \ref{f:sc} for Gaussian and
equiprobable binary inputs.

\subsection{The Vector Channel}
\label{s:vc}

Multiple-input multiple-output (MIMO) systems are frequently described
by the vector Gaussian channel:
\begin{equation}
  \Y = \sqrt{\snr} \, \H \, \X + \N
  \label{e:vch}
\end{equation}
where $\H$ is a deterministic $L\times K$ matrix and the noise $\N$
consists of independent standard Gaussian entries.  The input $\X$
(with distribution $P_\X$) and the output $\Y$ are column vectors of
appropriate dimensions.

The input and output are related by a Gaussian conditional probability
density:
\begin{equation}
  \vpyx = (2\pi)^{-\frac{L}{2}} \exph{
    \left\| \y-\sqrt{\snr}\,\H\x \right\|^2 },
  \label{e:vpyx}
\end{equation}
where $\|\cdot\|$ denotes the Euclidean norm of a vector.  The MMSE in
estimating $\H\X$ is
\begin{equation}
  \mmse(\snr) = \expect{ \left\|
    \H \, \X - \H \, \hvX(\Y;\snr) \right\|^2 },
%     \H \, \left( \X - \hvX(\Y;\snr) \right) \right\|^2 },
%  \mmse(\snr,\H) = \min_{\hvX} \expect{ \left\|
%    \H \, \left( \X - \hvX(\Y;\snr) \right) \right\|^2 },
  \label{e:eh}
\end{equation}
where $\hvX(\Y;\snr)$ is the conditional mean estimate.  A
generalization of Theorem \ref{th:di} is the following:
\begin{theorem} \label{th:dv}
  Let $\N$ be a vector with independent standard Gaussian components,
  independent of $\X$.  For every $P_\X$ satisfying\;
  $\Exp\|\X\|^2<\infty$,
  \begin{equation}    \label{e:vie}
    \pd{\snr} I\left(\X;\sqrt{\snr}\,\H\,\X+\N\right) = \half \mmse(\snr).
  \end{equation}
\end{theorem}
\begin{proof}
  See Section \ref{s:ic}.
  % See Section \ref{s:sic}.%  See Appendix \ref{a:dv}.
\end{proof}
A verification of \eref{e:vie} in the special case of Gaussian
input with positive definite covariance matrix $\mSigma$ is
straightforward.  The covariance of the conditional mean estimation
error is
\begin{equation}
  \expect{ \left(\X-\hvX\right) \tran{\left(\X-\hvX\right)} }
  = \inv{\left( \inv{\mSigma}+\snr\HT\H \right)},
\end{equation}
from which one can calculate:
\begin{equation}
  \expect{ \left\| \H\left(\X-\hvX\right) \right\|^2 }
  = \trace{ \H \inv{\left( \inv{\mSigma}+\snr\HT\H \right)}\! \HT }.
\end{equation}
The mutual information is \cite{Verdu86Allerton}:
\begin{equation}
  I(\X;\Y) = \half \log\det \left(
  \I+\snr\mSigma^\half\HT\H\mSigma^\half\right),
\end{equation}
where $\mSigma^\half$ is the unique positive semi-definite symmetric
matrix such that $(\mSigma^\half)^2 =\mSigma$.  Clearly,
\begin{eqnarray}
  && \nind \pd{\snr} I(\X;\Y) \nn \\
  &=& \half\, \trace{ \inv{\left(
    \I+\snr\,\mSigma^\half\HT\H\mSigma^\half\right)}
  \mSigma^\half\HT\H\mSigma^\half } \\
%  &=& \half \tr\left\{ \inv{\left(\inv{\mSigma}
%    +\snr\HT\H\right)} \HT\H \right\} \\
  &=& \half \,\expect{ \left\| \H\left(\X-\hvX\right) \right\|^2 }.
%  &=& \half \expect{ \left\| \H\,\X-\H\,\hvX \right\|^2 }.
%  \!\!\! \pd{\snr} I(\X;\Y)
%  &\!\!\!=\!\!\!&
%  \tr\left\{ \inv{\left(\I+\snr\HT\H\right)} \HT\H \right\} \\
% &\!\!\!=\!\!\!& \expect{ \left\| \H\, \left(\X-\hvX\right) \right\|^2 }.
  \label{e:ir}
\end{eqnarray}
\begin{comment}
Note that in the special case of independent Gaussian inputs
($\mSigma=\I$), the MMSE in estimating $\H\,\X$ can also be written
as a function of the MMSE in estimating $\X$:  %%SV22Mar
\begin{equation}
  \expect{ \left\| \H\,\X-\H\,\hvX \right\|^2 } =
  \oneon{\snr} \left( K - \expect{ \left\| \X-\hvX \right\|^2 } \right).
  \label{e:kx}
\end{equation}
Equation (\ref{e:kx}) does not hold in general for inputs not
consisting of independent Gaussian entries.
\end{comment}

% SV- Should explain the independent input case better.  Appendix maybe?
% parametric calculation (see 11/23/03 email)

\subsection{Incremental Channels}
\label{s:ic}

% DG041018: reworded to accommodate discrete-time result
The central relationship given in Sections \ref{s:sc} and \ref{s:vc}
can be proved in various, rather different, ways.  The most
enlightening proof is by considering what we call an incremental
channel.  A proof of Theorem \ref{th:di} using the SNR-incremental
channel is given next, while its generalization to the vector version
is omitted but straightforward.  Alternative proofs are relegated to
later sections.

%  A proof of Theorem \ref{th:in} using the ``time-incremental channel'' is given in Section \ref{s:tic}.  
% Alternative proofs are discussed in Section \ref{s:ap} .

% DG041018: commented.  reason: discussed in Section 2.4 \ref{s:ap}
\begin{comment}
In fact, five proofs for Theorems
\ref{th:di} and \ref{th:dv} are given in this paper, including two
direct proofs by differentiating the mutual information and a related
divergence respectively, a proof through the de Bruijn identity, and a
proof taking advantage of results in the continuous-time domain.
\end{comment}

% \subsubsection{The SNR-incremental Channel}
% \label{s:sic}

The key to the incremental-channel approach is to reduce the proof of
the relationship for all SNRs to that for the special case of
vanishing SNR, a domain in which we can capitalize on the following
result:
\begin{lemma}  \label{lm:id}
  As $\delta\rightarrow0$, the input-output mutual information of the
  canonical Gaussian channel:
  \begin{equation}
    Y = \sqrt{\delta}\, Z + U,
  \end{equation}
  where $\Exp Z^2<\infty$ and $U\sim\mathcal{N} (0,1)$ is independent
  of $Z$, is given by
  \begin{equation}
    I(Y;Z) = \frac{\delta}{2} \, \Exp( Z - \Exp Z )^2 + o(\delta).
%    \expect{ \left( Z - \Exp Z \right)^2 } + o(\delta).
  \end{equation}
\end{lemma}

Essentially, Lemma \ref{lm:id} states that the mutual information is
half the SNR times the variance of the input at the vicinity of zero
SNR, but insensitive to the shape of the input distribution otherwise.
Lemma \ref{lm:id} has been given in \cite[Lemma 5.2.1]{LapSha02IT} and
\cite[Theorem 4]{Verdu02IT} (also implicitly in
\cite{Verdu90IT}).\footnote{A proof of Lemma \ref{lm:id} is given in
  Appendix \ref{a:id} for completeness.}  Lemma \ref{lm:id} is the
special case of Theorem \ref{th:di} at vanishing SNR, which, by means
of the incremental-channel method, can be bootstrapped to a proof of
Theorem \ref{th:di} for all SNRs.
% DG041019: A brief proof is given here for completeness.

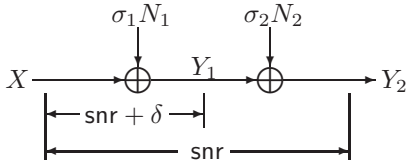
\begin{figure}
  \begin{center}
  \begin{picture}(150,60)(0,0)
    \put(0,27){$X$}
    \put(70,32){$Y_1$}
    \put(142,27){$Y_2$}
    \put(30,14){$\snr+\delta$}
    \put(70,0){$\snr$}
    \put(40,52){$\sigma_1 N_1$}
    \put(90,52){$\sigma_2 N_2$}
    \put(15,0){\line(0,1){25}}
    \put(75,12){\line(0,1){13}}
    \put(130,0){\line(0,1){25}}
    \put(50,50){\vector(0,-1){15}}
    \put(100,50){\vector(0,-1){15}}
    \put(10,30){\vector(1,0){35}}
    \put(55,30){\vector(1,0){40}}
    \put(105,30){\vector(1,0){35}}
    \put(50,30){\makebox(0,0){$\bigoplus$}}
    \put(100,30){\makebox(0,0){$\bigoplus$}}
    \put(28,17){\vector(-1,0){13}}
    \put(62,17){\vector(1,0){13}}
    \put(65,3){\vector(-1,0){50}}
    \put(85,3){\vector(1,0){45}}
  \end{picture}
\end{center}
\begin{comment}
  \begin{equation*}
    \begin{CD}
      & & \sigma_1 N_1 && \sigma_2 N_2 \\
      & & @VVV @VVV \\
      X @>>> \bigoplus @>{\displaystyle Y_1}>> \bigoplus @>>> Y_2
    \end{CD}
  \end{equation*}
\end{comment}
  \caption{An SNR-incremental Gaussian channel.}
  \label{f:ig}
\end{figure}

\begin{proof}[Theorem \ref{th:di}]
Fix arbitrary $\snr>0$ and $\delta>0$.  Consider a cascade of two
Gaussian channels as depicted in Figure \ref{f:ig}:
\begin{subequations}  \label{e:y12}%
\begin{eqnarray}
  Y_1 &=& X + \sigma_1 N_1, \label{e:y1} \\
  Y_2 &=& Y_1 + \sigma_2 N_2, \label{e:y2}
\end{eqnarray}
\end{subequations}
where $X$ is the input, and $N_1$ and $N_2$ are independent standard
Gaussian random variables.  Let $\sigma_1, \sigma_2 >0$ satisfy:
\begin{subequations}  \label{e:snr}
\begin{eqnarray}
  \sigma_1^2 &=& \oneon{\snr + \delta},  \\
  \sigma_1^2 + \sigma_2^2 &=& \oneon{\snr},
\end{eqnarray}
\end{subequations}
so that the signal-to-noise ratio of the first channel \eref{e:y1} is
$\snr+\delta$ and that of the composite channel is $\snr$.  Such a
system is referred to as an {\em SNR-incremental channel}\, since the
SNR increases by $\delta$ from $Y_2$ to $Y_1$.
% Note that we choose to
% scale the noise in Figure \ref{f:ig} for obvious reasons.

Theorem \ref{th:di} is equivalent to that, as $\delta \rightarrow0$,
\begin{eqnarray}
  I(X;Y_1)- I(X;Y_2)
  &=& I(\snr+\delta) - I(\snr) \\
  &=& \frac{\delta}{2} \, \mmse(\snr) + o(\delta).
  \label{e:ii}
\end{eqnarray}
% In the following we prove \eref{e:ii} for $\delta \rightarrow 0^+$ and
% the case of $\delta \rightarrow 0^-$ is trivial by swapping the roles
% of $Y_1$ and $Y_2$.  
Noting that $X$---$Y_1$---$Y_2$ is a Markov chain,
\begin{eqnarray}
  I(X;Y_1) - I(X;Y_2) &=& I(X;Y_1,Y_2) - I(X;Y_2)   \label{e:iii} \\
  &=& I( X; Y_1 | Y_2 ), \label{e:ix}
\end{eqnarray}
where \eref{e:ix} is the mutual information chain rule
\cite{CovTho91}.  
% Given $X$, the outputs $Y_1$ and $Y_2$ are jointly
% Gaussian.  Hence $Y_1$ is Gaussian conditioned on $X$ and $Y_2$.
A linear combination of \eref{e:y1} and \eref{e:y2} yields
\begin{eqnarray}
  %  (\snr+\delta)\, Y_1 = \snr \, Y_2 + \delta \, X + \sqrt{\delta} \, N
  (\snr+\delta)\, Y_1
  &=& \snr\, ( Y_2 - \sigma_2 N_2 ) + \delta\, ( X + \sigma_1 N_1 ) \\
  &=& \snr \, Y_2 + \delta \, X + \sqrt{\delta} \, N
  \label{e:yyxw}
\end{eqnarray}
where we have defined
\begin{equation}
  N = \oneon{\sqrt{\delta}} \, (
  \delta\, \sigma_1\, N_1 - \snr\, \sigma_2\, N_2 ). \label{e:nds}
\end{equation}
Clearly, the incremental channel \eref{e:y12} is equivalent to
\eref{e:yyxw} paired with \eref{e:y2}.  Due to \eref{e:snr} and mutual
independence of $(X,N_1,N_2)$, $N$ is a standard Gaussian random
variable independent of $X$.  Moreover, $(X,N,\sigma_1N_1
+\sigma_2N_2)$ are mutually independent since
\begin{equation}
  \expect{N (\sigma_1N_1+\sigma_2N_2)} = \oneon{\sqrt{\delta}} \, \left(
  \delta \, \sigma_1^2 - \snr \, \sigma_2^2 \right) = 0,
\end{equation}
also due to \eref{e:snr}.  Therefore $N$ is independent of $(X,Y_2)$
by \eref{e:y12}.  From \eref{e:yyxw}, it is clear that
\begin{eqnarray}
  %I(X;Y_1|Y_2=y_2) = I\left.\left(X; \delta\, X+\sqrt{\delta}\, N
  % \,\right|\, Y_2=y_2\right).
  && \nind I(X;Y_1|Y_2=y_2) \nn \\
  &=& I\left.\left(X; \snr\, Y_2 + \delta\, X+\sqrt{\delta}\, N \,\right|\, Y_2=y_2\right) \\
  &=& I\left.\left(X; \sqrt{\delta}\, X+ N \,\right|\, Y_2=y_2\right).
\end{eqnarray}
Hence given $Y_2=y_2$, \eref{e:yyxw} is equivalent to a Gaussian
channel with SNR equal to $\delta$ where the input distribution is
$P_{X|Y_2=y_2}$.  Applying Lemma \ref{lm:id} to such a channel
conditioned on $Y_2=y_2$, one obtains
\begin{equation}
  \begin{split}
    I(&X;Y_1|Y_2=y_2) = \\
    & \frac{\delta}{2} \, \expcnd{ \left( X -
        \expcnd{X}{Y_2=y_2} \right)^2 }{ Y_2=y_2 } + o(\delta). \label{e:ixyy}
  \end{split}
\end{equation}
Taking the expectation over $Y_2$ on both sides of \eref{e:ixyy}
yields
\begin{equation}
  I(X;Y_1|Y_2) = \frac{\delta}{2} \, \expect{ \left( X -
    \expcnd{X}{Y_2} \right)^2 } + o(\delta),
  \label{e:i12}
\end{equation}
which establishes \eref{e:ii} by \eref{e:ix} together with the fact
that
\begin{equation}
  \expect{ \left( X-\expcnd{X}{Y_2} \right)^2 } = \mmse(\snr).
\end{equation}
Hence the proof of Theorem \ref{th:di}.
\end{proof}

% \subsubsection{Mutual Information Chain Rule}
% \label{s:cr}

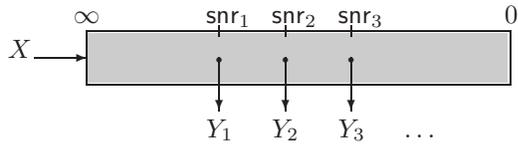
\begin{figure}
\begin{center}
\begin{picture}(200,45)(0,5)
  \put(0,30){$X$}
  \put(10,30){\vector(1,0){20}}
  \put(30,20){\framebox(160,20){}}
  \put(31,21){%
    \color{gray}%
    \begin{barenv}%
      \setnumberpos{empty}%
      \setwidth{158}%
      \bar{18}{8}[]
    \end{barenv}
  }
  \put(80,29){\circle*{2}}
  \put(105,29){\circle*{2}}
  \put(130,29){\circle*{2}}
  \put(80,30){\vector(0,-1){20}}
  \put(105,30){\vector(0,-1){20}}
  \put(130,30){\vector(0,-1){20}}
  \put(80,38){\line(0,1){4}}
  \put(105,38){\line(0,1){4}}
  \put(130,38){\line(0,1){4}}
  \put(25,43){$\infty$}
  \put(188,43){$0$}
%  \put(195,37){SNR}
%  \put(195,17){$\sigma^2$}
  \put(75,43){$\snr_1$}
  \put(100,43){$\snr_2$}
  \put(125,43){$\snr_3$}
%  \put(90,30){$\delta_1$}
%  \put(115,30){$\delta_2$}
  \put(75,0){$Y_1$}
  \put(100,0){$Y_2$}
  \put(125,0){$Y_3$}
  \put(150,0){$\dots$}
\end{picture}
\end{center}
  \caption{A Gaussian pipe where noise is added gradually.}
  \label{f:pipe}
\end{figure}

Underlying the incremental-channel proof of Theorem \ref{th:di} is the
chain rule for information:
\begin{equation}
  I(X;Y_1,\dots,Y_n) = \sum^n_{i=1} I(X;Y_i\,|\,Y_{i+1},\dots,Y_n).
  \label{e:cr}
\end{equation}
When $X$---$Y_1$---$\cdots$---$Y_n$ is a Markov chain, \eref{e:cr}
becomes
\begin{equation}
  I(X;Y_1) = \sum^n_{i=1} I(X;Y_i\,|\,Y_{i+1}),
  \label{e:ini}
\end{equation}
where we let $Y_{n+1}\equiv0$.  This applies to a train of outputs
tapped from a Gaussian pipe where noise is added gradually until the
SNR vanishes as depicted in Figure \ref{f:pipe}.  The sum in
\eref{e:ini} converges to an integral as $\{Y_i\}$ becomes a finer and
finer sequence of Gaussian channel outputs.  To see this note from
\eref{e:i12} that each conditional mutual information in \eref{e:ini}
corresponds to a low-SNR channel and is essentially proportional to
the MMSE times the SNR increment.  This viewpoint leads us to an
equivalent form of Theorem \ref{th:di}:
\begin{equation}
  I(\snr) = \half \int_0^\snr \mmse(\gamma) \,\intd \gamma.
  \label{e:iim}
\end{equation}
Therefore, as is illustrated by the curves in Figure \ref{f:sc}, the
mutual information is equal to an accumulation of the MMSE as a
function of the SNR due to the fact that an infinitesimal increase in
the SNR adds to the total mutual information an increase proportional
to the MMSE.

The infinite divisibility of Gaussian distributions, namely, the fact
that a Gaussian random variable can always be decomposed as the sum of
independent Gaussian random variables of smaller variances, is crucial
in establishing the incremental channel (or, the Markov chain).  This
property enables us to study the mutual information increase due to an
infinitesimal increase in the SNR, and thus obtain the differential
equations \eref{e:ie} and \eref{e:vie} in Theorems \ref{th:di} and
\ref{th:dv}.

The following corollaries are immediate from Theorem \ref{th:di}
together with the fact that $\mmse(\snr)$ is monotone decreasing.

\begin{corollary} \label{cr:cv}
  The mutual information $I(\snr)$ is a concave function in $\snr$.
\end{corollary}

\begin{corollary} \label{cr:im}
  The mutual information can be bounded as
  \begin{eqnarray}
    \expect{ \var{X|Y;\snr} } &=& \mmse(\snr) \\
    &\leq& \frac{2}{\snr} I(\snr) \\
    &\leq& \mmse(0) = \var{ X^2 }.
  \end{eqnarray}
  \begin{comment}
  \begin{equation}
    \begin{split}
      \expect{ \var{X|Y;\snr} } & = \mmse(\snr) \leq \frac{2}{\snr} I(\snr)\\
      & \leq \mmse(0) = \var{ X^2 }.
    \end{split}
  \end{equation}
  \end{comment}
\end{corollary}
% DG041018: rewording and new observation

% \subsection{Insights}
\subsection{Applications and Discussions}
\label{s:ad}

\begin{comment}
\subsubsection{Equivalent Forms of Theorem \ref{th:di}}

$Y=aX+N$

\begin{equation}
  \pd{a} I(a) = a \mmse(a).
\end{equation}

$Y=X+\sigma N$

\begin{equation}
  \pd{\sigma} I(\sigma) = \sigma^{-3} \mmse(\sigma).
\end{equation}
\end{comment}

\subsubsection{Some Applications of Theorems \ref{th:di} and \ref{th:dv}}

The newly discovered relationship between the mutual information and
MMSE finds one of its first uses in relating CDMA channel spectral
efficiencies (mutual information per dimension) under joint and
separate decoding in the large-system limit \cite{GuoVerITsub,
  Guo04PhD}.  Under an arbitrary finite-power input distribution,
Theorem \ref{th:di} is invoked in \cite{GuoVerITsub} to show that the
spectral efficiency under joint decoding is equal to the integral of
the spectral efficiency under separate decoding as a function of the
system load.  The practical lesson therein is the optimality in the
large-system limit of successive single-user decoding with
cancellation of interference from already decoded users, and an
individually optimal detection front end against yet undecoded users.
This is a generalization to arbitrary input signaling of previous
results that successive cancellation with a linear MMSE front end
achieves the CDMA channel capacity under Gaussian inputs
\cite{VarGue97Asilomar, VerSha99IT, GueVar04IT, Forney04Allerton}.

Relationships between information theory and estimation theory have
been identified occasionally, yielding results in one area taking
advantage of known results from the other.  This is exemplified by the
classical capacity-rate distortion relations, that have been used to
develop lower bounds on estimation errors \cite{ZakZiv72IT}.  The fact
that the mutual information and the MMSE determine each other by a
simple formula also provides a new means to calculate or bound one
quantity using the other.  An upper (resp.\ lower) bound for the
mutual information is immediate by bounding the MMSE for all SNRs
using a suboptimal (resp.\ genie aided) estimator.  Lower bounds on
the MMSE, e.g., \cite{BobZak76IT}, lead to new lower bounds on the
mutual information.
% \subsubsection{Extremality of Gaussian Inputs}

An important example of such relationships is the case of Gaussian
inputs.  Under power constraint, Gaussian inputs are most favorable
for Gaussian channels in information-theoretic sense (they maximize
the mutual information); on the other hand they are least favorable in
estimation-theoretic sense (they maximize the MMSE).  These well-known
results are seen to be immediately equivalent through Theorem
\ref{th:di} (or Theorem \ref{th:dv} for the vector case).  This also
points to a simple proof of the result that Gaussian inputs achieve
capacity by observing that the linear estimation upper bound for MMSE
is achieved for Gaussian inputs.\footnote{The observations here are
  also relevant to continuous-time Gaussian channels in view of
  results in Section \ref{s:ct}.}

Another application of the new results is in the analysis of
sparse-graph codes, where \cite{MeaUrb04ITW} has recently shown that
the so-called generalized extrinsic information transfer (GEXIT)
function plays a fundamental role.  This function is defined for
arbitrary codes and channels as minus the derivative of the
input-output mutual information per symbol with respect to a channel
quality parameter when the input is equiprobable on the codebook.
According to Theorem \ref{th:dv}, in the special case of the Gaussian
channel the GEXIT function is equal to minus one half of the average
MMSE of individual input symbols given the channel outputs.  Moreover,
\cite{MeaUrb04ITW} shows that \eref{e:pie} leads to a simple
interpretation of the ``area property'' for Gaussian channels (cf.\ 
\cite{AshKra04IT}).  Inspired by Theorem \ref{th:di},
\cite{BhaNar04Allerton} also advocated using the mean-square error as
the EXIT function for Gaussian channels.

\begin{comment}
More recently, an MSE-based transfer chart is proposed in light of
Theorem \ref{th:di} to analyze and design iterative decoding schemes
\cite{BhaNar04Allerton}.  Using Theorem \ref{th:di}, the ``area
property'' previously proved for erasure channels \cite{AshKra04IT} is
shown for Gaussian channels, which provides a theoretical
justification for designing good codes by matching the EXIT charts or
MSE charts when the log-likelihood ratios (LLR) are assumed to be
Gaussian \cite{BhaNar04Allerton}.

\end{comment}

As another application, the central theorems also provide an intuitive
proof of de Bruijn's identity as is shown next.

\subsubsection{De Bruijn's Identity}
\label{s:dbi}

An interesting insight is that Theorem \ref{th:dv} is equivalent to
the (multivariate) de Bruijn identity \cite{Stam59IC, Costa85IT}:
\begin{equation}
  \pd{t} h\left( \H\X+\sqrt{t}\,\N \right)
  = \half \trace{ \mat{J}\left( \H\X+\sqrt{t}\,\N \right) }
  \label{e:dbi}
\end{equation}
where $\N$ is a vector with independent standard Gaussian entries,
independent of $\X$.  Here, $h(\cdot)$ stands for the differential
entropy and $\mat{J} (\cdot)$ for Fisher's information matrix
\cite{Poor94}, which is defined as\footnote{The gradient operator can
  be written as $\nabla=\tran{\left[
      \frac{\partial}{\partial y_1}, \cdots, \frac{\partial}{\partial
        y_L} \right]}$ symbolically.  For any differentiable function $f:\;
  \mathbb{R}^L \rightarrow \mathbb{R}$, its gradient at any $\y$ is a column
  vector $\nabla f(\y)=\tran{\left[ \frac{\partial f}{\partial
        y_1}(\y), \cdots, \frac{\partial f}{\partial y_L}(\y)
    \right]}$.}
\begin{equation}
  \mat{J}(\y) = \expect{ \left[ \nabla \log p_\Y(\y) \right]
    \tran{\left[ \nabla \log p_\Y(\y) \right]} }.
  \label{e:dj}
\end{equation}
Let $\snr=1/t$ and $\Y = \sqrt{\snr} \, \H \, \X + \N$.  Then
\begin{equation}
  h\left( \H\X+\sqrt{t}\,\N \right)
  = I(\X;\Y) - \frac{L}{2} \log \frac{\snr}{2\pi e}.
  \label{e:hhi}
\end{equation}
%\begin{eqnarray}
%  h\left( \H\X+\sqrt{t}\,\N \right)
%  &=& h(\Y) - \frac{L}{2}\log\snr \\
%  &=& I(\X;\Y) - \frac{L}{2} \log \frac{\snr}{2\pi e}.
%  \label{e:hhi}
%\end{eqnarray}
%  h(\Y) = - \expect{ \log P_\Y(\Y) },
Meanwhile,
\begin{equation}
  \mat{J}\left( \H\X+\sqrt{t}\,\N \right)
  = \snr \, \mat{J}(\Y).
  \label{e:jy}
\end{equation}
\begin{comment}
\begin{eqnarray}
  \mat{J}\left( \H\X+\sqrt{t}\,\N \right)
  &=& \snr \, \mat{J}(\Y) \\
  &=& \snr \, \expect{ \left[ \ppd{\y} \log \vpY \right]
    \tran{\left[ \ppd{\y} \log \vpY \right]} }.
  \label{e:jy}
%  &=& \snr \, \expect{ \left[ \ppd{\y} \log p_\Y(\Y) \right]
%    \tran{\left[ \ppd{\y} \log p_\Y(\Y) \right]} }.
\end{eqnarray}
\begin{equation}
  \mat{J}(\Y) = \expect{ \left[ \ppd{\y} \log p_\Y(\Y) \right]
    \tran{\left[ \ppd{\y} \log p_\Y(\Y) \right]} }.
\end{equation}
\end{comment}
Note that
  \begin{equation}
    \vpy = \expect{ \vpyX },
    \label{e:vpy}
  \end{equation}
where $\vpyx$ is a Gaussian density \eref{e:vpyx}.  It can be
shown that
\begin{equation} \label{e:plug}
  \nabla \log \vpy = \sqrt{\snr} \, \H \hvX(\y;\snr) - \y.
\end{equation}
Plugging (\ref{e:plug}) into \eref{e:dj} and \eref{e:jy} gives
\begin{equation}
  \mat{J}(\Y) = \I - \snr\, \H\,\expect{\left(\X-\hvX\right)
    \tran{\left(\X-\hvX\right)}}\HT.
  \label{e:mj}
\end{equation}
Now de Bruijn's identity \eref{e:dbi} and Theorem \ref{th:dv} prove
each other by \eref{e:hhi} and \eref{e:mj}.  Noting this equivalence,
the incremental-channel approach offers an intuitive alternative to
the conventional technical proof of de Bruijn's identity obtained by
integrating by parts (e.g., \cite{CovTho91}).  Although equivalent to
de Bruijn's identity, Theorem \ref{th:dv} is important since mutual
information and MMSE are more canonical operational measures than
differential entropy and Fisher's information.

The Cram\'er-Rao bound states that the inverse of Fisher's information
is a lower bound on estimation accuracy.  The bound is tight for
Gaussian channels, where Fisher's information matrix and the
covariance of conditional mean estimation error determine each other
by \eref{e:mj}.  In particular, for a scalar channel,
\begin{equation}
  J\left( \sqrt{\snr}\, X+N \right)
  = 1 - \snr \cdot \mmse(\snr).
\end{equation}

\begin{comment}
Indeed, the differential entropy is a measure of the volume of the
typical set associated with a given probability density.  Thus its
derivative with respect to the variance of the Gaussian noise
perturbation is closely related to the surface area of the typical
set, which is the Fisher information.
\end{comment}

\subsubsection{Derivative of the Divergence}

Consider an input-output pair $(X,Y)$ connected through \eref{e:ch}.
The mutual information $I(X;Y)$ is the average value over the input
$X$ of the divergence $\divergence{ P_{Y|X=x} }{ P_Y }$.
%\begin{equation}
%  \divergence{ P_{Y|X=x} }{ P_Y } = \int \log \frac{
%    \intd P_{Y|X=x}(y) }{ \intd P_Y(y) } \,\intd P_{Y|X=x}(y).
%\end{equation}
Refining Theorem \ref{th:di}, it is possible to directly obtain the
derivative of the divergence given any value of the input:
\begin{theorem}  \label{th:cd}
  For every input distribution\, $P_X$ that satisfies\, $\Exp X^2
  <\infty$,
  \begin{equation}
    %    \pd{\snr} I(X;\sqrt{\snr}\cdot X+N\,|\,X=x) = \half \expcnd{
    % \pd{\snr} I(X;Y\,|\,X=x)
    \begin{split}
    \pd{\snr} \divergence{ P_{Y|X=x} }{ P_Y }
    =& \half \expcnd{ |X-X'|^2 }{X=x} \\
    & - \oneon{2\sqrt{\snr}} \expcnd{ X'\, N }{X=x},
    \label{e:cd}
    \end{split}
  \end{equation}
  where $X'$ is an auxiliary random variable which is independent
  identically distributed (i.i.d.) with $X$ conditioned on
  $Y=\sqrt{\snr}\,X+N$.
\end{theorem}

The auxiliary random variable $X'$ has an interesting physical
meaning.  It can be regarded as the output of the ``retrochannel''
\cite{GuoVerITsub, Guo04PhD}, which takes $Y$ as the input and
generates a random variable according to the posterior probability
distribution $p_{X|Y;\snr}$.  The joint distribution of $(X,Y,X')$ is
unique although the choice of $X'$ is not.

\begin{comment}
Using Theorem \ref{th:cd}, Theorem \ref{th:di} can be
recovered by taking expectation on both sides of \eref{e:cd}.  The
left hand side becomes the derivative of the mutual information.  The
right hand side becomes $1/2$ times the following:
\begin{eqnarray}
  \oneon{\sqrt{\snr}} \expect{ (X-X') (Y-\sqrt{\snr} X') }
  &=& \oneon{\sqrt{\snr}} \expect{XY-X'Y} + \expect{ (X')^2 - XX' } \\
  &=& \expect{ X^2 - \expcnd{XX'}{Y;\snr} } \label{e:xxy} \\
  &=& \expect{ X^2 - \left( \expcnd{X}{Y;\snr} \right)^2 }, \label{e:xyx}
\end{eqnarray}
which is the MMSE.  Note that \eref{e:xxy} and \eref{e:xyx} are since
$X'$ and $X$ are i.i.d.\ conditioned on $Y$.
\end{comment}

\subsubsection{Multiuser Channel}

A multiuser system in which users may transmit at different SNRs can
be modelled by:
\begin{equation}
%  \Y = \H \, \sqrt{\Snr} \, \X + \N
  \Y = \H \, \Snr \, \X + \N
  \label{e:kch}
\end{equation}
where $\H$ is deterministic $L\times K$ matrix known to the receiver,
$\Snr =\diag \{ \sqrt{\zSnr_1} ,\dots ,\sqrt{\zSnr_K}\}$ consists of
the square-root of the SNRs of the $K$ users, and $\N$ consists of
independent standard Gaussian entries.  The following theorem
addresses the derivative of the total mutual information with respect
to an individual user's SNR.
\begin{theorem} \label{th:k}
For every input distribution $P_\X$ that satisfies\; $\Exp\|\X\|^2
<\infty$,
\begin{equation}
  \begin{split}
  \ppd{\zSnr_k} & I(\X;\Y) \\
  =& \half
  \sum^K_{i=1} \sqrt{\frac{\zSnr_i}{\zSnr_k}} \, [\HT\H]_{ki} \,
  \expect{ \cov{X_k,X_i | \Y;\Snr} },
  \label{e:dk}
  \end{split}
\end{equation}
where $\cov{\cdot,\cdot |\cdot}$ denotes conditional covariance.
\end{theorem}

The proof of Theorem \ref{th:k} follows that of Theorem \ref{th:dv} in
Appendix \ref{a:dv} and is omitted.  Theorems \ref{th:di} and
\ref{th:dv} can be recovered from Theorem \ref{th:k} by setting
$\zSnr_k=\snr$ for all $k$.

\begin{comment}
Theorem \ref{th:di} can be recovered from Theorem \ref{th:k} by
setting $K=1$ and $\Snr=\sqrt{\snr}$, since
\begin{equation}
  \expect{ \cov{X,X | Y;\snr} } = \expect{ \var{X|Y;\snr} }
\end{equation}
is exactly the MMSE.  Theorem \ref{th:dv} can also be recovered by
letting $\zSnr_k = \snr$ for all $k$.
Then,
\begin{eqnarray}
  \pd{\snr} I(\X;\Y)
  &=& \sum^K_{k=1} \ppd{\zSnr_k} I(\X;\Y) \\
  &=& \half \sum^K_{k=1} \sum^K_{i=1} [\HT\H]_{ki}
    \, \expect{ \cov{X_k,X_i | \Y;\Snr} } \\
  &=& \half\, \expect{ \| \H\,\X- \H\,\expcnd{\X}{\Y;\Snr} \|^2 }.
%  &=& \half \Exp \| \H\, \left(\X-\expcnd{\X}{\Y;\Snr}\right) \|^2.
\end{eqnarray}
\end{comment}

\subsection{Alternative Proofs of Theorems \ref{th:di} and \ref{th:dv}}
\label{s:ap}

% DG041018: slight rewording of first sentence
%The incremental-channel proof of the central theorems given in Section
%\ref{s:ic} provides much information-theoretic insight into the
%results.  

In this subsection, we give an alternative proof of Theorem
\ref{th:dv}, which is based on the geometric properties of the
likelihood ratio between the output distribution and the noise
distribution.  This proof is a distilled version of the more general
result of Zakai \cite{Zakai04pre} (follow-up to this work) that uses
the Malliavin calculus and shows that the central relationship between
the mutual information and estimation error holds also in the abstract
Wiener space.  This alternative approach of Zakai makes use of
relationships between conditional mean estimation and likelihood
ratios due to Esposito \cite{Esposi68IC} and Hatsell and Nolte
\cite{HatNol71IT}.

As mentioned earlier, the central theorems also admit several other
proofs.  In fact, a third proof using the de Bruijn identity is
already evident in Section \ref{s:ad}.  A fourth proof of Theorems
\ref{th:di} and \ref{th:dv} by taking the derivative of the mutual
information is given in Appendices \ref{a:di} and \ref{a:dv}.  A fifth
proof taking advantage of results in the continuous-time domain is
relegated to Section \ref{s:ct}.
\begin{comment}
A fourth proof of
Theorems \ref{th:di} and \ref{th:dv} by taking the derivative of the
mutual information is given in Appendices \ref{a:di} and \ref{a:dv}
respectively.  A fifth proof taking advantage of results in the
continuous-time domain is relegated to Section \ref{s:dc}.
\end{comment}

It suffices to prove Theorem \ref{th:dv} assuming $\H$ to be the
identity matrix since one can always regard $\H\X$ as the input.  Let
$\Z=\sqrt{\snr}\,\X$.  Then the channel \eref{e:vch} is represented by
the canonical $L$-dimensional Gaussian channel:
\begin{equation} \label{e:yzn}
  \Y = \Z + \N.
\end{equation}
The mutual information, which is a conditional divergence, admits the
following decomposition \cite{Verdu90IT}:
\begin{eqnarray}
  I(\Y;\Z)
  &=& \cnddiv{P_{\Y|\Z}}{P_\Y}{P_\Z} \\
  &=& \cnddiv{P_{\Y|\Z}}{P_{\Y'}}{P_\Z} - \divergence{P_\Y}{P_{\Y'}}
  \label{e:svf}
\end{eqnarray}
\begin{comment}
\begin{equation}
  I(\Y;\Z) = \cnddiv{P_{\Y|\Z}}{P_\Y}{P_\Z}
  = \cnddiv{P_{\Y|\Z}}{P_{\Y'}}{P_\Z} - \divergence{P_\Y}{P_{\Y'}},
  \label{e:svf}
\end{equation}
\end{comment}
where $P_{\Y'}$ is an arbitrary distribution as long as the two
divergences on the right hand side of \eref{e:svf} are well-defined.
Choose $\Y'=\N$.  Then the mutual information can be expressed in
terms of the divergence between the unconditional output distribution
and the noise distribution:
\begin{equation}
  I(\Y;\Z) %&=& \cnddiv{P_{\Y|\Z}}{P_\N}{P_\Z} - \divergence{P_\Y}{P_\N} \\
  = \half \Exp \|\Z\|^2 - \divergence{P_\Y}{P_\N}.
\end{equation}
Hence Theorem \ref{th:dv} is equivalent to the following:
\begin{theorem} \label{th:dd}
  For every $P_\X$ satisfying\; $\Exp\|\X\|^2<\infty$,
  \begin{equation}
    \begin{split}
    \pd{\snr} & \divergence{P_{\sqrt{\snr}\,\X+\N}}{P_\N} \\
    & = \half
    \expect{ \left\| \expcnd{\X}{\sqrt{\snr}\,\X+\N} \right\|^2 }.
%    \pd{\snr} \divergence{P_{\sqrt{\snr}\,\H\X+\N}}{P_\N} = \half
%    \expect{ \left\| \expcnd{\H\X}{\sqrt{\snr}\,\H\X+\N} \right\|^2 }.
    \end{split}
  \end{equation}
\end{theorem}
Theorem \ref{th:dd} can be proved using geometric properties of the
likelihood ratio
\begin{equation}
  l(\y) = \frac{ p_\Y(\y) }{ p_\N(\y) }.
\end{equation}
The following lemmas are important steps.
\begin{lemma}[Esposito \cite{Esposi68IC}] \label{lm:esp}
  The gradient of the log-likelihood ratio gives the conditional mean
  estimate:
  \begin{equation} \label{e:esp}
    \nabla \log l(\y) = \expcnd{\Z}{\Y=\y}.
  \end{equation}
\end{lemma}
\begin{lemma}[Hatsell and Nolte \cite{HatNol71IT}] \label{lm:hn}
  The log-likelihood ratio satisfies Poisson's equation:\footnote{For
  any differentiable $\f:\; \mathbb{R}^L \rightarrow \mathbb{R}^L$,
  $\nabla \cdot \f = \sum^L_{l=1}
  \frac{\partial f_l}{\partial y_l}$.  If $\f$ is doubly
  differentiable, its Laplacian is defined as $\nabla^2 f=\nabla
  \cdot (\nabla f) =\sum^L_{l=1} \frac{\partial^2 f}{\partial y_l^2}$.}
  \begin{equation} \label{e:hn}
    \nabla^2 \log l(\y) =
    % \trace{ \cov{\Z,\Z\,|\,\Y=\y} }.
    \expcnd{\|\Z\|^2}{\Y=\y} - \left\| \expcnd{\Z}{\Y=\y} \right\|^2.
  \end{equation}
\end{lemma}

From Lemmas \ref{lm:esp} and \ref{lm:hn},
\begin{equation}
  \expcnd{\|\Z\|^2}{\Y=\y}
  = \nabla^2 \log l(\y) + \|\nabla \log l(\y)\|^2.
\end{equation}
The following result is immediate. 
\begin{comment}
\begin{eqnarray}
  && \nind \expcnd{\|\Z\|^2}{\Y=\y} \nn \\
  \nsp{2}&=&\nsp{2} \nabla^2 \log l(\y) + \|\nabla \log l(\y)\|^2 \\
  \nsp{2}&=&\nsp{2} l^{-2}(\y)
  \left[ l(y) \nabla^2 l(\y) - \|\nabla l(\y)\|^2
        + \|\nabla l(\y)\|^2 \right].
      %&=& \frac{ l(y) \nabla^2 \log l(\y) - \|\nabla l(\y)\|^2
       % + \|\nabla l(\y)\|^2 }{ l^2(\y) }.
\end{eqnarray}
Thus we have proved
\end{comment}
\begin{lemma} \label{lm:ell}
  \begin{equation} \label{e:n2l}
    \expcnd{\|\Z\|^2}{\Y=\y} = l^{-1}(\y) \nabla^2 l(\y).
    % \expcnd{\|\Z\|^2}{\Y=\y} = \frac{ \nabla^2 l(\y) }{ l(\y) }.
  \end{equation}
\end{lemma}
\begin{comment}
\begin{proof}
By \eref{e:esp} and \eref{e:hn},
\begin{eqnarray}
  \expcnd{\|\Z\|^2}{\Y=\y} 
  &=& \nabla^2 \log l(\Y) + \left\| \expcnd{\Z}{\Y=\y} \right\|^2 \\
  &=& \frac{ \nabla^2 l(\y) l(\y) - \|\nabla l(\y)\|^2 }{ l^2(\y) }
  + \left\| \nabla \log l(\y) \right\|^2 \\
  &=& \frac{ \nabla^2 l(\y) }{ l(\y) }.
\end{eqnarray}
\end{proof}
\end{comment}

A proof of Theorem \ref{th:dd} is obtained by taking the derivative
directly.

\begin{proof}[Theorem \ref{th:dd}]
  Note that the likelihood ratio can be expressed as
  \begin{eqnarray}
    l(\y) &=& \frac{ \expect{\vpyX} }{ p_\N(\y) } \\
    &=& \expect{ \expb{ \sqrt{\snr} \,\yT\X
        - \frac{\snr}{2} \|\X\|^2 } }.
  \end{eqnarray}
  Also, for any function $f(\cdot)$,
  \begin{equation}
    \begin{split}
      & \expect{ f(\X)\, \expb{ \sqrt{\snr} \,\yT\X
        - \frac{\snr}{2} \|\X\|^2 } } \\
    & \quad = l(\y) \, \expcnd{ f(\X) }{ \Y=\y }.
    \end{split}
  \end{equation}
  Hence,
  \begin{eqnarray}
    \pd{\snr} l(\y)
%    &=& \half \expect{ \left( \oneon{\sqrt{\snr}} \yT \X
%      - \|\X\|^2 \right) \, \expb{ \sqrt{\snr} \,\yT\X
%        - \frac{\snr}{2} \|\X\|^2 } } \\
    \nsp{1} &=& \nsp{1} \half l(\y) \biggl[ \oneon{\sqrt{\snr}} \yT\,\expcnd{\X}{\Y=\y} \nn \\
      && \qquad\qquad - \expcnd{\|\X\|^2}{\Y=\y} \biggr] \\
    \nsp{1} &=& \nsp{1} \oneon{2\snr} \left[ l(\y) \, \yT\, \nabla \log l(\y) -
      \nabla^2 \log l(\y) \right] \label{e:dsl}
  \end{eqnarray}
  % DG041018: added 'where ...' and slightly reworded
  where \eref{e:dsl} is due to Lemmas \ref{lm:esp} and \ref{lm:ell}.
  Note that the order of expectation with respect to $P_\X$ and the
  derivative with respect to the SNR can be exchanged as long as the
  input has finite power by Lebesgue's (Dominated) Convergence Theorem
  \cite{Royden88, Rudin76} (see also Lemma \ref{lm:ds} in Appendix
  \ref{a:dv}).

  The divergence can be written as
  \begin{eqnarray}
    \divergence{P_\Y}{P_\N}
    &=& \int p_\Y(\y) \log \frac{p_\Y(\y)}{p_\N(\y)} \,\intd \y \\
    &=& \expect{ l(\N) \log l(\N) },
  \end{eqnarray}
  and its derivative
  \begin{equation} \label{e:dsd}
    \pd{\snr} \divergence{P_\Y}{P_\N} = \expect{ \log l(\N) \pd{\snr}
    l(\N) }.
  \end{equation}
  Again, the order of derivative and expectation can be exchanged by
  the Lebesgue Convergence Theorem.  By \eref{e:dsl}, the derivative
  \eref{e:dsd} can be evaluated as
  \begin{eqnarray}
%    &=& \half \expect{ \log l(\N) \cdot l(\N) \left[
%       \frac{\tran{\N}}{\sqrt{\snr}}
%       \expcnd{\H\X}{\Y=\N} - \expcnd{\|\H\X\|^2}{\Y=\N} \right] } \\
    && \nind \oneon{2\snr}
    \expect{ l(\N)\log l(\N)\,\N\cdot\nabla\log l(\N) } \nn\\
    && \qquad \qquad
    - \oneon{2\snr} \expect{ \log l(\N) \, \nabla^2 l(\N) } \nn \\
    &=& \oneon{2\snr} \Exp \bigl\{ \nabla \cdot [ l(\N)\log l(\N) \nabla
      \log l(\N) ] \nn \\
      && \qquad\qquad - \log l(\N) \, \nabla^2 l(\N) \bigr\} \label{e:ndl} \\
%    &=& \oneon{2\snr} \expect{ \nabla \log l(\N) \nabla l(\N) + \log l(\N)
%      \nabla^2 l(\N) } - \oneon{2\snr} \expect{ \log l(\N) \nabla^2 l(\N) } \\
    &=& \oneon{2\snr} \expect{ l(\N) \,\| \nabla \log l(\N) \|^2 } \\
    &=& \oneon{2\snr} \Exp \left\| \nabla \log l(\Y) \right\|^2 \\
    &=& \half \Exp \| \expcnd{ \X }{\Y} \|^2,
  \end{eqnarray}
  where to write \eref{e:ndl} we used the following relationship
  (which can be checked by integration by parts) satisfied by a
  standard Gaussian vector $\N$:
  \begin{equation}
    \expect{ \tran{\N} \f(\N) } = \expect{ \nabla \cdot \f(\N) }
  \end{equation}
  % DG041018: added 'vector-valued function'
  for every vector-valued function $\f:$ $\mathbb{R}^L\rightarrow\mathbb{R}^L$
  that satisfies $f_i(\n) e^{-\half n_i^2} \rightarrow 0$ as
  $n_i\rightarrow \infty$, $i=1,\dots,L$.
\end{proof}

\subsection{Asymptotics of Mutual Information and MMSE}

In the following, the asymptotics of the mutual information and MMSE
at low and high SNRs are studied mainly for the scalar Gaussian
channel.

The Lebesgue Convergence Theorem guarantees continuity of the MMSE
estimate:
\begin{equation}
  \limzero{\snr} \expcnd{X}{Y;\snr} = \Exp X,
\end{equation}
and hence
\begin{equation}
  %  \limzero{\snr} \mmse(\snr) = \mmse(0) = \var X.
  \limzero{\snr} \mmse(\snr) = \mmse(0) = \sigma_X^2
  \label{e:lme}
\end{equation}
where $\sigma^2_{(\cdot)}$ denotes the variance of a random variable.
% DG041122: sigma_()
It has been shown in \cite{Verdu02IT} that symmetric (proper-complex
in the complex case) signaling is second-order optimal in terms of
mutual information for in the low SNR regime.

A more refined study of the asymptotics is possible by examining the
Taylor series expansion of a family of well-defined functions:
\begin{equation}
  \pyi{i} = \expect{ X^i p_{Y|X;\snr} (y \,|\, X;\snr ) },
  \; i=0,1,\dots
  \label{e:pyi}
\end{equation}
Clearly, $p_{Y;\snr}(y;\snr) = \pyi{0}$, and the conditional mean
estimate is expressed as
\begin{equation}  \label{e:x2}
  \expcnd{ X }{ Y=y;\snr } = \frac{ q_1(y;\snr) }{ q_0(y;\snr) }.
\end{equation}
Meanwhile, by definition \eref{e:ixy} and noting that $p_{Y|X;\snr}$
% the input-output conditional probability density function \eref{e:pyx}
is Gaussian, one has
\begin{equation} \label{e:ip0}
  I(\snr) = -\half \log(2\pi e) - \int \pyi{0} \log \pyi{0} \,\intd y.
\end{equation}
%\begin{equation} \label{e:p0}
%   p_{Y;\snr}(y;\snr) = \pyi{0}.
%\end{equation}
As $\snr\rightarrow0$,
\begin{equation}  \label{e:qiy}
  \begin{split}
    q_i(&y;\snr) \\
    =& \oneon{\sqrt{2\pi}} e^{-\frac{y^2}{2}} \, \Exp
    \biggl\{ X^i \, \biggl[ 1+Xy\snr^\half +\frac{X^2}{2}(y^2-1)\snr \\
    & + \frac{X^3}{6}(y^2-3)y\snr^{\frac{3}{2}}
    + \frac{X^4}{24}(y^4-6y^2+3)\snr^2 \\
    & + \frac{X^5}{120}(15y-10y^3+y^5) \snr^\frac{5}{2} \\
    & + \frac{X^6}{720}(y^6-15y^4+45y^2-15) \snr^3
     + \mathcal{O}(\snr^{\frac{7}{2}}) \biggr] \biggr\}.
%     + \mathcal{O}\left(\snr^{\frac{7}{2}}\right) \biggr] \biggr\}.
  \end{split}
  \begin{comment}
    q_i(&y;\snr) \\
    =& \oneon{\sqrt{2\pi}} e^{-\frac{y^2}{2}} \, \Exp
    \biggl\{ X^i \, \biggl[ 1+Xy\snr^\half +\half(y^2-1)X^2\snr \\
    & \; + \oneon{6}(y^2-3)yX^3\snr^{\frac{3}{2}}
    + \oneon{24}(y^4-6y^2+3)X^4\snr^2 \\
    & \; + \oneon{120}(15y-10y^3+y^5)X^5 \snr^\frac{5}{2} \\
    & \; + \oneon{720}(y^6-15y^4+45y^2-15)X^6 \snr^3
     + \mathcal{O}\left(\snr^{\frac{7}{2}}\right) \biggr] \biggr\}.
  \end{comment}
\end{equation}
Without loss of generality, it is assumed that the input $X$ has zero mean
and unit variance.  
Using \eref{e:x2}--\eref{e:qiy}, a finer characterization of the MMSE
and mutual information is obtained as
% \eref{e:lme} is obtained by definition \eref{e:ei} as
\begin{equation}
  \begin{split}
    \mmse(\snr) = & 1 - \snr + \snr^2
    - \oneon{6}\Bigl[ \left(\Exp X^4\right)^2 - 6 \Exp X^4 \\
    & \qquad - 2\left(\Exp X^3\right)^2 + 15 \Bigr] \snr^3
    + \mathcal{O}\left(\snr^4\right),
  \label{e:ez}
  \end{split}
\end{equation}
\begin{comment}
\begin{equation}
  \mmse(\snr) = 1 - \snr + \snr^2
  - \oneon{6}\left[ \left(\Exp X^4\right)^2 - 6 \Exp X^4
  - 2\left(\Exp X^3\right)^2 + 15 \right] \snr^3
  + \mathcal{O}\left(\snr^4\right),
  \label{e:ez}
\end{equation}
\end{comment}
% DG041018
% Note that the expression \eref{e:iz} for the mutual information can
% also be refined either by noting that
%The mutual information can be obtained either by noting that
%\begin{equation}
%  I(\snr) = -\half \log(2\pi e) - \expect{ \log q_0(Y;\snr) },
  % I(\snr) = -\half \log(2\pi e) - \expect{ \log p_{Y;\snr}(Y;\snr) },
%\end{equation}
% or integrating both sides of \eref{e:ez} and invoking Theorem \ref{th:di}:
%Using \eref{e:qiy},
and
\begin{equation}
  \begin{split}
    I(\snr) = & \half \snr - \oneon{4} \snr^2 + \oneon{6} \snr^3
    - \oneon{48}\Bigl[ \left(\Exp X^4\right)^2 \\
      & \; - 6 \Exp X^4 - 2\left(\Exp X^3\right)^2 + 15 \Bigr]
    \snr^4 + \mathcal{O}\left(\snr^5\right)
  \label{e:izf}
  \end{split}
\end{equation}
respectively.  It is interesting to note that that higher order
moments than the mean and variance have no impact on the mutual
information to the third order of the SNR.

%In view of the definition of higher-order optimality due to
%\cite{Verdu02IT}, all zero-mean distributions are second- and
%third-order optimal for real-valued Gaussian channels.

% DG041018: rewording
% The smoothness of the mutual information and MMSE carries over to the
% vector channel model \eref{e:vch} for finite-power inputs.  
% also have their counterparts.  
The asymptotic properties carry over to the vector channel model
\eref{e:vch} for finite-power inputs.  The MMSE of a real-valued
vector channel is obtained to the second order as:
\begin{equation}
  \begin{split}
  \mmse(\snr) = & \trace{\H\mSigma\HT} \\
  & \; - \snr\cdot \trace{ \H\mSigma\HT\H\mSigma\HT } +
  \mathcal{O}(\snr^2)
  \end{split}
\end{equation}
\begin{comment}
\begin{equation}
  \mmse(\snr)
  = \trace{\H\mSigma\HT}
  - \snr\cdot \trace{ \H\mSigma\HT\H\mSigma\HT } +
  %  - \snr\cdot \trace{ \left( \H\cov{\X}\tran{\H} \right)^2 } +
  \mathcal{O}(\snr^2)
\end{equation}
\end{comment}
where $\mSigma$ is the covariance matrix of the input vector.  The
input-output mutual information is straightforward by Theorem
\ref{th:dv} (see also \cite{PreVer04IT}).
% DG041019: commented
\begin{comment}
\begin{equation}
  I(\X;\sqrt{\snr}\,\H\,\X + \N)
  = \frac{\snr}{2} \trace{ \H\mSigma\HT } - \frac{\snr^2}{4}
  \trace{ \H\mSigma\HT\H\mSigma\HT } +
%  = \frac{\snr}{2} \trace{ \H\cov{\X}\tran{\H} } - \frac{\snr^2}{4}
%  \trace{ \left( \H\cov{\X}\tran{\H} \right)^2 } +
  \mathcal{O}(\snr^3).
  % o(\snr^2).
\end{equation}
\end{comment}
The asymptotics can be refined to any order of the SNR using the
Taylor series expansion.
% DG041018
% above analysis.

At high SNRs, the mutual information is upper bounded for
finite-alphabet inputs such as the binary one \eref{e:ib}, whereas it
can increase at the rate of $\half\log\snr$ for Gaussian inputs.  By
Shannon's entropy power inequality \cite{Shanno48BSTJ, CovTho91},
given any symmetric input distribution with a density, there exists an
$\alpha \in(0,1]$ such that the mutual information of the scalar
channel is bounded:
\begin{equation}
  \half \log( 1+\alpha\,\snr ) \le I(\snr) \le \half \log( 1+\snr ).
\end{equation}

The MMSE behavior at high SNR depends on the input distribution.  The
decay can be as slow as $\mathcal{O} (1/\snr)$ for Gaussian input,
whereas for binary input, the MMSE decays as $e^{-2\snr}$.  In fact,
the MMSE can be made to decay faster than any given exponential
for sufficiently skewed binary inputs \cite{Guo04PhD}.

%%%%%%%%%%%%%%%%%%%%%%%%%%%%%%%%%%%%%%%%%%%%%%%%%%%%%%%%%%%%

\section{Continuous-time Gaussian Channels}
\label{s:ct}

The success in the discrete-time Gaussian channel setting in Section
\ref{s:dt} can be extended to more technically challenging
continuous-time models.  Consider the following continuous-time
Gaussian channel:
\begin{equation}
  R_t = \sqrt{\snr} \, X_t + N_t, \quad t\in[0,T],
  \label{e:xn}
\end{equation}
where $\rp{X}$ is the input process, and $\rp{N}$ a white Gaussian
noise with a flat double-sided power spectrum density of unit height.
Since $\rp{N}$ is not second-order, it is mathematically more
convenient to study an equivalent model obtained by integrating the
observations in \eref{e:xn}. In a concise form, the input and output
processes are related by a standard Wiener process $\rp{W}$
independent of the input \cite{Oksend03, LipShi01I}:
\begin{equation}
  \intd Y_t = \sqrt{\snr} \, X_t\, \intd t + \intd W_t, \quad
  t\in[0,T].
  \label{e:ct}
\end{equation}
Also known as Brownian motion, $\rp{W}$ is a continuous Gaussian
process that satisfies
\begin{equation}
  \expect{ W_t W_s } = \min(t,s), \quad \forall t,s.
\end{equation}
Instead of scaling the Brownian motion (as is customary in the
literature), we choose to scale the input process so as to minimize
notation in the analysis and results.
\begin{comment}
% An example of the sample paths of the signals is shown in Figure \ref{f:ct}.  
  The additive Brownian motion model is fundamental in many
  applications and is central in many textbooks (see e.g.
  \cite{LipShi01I}).  Note that the continuous-time problem is almost
  always modelled by scaling the Wiener process rather than the input
  process as in~\eref{e:ct}.  This is one of the reasons that such a
  simple relationship as~\eref{e:ims} has never been noticed by
  previous authors.

\begin{figure}
  % Figure generated using /matlab/misc/plot_brownian.m
  \begin{center}
    \includegraphics[width=\myfigwidth]{brownian}
%    \includegraphics{brownian}
    \caption{Sample paths of the input, noise, and output processes
      associated with the model \eref{e:ct}.  The input $\rp{X}$ is a
      random telegraph waveform with unit transition rate.  The
      signal-to-noise ratio is 5 dB.}
  \label{f:ct}
  \end{center}
\end{figure}
\end{comment}

\subsection{Mutual Information and MMSEs}

We are concerned with three quantities associated with the model
\eref{e:ct}, namely, the causal MMSE achieved by optimal filtering,
the noncausal MMSE achieved by optimal smoothing, and the mutual
information between the input and output processes.  As a convention,
let $X_a^b$ denote the process $\rp{X}$ in the interval $[a,b]$.
Also, let $\mu_X$ denote the probability measure induced by $\rp{X}$
in the interval of interest, which, for concreteness we assume to be
$[0,T]$.  The input-output mutual information is defined by
\cite{Kolmog56IT, Pinske64}:
\begin{equation}
  I\left(X_0^T;Y_0^T\right) = \int \log \Phi \,\intd \mu_{XY}
%    I(Z_0^T;Y_0^T) = \int \log \frac{ \intd \mu_{YZ} }{ \intd \mu_Y
%      \intd \mu_Z } \,\intd \mu_{YZ}.
  \label{e:iT}
\end{equation}
if the Radon-Nikodym derivative
\begin{equation}
  \Phi = \frac{ \intd \mu_{XY} }{ \intd \mu_X \intd \mu_Y }
  \label{e:rnd}
\end{equation}
exists.  The causal and noncausal MMSEs at any time $t\in[0,T]$ are
defined in the usual way:
\begin{equation}
  \cmmse(t,\snr) = \expect{ \left( X_t - \expcnd{X_t}{Y_0^t;\snr} \right)^2 },
  \label{e:cmt}
\end{equation}
and
\begin{equation}
  \mmse(t,\snr) = \expect{ \left( X_t - \expcnd{X_t}{Y_0^T;\snr} \right)^2 }.
  \label{e:mt}
\end{equation}

Recall the {\em mutual information rate} (mutual information per unit
time) defined as:
\begin{equation}  \label{e:isnr}
  I(\snr) = \oneon{T} I\left( X^T_0;Y^T_0 \right).
\end{equation}
Similarly, the average causal and noncausal MMSEs (per unit time) are
defined as
\begin{equation} \label{e:cs}
  \cmmse(\snr) = \oneon{T} \int_0^T \cmmse(t,\snr) \, \intd t
% \expect{ \left( X_t - \expcnd{X_t}{Y^t_{-\infty};\snr} \right)^2 }, \label{e:cms}
\end{equation}
and
\begin{equation} \label{e:ms}
  \mmse(\snr) = \oneon{T} \int_0^T \mmse(t,\snr) \, \intd t
%  \mmse(\snr) = \expect{ \left( X_t - \expcnd{X_t}{Y^\infty_{-\infty};\snr} \right)^2 }
\end{equation}
respectively.

To start with, let $T\rightarrow\infty$ and assume that the input to
the continuous-time model \eref{e:ct} is a stationary\footnote{For
stationary input it would be more convenient to shift $[0,T]$ to
$[-T/2,T/2]$ and then let $T\rightarrow\infty$ so that the causal and
noncausal MMSEs at any time $t\in(-\infty,\infty)$ is independent of
$t$.  We stick to $[0,T]$ in this paper for notational simplicity in
case of general inputs.} Gaussian process with power spectrum
$S_X(\omega)$.  The mutual information rate was obtained by Shannon
\cite{Shanno49IRE}:
\begin{equation}
  I(\snr) = \half \int^\infty_{-\infty} \log \left(
    1+\snr \,S_X(\omega) \right) \frac{ \intd \omega }{2\pi}.
  \label{e:irs}
\end{equation}
With Gaussian input, both optimal filtering and smoothing are linear.
The noncausal MMSE is due to Wiener \cite{Wiener49},
\begin{equation}
  \mmse(\snr) = \int^\infty_{-\infty}
  \frac{S_X(\omega)}{1+\snr\, S_X(\omega)} \frac{ \intd \omega }{2\pi},
  \label{e:lm}
\end{equation}
and the causal MMSE is due to Yovits and Jackson \cite{YovJac55IRE}:
\begin{equation}
  \cmmse(\snr) = \oneon{\snr} \int^\infty_{-\infty} \log \left(
    1+\snr \,S_X(\omega) \right) \frac{ \intd \omega }{2\pi}.
    \label{e:lcm}
\end{equation}
From \eref{e:irs} and \eref{e:lm}, it is easy to see that the
derivative of the mutual information rate is equal to half the
noncausal MMSE, i.e., the central formula \eref{e:pie} holds literally
in this case.  Moreover, \eref{e:irs} and \eref{e:lcm} show that the
mutual information rate is equal to the causal MMSE scaled by half the
SNR, although, interestingly, this connection escaped Yovits and
Jackson \cite{YovJac55IRE}.

In fact, these relationships are true not only for Gaussian inputs.
% DG041019
%Theorem \ref{th:di} can be generalized to the continuous-time model
%with an arbitrary input process:
\begin{theorem}  \label{th:i}
  % DG041019 rewording
  For every input process $\rp{X}$ to the Gaussian channel \eref{e:ct}
  with finite average power, i.e.,
  \begin{equation}
    \int_0^T \Exp X^2_t \,\intd t < \infty,
    \label{e:tx}
  \end{equation}
  the input-output mutual information rate and the average noncausal
  MMSE are related by
  \begin{equation}
    \pd{\snr} I(\snr) = \half \mmse(\snr).
    \label{e:ims}
  \end{equation}
\end{theorem}
\begin{proof}
  See Section \ref{s:st}.
\end{proof}

% DG041019
% What is special for the continuous-time model is the relationship
% between the mutual information rate and the causal MMSE.
\begin{theorem}[Duncan \cite{Duncan70JAM}]  \label{th:dun}
  For any input process with finite average power,
  \begin{equation}
    I(\snr) = \frac{\snr}{2} \cmmse(\snr).
    %    \frac{\snr}{2T} \int_0^T \cmmse(t,\snr) \,\intd t.
    \label{e:dun}
  \end{equation}
\end{theorem}

Together, Theorems \ref{th:i} and \ref{th:dun} show that the mutual
information, the causal MMSE and the noncausal MMSE satisfy a triangle
relationship.  In particular, using the information rate as a bridge,
the causal MMSE is found to be equal to the noncausal MMSE averaged
over SNR.
\begin{theorem}  \label{th:e}
  For any input process with finite average power,
  \begin{equation}
    \cmmse(\snr) = \oneon{\snr} \int_0^\snr \mmse(\gamma) \,\intd \gamma.
%    \int_0^T \cmmse(t,\snr) \frac{\intd t}{T} = \int_0^\snr
%    \int_0^T \mmse(t,\gamma) \frac{\intd t}{T} \frac{\intd \gamma}{\snr}.
%    \int_0^T \cmmse(t,\snr) \,\intd t = \oneon{\snr} \int_0^\snr
%    \int_0^T \mmse(t,\gamma) \,\intd t\,\intd \gamma.
    \label{e:e}
  \end{equation}
\end{theorem}
Equality \eref{e:e} is a surprising fundamental relationship between
causal and noncausal MMSEs.  It is quite remarkable considering the
fact that nonlinear filtering is usually a hard problem and few
analytical expressions are known for the optimal estimation errors.

% DG041018: commented
\begin{comment}
Note that \eref{e:e} can be rewritten as
\begin{equation}
  \cmmse(\snr) - \mmse(\snr) = - \snr \pd{\snr} \cmmse(\snr),
  \label{e:dm}
\end{equation}
which quantifies the increase of the minimum estimation error due to
the causality constraint.  
\end{comment}
% DG041018: rewording
Although in general the optimal anti-causal filter is different from
the optimal causal filter, an interesting observation that follows
from Theorem \ref{th:e} is that for stationary inputs the average
anti-causal MMSE per unit time is equal to the average causal one.  To
see this, note that the average noncausal MMSE remains the same in
reversed time and that white Gaussian noise is reversible.
%% DG: shlomo seems to question the wording!!!

It is worth pointing out that Theorems \ref{th:i}--\ref{th:e} are
still valid if the time averages in \eref{e:isnr}--\eref{e:ms} are
replaced by their limits as $T\rightarrow\infty$.  This is
particularly relevant to the case of stationary inputs.

% DG041018 forward->causal  backward->anti-causal
\begin{figure}
  % Figure generated using /matlab/filtering/telegraph/tsttelegraph.m
  \begin{center}
    \includegraphics[width=\myfigwidth]{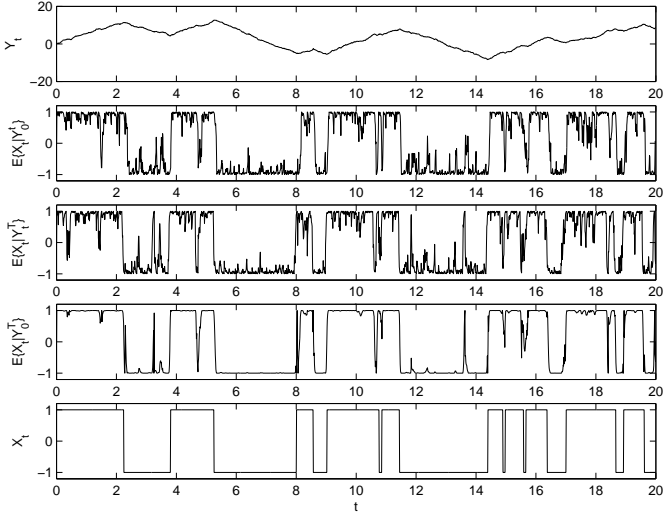}
    \caption{Sample path of the input and output processes of an
      additive white Gaussian noise channel, the output of the optimal
      causal and anti-causal filters, as well as the output of the
      optimal smoother.  The input $\rp{X}$ is a random telegraph
      waveform with unit transition rate.  The SNR is 15 dB.}
  \label{f:ct}
  \end{center}
\end{figure}

% DG081018: section title removed
% \subsubsection{Random Telegraph Input}

Besides Gaussian inputs, another example of the relation in Theorem
\ref{th:e} is an input process called the random telegraph waveform,
where $\rp{X}$ is a stationary Markov process with two equally
probable states ($X_t=\pm1$).  See Figure \ref{f:ct} for an
illustration.  Assume that the transition rate of the input Markov
process is $\nu$, i.e., for sufficiently small $h$,
\begin{equation}
  \Prob\{ X_{t+h}=X_t \} = 1 - \nu h + o(h),
\end{equation}
the expressions for the MMSEs achieved by optimal filtering and
smoothing are obtained as \cite{Wonham65JSC, Yao85IT}:
\begin{equation}
  \cmmse(\snr) = \frac{\int_1^\infty u^{-\half} (u-1)^{-\half}
    e^{-\frac{2\nu u}{\snr}} \, \intd u}{ \int_1^\infty u^\half
    (u-1)^{-\half} e^{-\frac{2\nu u}{\snr}} \,\intd u},
  \label{e:rtc}
\end{equation}
and
\begin{equation}
  \mmse(\snr) = \frac{ \int^1_{-1} \int^1_{-1} \frac{ (1+xy)
      \expb{ -\frac{2\nu}{\snr} \left( \oneon{1-x^2} + \oneon{1-y^2} \right) }
    }{-(1-x)^3(1-y)^3(1+x)(1+y)} \, \intd x \intd y }{
    \left[ \int_1^\infty u^\half
      (u-1)^{-\half} e^{-\frac{2\nu u}{\snr}} \,\intd u \right]^2}
  \label{e:rtn}
\end{equation}
% DG041019 refer to new appendix
respectively.  The relationship \eref{e:e} is verified in Appendix
\ref{a:rt}.  The MMSEs are plotted in Figure \ref{f:tg} as functions
of the SNR for unit transition rate.

\begin{figure}
  % Figure generated using /math/I-mmse/telegraph.nb
  \begin{center}
    \includegraphics[width=\myfigwidth]{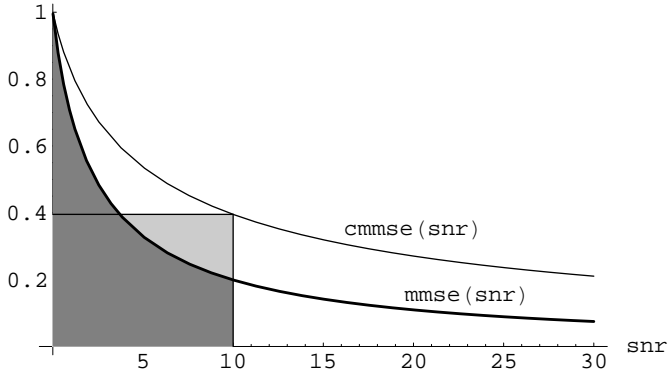}
    \caption{The causal and noncausal MMSEs of continuous-time
    Gaussian channel with the random telegraph waveform input.
    The rate $\nu=1$.  The two shaded regions have the same area
    due to Theorem \ref{th:e}.}
    \begin{comment}
    \caption{The mutual information (in nats) and causal and
    noncausal MMSEs of continuous-time Gaussian channel with
    random telegraph input.  The rate $\nu=1$.  The derivative of
    the mutual information, shown in thick dashed line, coincides
    with half the noncausal MMSE.}
    \end{comment}
  \label{f:tg}
  \end{center}
\end{figure}

Figure \ref{f:ct} shows experimental results of the filtering and
smoothing of the random telegraph signal corrupted by additive white
Gaussian noise.  The optimal causal filter follows Wonham
\cite{Wonham65JSC}:
\begin{equation}
  \begin{split}
  \intd \hX_t =
  & -\left[ 2\nu \hX_t + \snr \,\hX_t \left(1-\hX_t^2\right) \right] \intd t\\
  & \quad  + \sqrt{\snr} \,\left(1-\hX_t^2\right) \intd Y_t,
  \end{split}
%  \intd \hX_t = - \left[ 2\nu \hX_t + \snr \,\hX_t \left(1-\hX_t^2\right)
%    \right] \intd t + \sqrt{\snr} \,\left(1-\hX_t^2\right) \intd Y_t,
\end{equation}
where
\begin{equation}
  \hX_t = \expcnd{ X_t }{ Y_0^t }.
\end{equation}
The anti-causal filter is merely a time reversal of the filter of the
same type.  The smoother is due to Yao \cite{Yao85IT}:
\begin{equation}
  \expcnd{X_t}{Y_0^T} = \frac{ \expcnd{X_t}{Y_0^t} + \expcnd{X_t}{Y_t^T} }
  { 1 + \expcnd{X_t}{Y_0^t} \expcnd{X_t}{Y_t^T} }.
\end{equation}

\subsection{Low- and High-SNR Asymptotics}

Based on Theorem \ref{th:e}, one can study the asymptotics of the
mutual information and MMSE under low SNRs.  The causal and noncausal
MMSE relationship implies that
\begin{equation}
  \limzero{\snr} \frac{ \mmse(0) - \mmse(\snr) }{
    \cmmse(0) - \cmmse(\snr) } = 2
%  \mmse(0) - \mmse(\snr) = 2 [ \cmmse(0) - \cmmse(\snr)] + o(\snr)
  \label{e:f2}
\end{equation}
where
\begin{equation}
  \cmmse(0) = \mmse(0) = \oneon{T} \int_0^T \Exp X_t^2\, \intd t.
\end{equation}
% DG041019 rewording
Hence the initial rate of decrease (with $\snr$) of the noncausal MMSE is
twice that of the causal MMSE.

In the high-SNR regime, there exist inputs that make the MMSE
exponentially small.  However, in case of Gauss-Markov input
processes, Steinberg \etal\ \cite{SteBob01JAM} observed that the
causal MMSE is asymptotically twice the noncausal MMSE, as long as the
input-output relationship is described by
\begin{equation}
  \intd Y_t = \sqrt{\snr} \, h(X_t) \,\intd t + \intd W_t
\end{equation}
where $h(\cdot)$ is a differentiable and increasing function.  In the
special case where $h(X_t)=X_t$, Steinberg \etal's observation can be
justified by noting that in the Gauss-Markov case, the smoothing MMSE
satisfies \cite{BobZak82IT}:
\begin{equation}
  \mmse(\snr) = \frac{c}{\sqrt{\snr}} + o\left(\oneon{\snr}\right),
%  \mmse(\snr) = c \, \snr^{-\half} + o\left(\snr^{-1}\right),
\end{equation}
which implies according to \eref{e:e} that
\begin{equation}
  \liminfty{\snr} \frac{ \cmmse(\snr) }{ \mmse(\snr) } = 2.
%  \cmmse(\snr) = 2 \, c \, \snr^{-\half} + o\left(\snr^{-1}\right).
  \label{e:fs2}
\end{equation}
Unlike the universal factor of 2 result in \eref{e:f2} for the low SNR
regime, the 3 dB loss incurred by the causality constraint fails to
hold in general in the high SNR asymptote.  For example, for the
random telegraph waveform input, the causality penalty increases in
the order of $\log \snr$ \cite{Yao85IT}. % DG041019 ?????? why?

\begin{comment}
    {\bf SV- Need to give the most general result for a class of diffusions
that satisfy
\begin{equation}
  \liminfty{\snr} \frac{ \cmmse(\snr) }{ \mmse(\snr) } = 2.
\end{equation}
Bobrovski-Zakai, Steinberg ...
}
\end{comment}

\subsection{The SNR-Incremental Channel}
\label{s:st}

% DG041018: incremental channel figure removed (similar to the scalar case)
\begin{comment}
\begin{figure}
  \begin{center}
  \begin{picture}(150,60)(-15,0)
    \put(-15,27){$X_t\,\intd t$}
    \put(67,33){$\intd Y_{1t}$}
    \put(142,27){$\intd Y_{2t}$}
    \put(30,14){$\snr+\delta$}
    \put(70,0){$\snr$}
    \put(37,52){$\sigma_1 \intd W_{1t}$}
    \put(87,52){$\sigma_2 \intd W_{2t}$}
    \put(15,0){\line(0,1){25}}
    \put(75,12){\line(0,1){13}}
    \put(130,0){\line(0,1){25}}
    \put(50,50){\vector(0,-1){15}}
    \put(100,50){\vector(0,-1){15}}
    \put(12,30){\vector(1,0){33}}
    \put(55,30){\vector(1,0){40}}
    \put(105,30){\vector(1,0){35}}
    \put(50,30){\makebox(0,0){$\bigoplus$}}
    \put(100,30){\makebox(0,0){$\bigoplus$}}
    \put(28,17){\vector(-1,0){13}}
    \put(62,17){\vector(1,0){13}}
    \put(65,3){\vector(-1,0){50}}
    \put(85,3){\vector(1,0){45}}
  \end{picture}
\end{center}
  \caption{A continuous-time incremental Gaussian channel.}
  \label{f:igc}
\end{figure}
\end{comment}
\begin{comment}
  \begin{equation*}
    \begin{CD}
      & & \sigma_1 \, \intd W_{1t} && \sigma_2 \, \intd W_{2t} \\
      & & @VVV @VVV \\
      X_t \intd t @>>> \bigoplus @>{\displaystyle \intd
        Y_{1t}}>> \bigoplus @>>> \intd Y_{2t}
    \end{CD}
  \end{equation*}
\end{comment}

Theorem \ref{th:i} can be proved using the SNR-incremental channel
approach developed in Section \ref{s:dt}.  Consider a cascade of two
Gaussian channels with independent noise processes:
% DG041018: as depicted in Figure \ref{f:igc}:
\begin{subequations} \label{e:icc}
\begin{eqnarray}
  \intd Y_{1t} &=& X_t \, \intd t + \sigma_1\, \intd W_{1t},
    \label{e:y1t}\\
  \intd Y_{2t} &=& \intd Y_{1t} + \sigma_2 \, \intd W_{2t},
\end{eqnarray}
\end{subequations}
where $\{W_{1t}\}$ and $\{W_{2t}\}$ are independent standard Wiener
processes also independent of $\rp{X}$, and $\sigma_1$ and $\sigma_2$
satisfy \eref{e:snr} so that the signal-to-noise ratios of the first
channel and the composite channel are $\snr+\delta$ and $\snr$
respectively.
\begin{comment}
The following relation is satisfied:
\begin{equation}
  I(X^T_0;Y^T_{1,0}) - I(X^T_0;Y^T_{2,0})
  = I( X^T_0 ; Y^T_{1,0} | Y^T_{2,0} ).
  \label{e:iiic}
\end{equation}
\end{comment}
Following steps similar to those that lead to \eref{e:yyxw}, it can be
shown that
\begin{equation}
  (\snr+\delta) \, \intd Y_{1t} = \snr \, \intd Y_{2t} +
  \delta \, X_t \intd t + \sqrt{\delta} \, \intd W_t,
  \label{e:yyx}
\end{equation}
where $\rp{W}$ is a standard Wiener process independent of $\rp{X}$
and $\{Y_{2t}\}$.  Hence conditioned on the process $\{Y_{2t}\}$ in
$[0,T]$, \eref{e:yyx} can be regarded as a Gaussian channel with an
SNR of $\delta$.
%DN: Need to be very careful about what it means by conditioning on $Y_2$.
% It is in fact conditioning on the filtration generated by $\{Y_{2t}\}$.)
Similar to Lemma \ref{lm:id}, the following result holds.
\begin{lemma}  \label{lm:dx}
  As $\delta\rightarrow0$, the input-output mutual information of the
  following Gaussian channel:
  \begin{equation}
    \intd Y_t = \sqrt{\delta} \, Z_t \intd t + \intd W_t,
    \quad t\in[0,T],
  \end{equation}
  where $\rp{W}$ is standard Wiener process independent of the input
  $\rp{Z}$, which satisfies
  \begin{equation}
%    \oneon{T} \int_0^T \Exp Z_t^2 \,\intd t < \infty,
    \int_0^T \Exp Z_t^2 \,\intd t < \infty,
    \label{e:z2}
  \end{equation}
  is given by the following:
  \begin{equation}
    \limzero{\delta} \oneon{\delta}\, I\left(Z_0^T;Y_0^T\right)
    = \half \int_0^T \Exp \left( Z_t - \Exp Z_t \right)^2 \,\intd t.
    % I\left(Z_0^T;Y_0^T\right) = \frac{\delta}{2} \int_0^T \Exp
    % \left( Z_t - \Exp Z_t \right)^2 \,\intd t + o(\delta).
  \end{equation}
\end{lemma}
\begin{proof}
  See Appendix \ref{a:dx}.
\end{proof}
\begin{comment}
Essentially, the lemma states that regardless of the input
distribution, the mutual information is approximately equal to half
the input power times the signal-to-noise ratio (in this case
$\delta$) to the first order at the vicinity of zero signal-to-noise
ratio.
\end{comment}

Applying Lemma~\ref{lm:dx} to the Gaussian channel~\eref{e:yyx}
conditioned on $\{Y_{2t}\}$ in $[0,T]$, one has
\begin{equation}
  \begin{split}
  I & \left( X^T_0 ; Y^T_{1,0} | Y^T_{2,0} \right) \\
  & = \frac{\delta}{2} \int_0^T \expect{ \left( X_t -
    \expcnd{X_t}{Y^T_{2,0}} \right)^2 } \intd t + o(\delta).
  \end{split}
  \begin{comment}
  \frac{\delta}{2} \int_0^T \expect{ \left( X_t -
    \expcnd{X_t}{Y^T_{2,0}} \right)^2 } \intd t + o(\delta).
  \end{comment}
  \label{e:ido}
\end{equation}
Since $\rp{X}$---$\{Y_{1t}\}$---$\{Y_{2t}\}$ is a Markov chain, the
left hand side of~\eref{e:ido} is recognized as the mutual information
increase:
\begin{eqnarray}
  I\left( X^T_0 ; Y^T_{1,0} \,|\, Y^T_{2,0} \right)
  \nsp{1} &=&\nsp{1} I\left(X^T_0;Y^T_{1,0}\right)
  - I\left(X^T_0;Y^T_{2,0}\right) \\
  \nsp{1}&=&\nsp{1} T\,[I(\snr+\delta)-I(\snr)].
  \begin{comment}
  I(\snr+\delta)-I(\snr)
  &=& I(X^T_0;Y^T_{1,0}) - I(X^T_0;Y^T_{2,0}) \\
  &=& I( X^T_0 ; Y^T_{1,0} | Y^T_{2,0} ).
  \end{comment}
  \label{e:iiic}
\end{eqnarray}
By \eref{e:iiic} and definition of the noncausal MMSE \eref{e:mt},
\eref{e:ido} can be rewritten as
\begin{equation}
  I(\snr+\delta)-I(\snr)
  = \frac{\delta}{2T} \int_0^T \mmse(t,\snr) \,\intd t + o(\delta).
  \label{e:idm}
\end{equation}
Hence the proof of Theorem \ref{th:i}.

\begin{comment}
, it is clear that
\begin{equation}
  I(\snr+\delta)-I(\snr)
%  &=& I( X^T_0 ; Y^T_{1,0} | Y^T_{2,0} ) \\
  = \delta \, \int_0^T \mmse(t,\snr) \,\intd t + o(\delta).
%  \expect{ \left( X_t - \expcnd{X_t}{Y^\infty_{-\infty}} \right)^2 }
  \label{e:idm}
\end{equation}
Theorem \ref{th:i} is proved by noting that the mutual information
vanishes trivially at zero SNR.

In fact any noisy channel with independent increments property will be
favored by Theorem~\ref{th:i} or likewise, in particular Poisson
channels, which will be the subject of another paper.
\end{comment}

The property that independent Wiener processes sum up to a Wiener
process is essential in the above proof.  The incremental channel
device is very useful in proving integral equations such as in
Theorem \ref{th:i}.
% DG041019 the following is moved to gaussian pipe section
\begin{comment}
Indeed, by the SNR-incremental channel it has
been shown that the mutual information at a given SNR is an
accumulation of the MMSEs of degraded channels due to the fact that an
infinitesimal increase in the signal-to-noise ratio adds to the total
mutual information an increase proportional to the MMSE.
\end{comment}

\subsection{The Time-Incremental Channel}
\label{s:ti}

Note Duncan's Theorem (Theorem \ref{th:dun}) that links the mutual
information and the causal MMSE is also an integral equation, although
implicit, where the integral is with respect to time on the right hand
side of \eref{e:dun}.  Analogous to the SNR-incremental channel, one
can investigate the mutual information increase due to an
infinitesimal additional observation time of the channel output using
a ``time-incremental channel''.  This approach leads to a more
intuitive proof of Duncan's Theorem than the original one in
\cite{Duncan70JAM}, which relies on intricate properties of likelihood
ratios and stochastic calculus.

Duncan's Theorem is equivalent to
\begin{equation}
  \begin{split}
  I &\left( X_0^{t+\delta};Y_0^{t+\delta}\right)
  - I\left(X_0^t;Y_0^t\right) \\
  & = \delta\,\frac{\snr}{2} \,
  \expect{ \left( X_t - \expcnd{X_t}{Y_0^t} \right)^2 } + o(\delta),
  \end{split}
  \label{e:iids}
\end{equation}
which is to say the mutual information increase due to the extra
observation time is proportional to the causal MMSE.  The left hand
side of \eref{e:iids} can be written as
\begin{eqnarray}
  && \nind I\left(X_0^{t+\delta};Y_0^{t+\delta}\right)
  - I\left(X_0^t;Y_0^t\right) \nn \\
  &=& I\left(X_0^t,X_t^{t+\delta};Y_0^t,Y_t^{t+\delta}\right)
  - I\left(X_0^t;Y_0^t\right) \\
  &=& I\left( X_t^{t+\delta};Y_t^{t+\delta}\,|\,Y_0^t \right)
  + I\left( X_0^t;Y_t^{t+\delta} \,|\, X_t^{t+\delta},Y_0^t \right) \nn\\
  && \quad + I\left( X_0^t,X_t^{t+\delta};Y_0^t \right)
  - I\left( X_0^t;Y_0^t \right) \\
  &=& I\left( X_t^{t+\delta};Y_t^{t+\delta}\,|\,Y_0^t \right)
  + I\left( X_0^t;Y_t^{t+\delta} \,|\, X_t^{t+\delta},Y_0^t \right) \nn\\
  && \quad + I\left( X_t^{t+\delta};Y_0^t \,|\, X_0^t \right). \label{e:i0}
\end{eqnarray}
Since $Y_0^t$---$X_0^t$---$X_t^{t+\delta}$---$Y_t^{t+\delta}$ is a
Markov chain, the last two mutual informations in \eref{e:i0} vanish
due to conditional independence.  Therefore,
\begin{equation}  \label{e:ity}
  I\left(X_0^{t+\delta};Y_0^{t+\delta}\right)
  - I\left(X_0^t;Y_0^t\right)
  = I\left(X_t^{t+\delta};Y_t^{t+\delta} \,|\, Y_0^t\right),
\end{equation}
\begin{comment}
\begin{eqnarray}
  I(X_0^{t+\delta}\!\!\!\!\!\!&;&\!\!\!\!\!Y_0^{t+\delta})
  - I(X_0^t;Y_0^t) \nn \\
  &=& I(X_0^t,X_t^{t+\delta};Y_0^t,Y_t^{t+\delta}) - I(X_0^t;Y_0^t) \\
  &=& I(X_t^{t+\delta};Y_t^{t+\delta}|Y_0^t),
  \label{e:ity}
\end{eqnarray}
\end{comment}
i.e., the increase in the mutual information is the conditional mutual
information between the input and output during the extra time
interval given the past observation.  Note that conditioned on
$Y_0^t$, the probability law of the channel in $(t,t+\delta)$ remains
the same but with different input statistics due to conditioning on
$Y_0^t$.  Let us denote this new channel by
\begin{equation}
  \intd \tilde{Y}_t = \sqrt{\snr} \, \tilde{X}_t \intd t
  + \intd W_t, \quad t\in[0,\delta],
  \label{e:yzd}
\end{equation}
where the time duration is shifted to $[0,\delta]$, and the input
process $\tilde{X}_0^\delta$ has the same law as $X_t^{t+\delta}$
conditioned on $Y_0^t$.  Instead of looking at this new problem of an
infinitesimal time interval $[0,\delta]$, we can convert the problem
to a familiar one by an expansion in the time axis.  Since
$\sqrt{\delta}\, W_{t/\delta}$ is also a standard Wiener process, the
channel \eref{e:yzd} in $[0,\delta]$ is equivalent to a new channel
described by
\begin{equation}
  \intd \tilde{\tilde{Y}}_\tau = \sqrt{\delta\,\snr} \,
  \tilde{\tilde{X}}_\tau \intd \tau + \intd W'_\tau,  \quad \tau\in[0,1],
  \label{e:yzb}
\end{equation}
where $\tilde{\tilde{X}}_\tau =\tilde{X}_{\tau\delta}$, and $\rp{W'}$
is a standard Wiener process.  The channel \eref{e:yzb} is of (fixed)
unit duration but a diminishing signal-to-noise ratio of
$\delta\,\snr$.  It is interesting to note that the trick here
performs a ``time-SNR'' transform.  By Lemma \ref{lm:dx}, the mutual
information is
\begin{eqnarray}
  && \nind I \left(X_t^{t+\delta};Y_t^{t+\delta}|Y_0^t\right) \nn\\
  &=& I\left( \tilde{\tilde{X}}_0^1; \tilde{\tilde{Y}}_0^1 \right) \\
  &=& \frac{\delta\,\snr}{2} \int_0^1 \Exp(\tilde{\tilde{X}}_\tau
  - \Exp \tilde{\tilde{X}}_\tau)^2 \,\intd\tau + o(\delta) \\
  &=& \frac{\delta\,\snr}{2} \int_0^1 \expect{\left( X_{t+\tau\delta} -
    \expcnd{X_{t+\tau\delta}}{Y_0^t;\snr} \right)^2}\intd\tau \nn\\
        && \qquad + o(\delta) \\
  &=& \frac{\delta\,\snr}{2}
    \expect{\left( X_t - \expcnd{X_t}{Y_0^t;\snr} \right)^2} + o(\delta),
    \label{e:tdc}
\end{eqnarray}
where \eref{e:tdc} is justified by the continuity of the MMSE.  The
relation \eref{e:iids} is then established by \eref{e:ity} and
\eref{e:tdc}, and hence the proof of Duncan's Theorem.
%%% DG: justify continuity of the MMSE?

Similar to the discussion in Section \ref{s:ic}, the integral
equations in Theorems \ref{th:i} and \ref{th:dun} proved by using the
SNR- and time-incremental channels are also consequences of the mutual
information chain rule applied to a Markov chain of the channel input
and degraded versions of channel outputs.  The independent-increment
properties of Gaussian processes both SNR-wise and time-wise are
quintessential in establishing the results.

\subsection{A Fifth Proof of Theorem \ref{th:di}}
\label{s:5th}

A fifth proof of the mutual information and MMSE relation in the
random variable/vector model can be obtained using continuous-time
results.  For simplicity Theorem \ref{th:di} is proved using Theorem
\ref{th:dun}.  The proof can be easily modified to show Theorem
\ref{th:dv}, using the vector version of Duncan's Theorem
\cite{Duncan70JAM}.

A continuous-time counterpart of the model \eref{e:ch} can be
constructed by letting $X_t\equiv X$ for $t\in[0,1]$ where $X$ is a
random variable independent of $t$:
\begin{equation}
  \intd Y_t = \sqrt{\snr} \, X \,\intd t + \intd W_t.
  \label{e:ctx}
\end{equation}
For every $u\in[0,1]$, $Y_u$ is a sufficient statistic of the
observation $Y_0^u$ for $X$ (and $X_0^u$).  This is because that the
process $\{Y_t-(t/u)Y_u\}$, $t\in[0,u]$, is independent of $X$ (and
$X_0^u$).  Therefore, the input-output mutual information of the
scalar channel \eref{e:ch} is equal to the mutual information of the
continuous-time channel \eref{e:ctx}:
\begin{equation}
  I(\snr) = I(X;Y_1) = I\left(X_0^1;Y_0^1\right).
  \label{e:iei}
\end{equation}
Integrating both sides of \eref{e:ctx}, one has
\begin{equation}
  Y_u = \sqrt{\snr} \, u \, X + W_u, \quad u\in[0,1],
  \label{e:yu}
\end{equation}
where $W_u\sim \mathcal{N} (0,u)$.  Note that \eref{e:yu} is a scalar
Gaussian channel with a time-varying SNR which grows linearly from 0
to $\snr$.  Due to the sufficiency of $Y_u$, the MMSE of the
continuous-time model given the observation $Y_0^u$, i.e., the causal
MMSE at time $u$, is equal to the MMSE of a scalar Gaussian channel
with an SNR of $u\,\snr$:
\begin{equation} \label{e:cuu}
  \cmmse(u,\snr) = \mmse(u\,\snr).
\end{equation}
By Duncan's Theorem, the mutual information can be written as
\begin{eqnarray}
  I(X_0^1;Y_0^1)
  &=& \frac{\snr}{2} \int_0^1 \cmmse(u,\snr) \,\intd u \\
  &=& \frac{\snr}{2} \int_0^1 \mmse(u\,\snr) \,\intd u \\
  &=& \half \int_0^\snr \mmse(\gamma) \,\intd \gamma.
\end{eqnarray}
Thus Theorem \ref{th:di} follows by also noticing \eref{e:iei}.

Note that for constant input applied to a continuous-time Gaussian
channel, the noncausal MMSE at any time $t$ \eref{e:mt} is equal to
the MMSE of a scalar channel with the same SNR:
\begin{equation} \label{e:utu}
  \mmse(t,u\,\snr) = \mmse(u\,\snr), \quad \forall t\in[0,T].
\end{equation}
Together, \eref{e:cuu} and \eref{e:utu} yield \eref{e:e} for constant
input by averaging over time $u$.  Indeed, during any observation time
interval of the continuous-time channel output, the SNR of the desired
signal against noise is accumulated over time.  The integral over time
and the integral over SNR are interchangeable in this case.  This is
another example of the ``time-SNR'' transform which appeared in
Section \ref{s:ti}.
% DG041019: rewording
\begin{comment}
Indeed, for an observation time duration $[0,u]$ of the
continuous-time channel output, the corresponding signal-to-noise
ratio is $u\,\snr$ in the equivalent scalar channel model; or in
other words, 
\end{comment}

Regarding the above proof, note that the constant input can be
replaced by a general form of $X\, h(t)$, where $h(t)$ is a
deterministic signal.

%%%%%%%%%%%%%%%%%%%%%%%%%%%%%%%%%%%%%%%%%%%%%%%%%%%%%%%%%%%%

\section{Discrete-time Gaussian Channels}
\label{s:dc}

\subsection{Mutual Information and MMSE}

Consider a real-valued discrete-time Gaussian-noise channel of the
form
\begin{equation}
  Y_i = \sqrt{\snr} \, X_i + N_i, \quad i=1,2,\dots,
  \label{e:chn}
\end{equation}
where the noise $\{N_i\}$ is a sequence of independent standard
Gaussian random variables, independent of the input process $\{X_i\}$.
Let the input statistics be fixed and not dependent on $\snr$.

\begin{comment}
  Given that we have already treated the case of a finite-dimensional
  vector channel, an advantageous analysis of \eref{e:chn} consists of
  treating the finite-horizon case $i=1,\dots,n$ and then taking the
  limit as $n\rightarrow\infty$.
\end{comment}

The finite-horizon version of \eref{e:chn} corresponds to the vector
channel \eref{e:vch} with $\H$ being the identity matrix.  Let $\X^n=
\tran{[ X_1,\dots, X_n]}$, $\Y^n= \tran{[ Y_1,\dots, Y_n]}$, and
$\N^n= \tran{[ N_1,\dots, N_n]}$.  The relation \eref{e:vie} between
the mutual information and the MMSE holds due to Theorem \ref{th:dv}.
\begin{corollary}  \label{cr:dn}
  If\, $\sum^n_{i=1} \Exp X_i^2<\infty$, then
  \begin{equation} \label{e:dn}
    \pd{\snr} I\left( \X^n; \sqrt{\snr}\, \X^n+\N^n \right)
    = \half \sum^n_{i=1} \mmse(i,\snr),
%    I(X^L;Y^n) = \int_0^\snr \sum^n_{i=1}
%    \mmse(\gamma,i,n) \,\intd \gamma,
  \end{equation}
  where
  \begin{equation} \label{e:ei}
    \mmse(i,\snr) = \expect{ \left( X_i
      - \expcnd{X_i}{\Y^n;\snr} \right)^2 }
  \end{equation}
  is the noncausal MMSE at time $i$ given the entire observation
  $\Y^n$.
\end{corollary}

It is also interesting to consider optimal filtering and prediction in
this setting.  Denote the filtering MMSE as
\begin{equation}
%  \cmmse(i,\snr) = \Exp{ \left[ X_i
%    - \expcnd{X_i}{\Y^i;\snr} \right]^2 },
  \cmmse(i,\snr) = \expect{ \left( X_i
    - \expcnd{X_i}{\Y^i;\snr} \right)^2 },
\end{equation}
and one-step prediction MMSE as
\begin{equation}
  \pmmse(i,\snr) = \expect{ \left( X_i
    - \expcnd{X_i}{\Y^{i-1};\snr} \right)^2 }.
%  \pmmse(i,\snr) = \Exp{ \left[ X_i
%    - \expcnd{X_i}{\Y^{i-1};\snr} \right]^2 }.
\end{equation}
% We have the following result:
\begin{theorem}  \label{th:in}
  The input-output mutual information satisfies:
  \begin{subequations}    \label{e:in}
    \begin{eqnarray}
%    \sum_{i=1}^n \cmmse(i,\snr) \leq \frac{I(\X^n;\Y^n)}{\snr}
%    \leq \sum_{i=1}^n \pmmse(i,\snr).
      \frac{\snr}{2} \sum_{i=1}^n \cmmse(i,\snr)
      \nsp{1} &\leq& \nsp{1} I\left(\X^n;\Y^n\right) \\
      \nsp{1} &\leq& \nsp{1} \frac{\snr}{2} \sum_{i=1}^n \pmmse(i,\snr).
    \end{eqnarray}
  \end{subequations}
  \begin{comment}
  \begin{equation}
    \begin{split}
      \snr \sum_{i=1}^n \cmmse(i,\snr) & \leq I(\X^n;\Y^n) \\
    & \leq \snr \sum_{i=1}^n \pmmse(i,\snr).
    \end{split}
  \end{equation}
  \end{comment}
\end{theorem}

\begin{proof}
  We study the increase in the mutual information due to an extra
  sample of observation by considering a conceptual time-incremental
  channel.  Since $\Y^i$---$\X^i$---$X_{i+1}$---$Y_{i+1}$ is a Markov
  chain, the mutual information increase is equal to
  \begin{equation}
    I\left(\X^{i+1};\Y^{i+1}\right) - I\left(\X^i;\Y^i\right)
    = I\left(X_{i+1};Y_{i+1} \,|\, \Y^i\right)
    \label{e:din}
  \end{equation}
  using an argument similar to the one that leads to \eref{e:ity}.
  This conditional mutual information can be regarded as the
  input-output mutual information of the simple scalar channel
  \eref{e:ch} where the input distribution is replaced by the
  conditional distribution $P_{X_{i+1}|\Y^i}$.  By Corollary
  \ref{cr:im},
  \begin{eqnarray}
    \nind \expect{ \var{X_{i+1}|\Y^{i+1};\snr} }
    \nsp{1}&\leq&\nsp{1}
    \frac{2}{\snr} I\left(X_{i+1};Y_{i+1} \,|\, \Y^i\right) \\
    \nsp{1}&\leq&\nsp{1} \expect{ \var{X_{i+1}|\Y^i;\snr} },
  \end{eqnarray}
  or equivalently,
  \begin{equation} \label{e:inn}
    \frac{\snr}{2} \cmmse(i,\snr) 
    \leq I\left(X_{i+1};Y_{i+1} \,|\, \Y^i\right)
    \leq \frac{\snr}{2} \pmmse(i,\snr).
  \end{equation}
  Finally, we obtain the desired bounds in Theorem \ref{th:in} summing
  \eref{e:inn} over $n$ and using \eref{e:din}.
\end{proof}

Corollary \ref{cr:dn} and Theorem \ref{th:in} are still valid if all
sides are normalized by $n$ and we then take the limit of
$n\rightarrow \infty$.  As a result, the derivative of the mutual
information rate (average mutual information per sample) is equal to
half the average noncausal MMSE per symbol.  Also, the mutual
information rate is sandwiched between half the SNR times the average
causal and prediction MMSEs per symbol.

\subsection{Discrete-time vs.\ Continuous-time}

% DG041018: rewording
In previous sections, the mutual information and the estimation errors
have been shown to satisfy similar relations in both discrete- and
continuous-time random process models.  Indeed, discrete-time
processes and continuous-time processes are related fundamentally.
For example, discrete-time process can be regarded as the result of
integrate-and-dump sampling of the continuous-time one.

It is straightforward to recover the discrete-time results using the
continuous-time ones by considering an equivalent of the discrete-time
model \eref{e:chn} as a continuous-time one with piecewise constant
input:
\begin{equation} \label{e:pct}
  \intd Y_t = \sqrt{\snr} \, X_{\lceil t\rceil}\, \intd t
  + \intd W_t, \quad t\in[0,\infty).
\end{equation}
During the time interval $(i-1,i]$ the input to the continuous-time
model is equal to the random variable $X_i$.  The samples of $\rp{Y}$
at natural numbers are sufficient statistics for the input process
$\{X_n\}$.  Thus, Corollary \ref{cr:dn} follows directly from Theorem
\ref{th:i}.  Analogously, Duncan's Theorem can be used to prove
Theorem \ref{th:in} \cite{Guo04PhD}.

Conversely, for sufficiently smooth input processes, the
continuous-time results (Theorem \ref{th:i} and Duncan's Theorem) can
be derived from the discrete-time ones (Corollary \ref{cr:dn} and
Theorem \ref{th:in}).  This can be accomplished by sampling the
continuous-time channel outputs and taking the limit of all sides of
\eref{e:in} with vanishing sampling interval.  However, in their full
generality, the continuous-time results are not a simple extension of
the discrete-time ones.  A complete analysis of the continuous-time
model involves stochastic calculus as developed in Section \ref{s:ct}.

\section{Generalizations}
\label{s:g}

\subsection{General Additive-noise Channel}
\label{s:pi}

% DG041018: rewording
% Theorems \ref{th:di} and \ref{th:dv} 
%Sections \ref{s:dt} and \ref{s:ct} show the relationship between the
%mutual information and the MMSE as long as the signal is observed
%embedded in Gaussian noise.  Let us now 

\begin{figure}
  \begin{center}
    \begin{picture}(180,75)(0,5)
      \put(0,32){$X$}
%      \put(0,20){$\sim P_X$}
      \put(12,35){\vector(1,0){23}}
      \put(35,25){\framebox(30,20){$P_{Z|X}$}}
      \put(75,37){$Z$}
      \put(65,35){\vector(1,0){30}}
      \put(100,35){\makebox(0,0){$\bigotimes$}}
      \put(85,5){ $\sqrt{\snr}$ }
      \put(100,15){\vector(0,1){15}}
      \put(105,35){\vector(1,0){25}}
      \put(135,35){\makebox(0,0){$\bigoplus$}}
      \put(135,55){\vector(0,-1){15}}
      \put(130,60){$N$} %  \put(112,60){$\N\sim\mathcal{N}(0,\I)$}
      \put(140,35){\vector(1,0){25}}
      \put(167,32){$Y$}
    \end{picture}
  \end{center}
  \caption{General additive-noise channel.}
  \label{f:pp}
\end{figure}
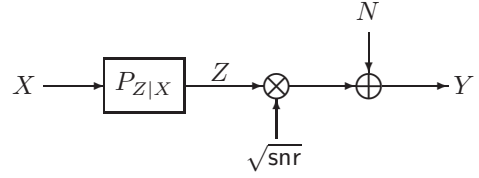

Consider a general setting where the input is preprocessed arbitrarily
before contamination by additive Gaussian noise.  The scalar channel
setting as depicted in Figure \ref{f:pp} is first considered for
simplicity.

Let $X$ be a random object jointly distributed with a real-valued
random variable $Z$.  The channel output is expressed as
\begin{equation}
  Y = \sqrt{\snr} \, Z + N,
  \label{e:yz}
\end{equation}
where the noise $N\sim\mathcal{N}(0,1)$ is independent of $X$ and $Z$.
The preprocessor can be regarded as a channel with arbitrary
conditional probability distribution $P_{Z|X}$.  Since $X$---$Z$---$Y$
is a Markov chain,
\begin{equation}
\label{e:chr}
  I(X;Y) = I(Z;Y) - I(Z;Y\,|\,X).
\end{equation}
Note that given $(X,Z)$, the channel output $Y$ is Gaussian.  Two
applications of Theorem \ref{th:di} to the right hand side of
\eref{e:chr} give the following:
\begin{theorem}  \label{th:pp}
  Let $X$---$Z$---$Y$ be a Markov chain and $Y=\sqrt {\snr} Z+N$.
  If\, $\Exp Z^2<\infty$, then
  \begin{equation}
    \begin{split}
    \pd{\snr} I(X;Y) =&
    \half \expect{ \left( Z - \expcnd{Z}{Y;\snr} \right)^2 } \\
    & \; - \half \expect{ \left( Z - \expcnd{Z}{Y,X;\snr} \right)^2 }.
    \end{split}
    \label{e:pp}
  \end{equation}
\end{theorem}

The special case of this result for vanishing SNR is given by Theorem
1 of \cite{PreVer04IT}.  As a simple illustration of Theorem
\ref{th:pp}, consider a scalar channel where $X\sim\mathcal {N}\left
  (0,\sigma_X^2\right)$ and $P_{Z|X}$ is a Gaussian channel with noise
variance $\sigma^2$.  Then straightforward calculations yield
\begin{equation}
  I(X;Y) = \half \log \left(
  1 + \frac{\snr\,\sigma_X^2}{1+\snr\,\sigma^2} \right),
  % \sqrt{\snr}\, Z+N
\end{equation}
the derivative of which is equal to half the difference of the two
MMSEs:
\begin{equation}
  \half \left[ \frac{ \sigma_X^2 + \sigma^2 }{ 1 + \snr \left(
    \sigma_X^2+\sigma^2\right) }
  - \frac{ \sigma^2 }{ 1 + \snr\, \sigma^2 } \right].
\end{equation}

% DG041018: rewording
\begin{comment}
and
\begin{eqnarray}
  \expect{ \left( Z - \expcnd{Z}{Y;\snr} \right)^2 }
  &=& \frac{ \sigma_X^2 + \sigma^2 }{ 1 + \snr \left(
    \sigma_X^2+\sigma^2\right) },\\
  \expect{ \left( Z - \expcnd{Z}{Y,X;\snr} \right)^2 }
  &=& \frac{ \sigma^2 }{ 1 + \snr\, \sigma^2 }.
\end{eqnarray}
The relationship \eref{e:pp} is easy to check.
\end{comment}

In the special case where the preprocessor is a deterministic function
of the input, e.g., $Z=g(X)$ where $g(\cdot)$ is an arbitrary
deterministic mapping, the second term on the right hand side of
\eref{e:pp} vanishes.  If, furthermore, $g(\cdot)$ is a one-to-one
transformation, then $I(X;Y)=I(g(X);Y)$, and
\begin{equation}
  \begin{split}
    \pd{\snr} & I(X;\sqrt{\snr}\, g(X)+N) \\
    & = \half \expect{ \left( g(X) - \expcnd{g(X)}{Y;\snr} \right)^2 }.
  \label{e:pg}
  \end{split}
\end{equation}
Hence \eref{e:ie} holds verbatim where the MMSE in this case is
defined as the minimum error in estimating $g(X)$.  Indeed, the vector
channel in Theorem \ref{th:dv} is merely a special case of the vector
version of this general result.
%%%%%%%%%%%%%%%%%%%%SVMAR30

One of the many scenarios in which the general result can be useful is
the intersymbol interference channel.  The input $Z_i$ to the Gaussian
channel is the desired symbol $X_i$ corrupted by a function of the
previous symbols $(X_{i-1},X_{i-2},\dots)$.  Theorem \ref{th:pp} can
possibly be used to calculate (or bound) the mutual information given
a certain input distribution.  Another domain of applications of
Theorem \ref{th:pp} is the case of fading channels known or unknown at
the receiver, e.g., the channel input $Z=AX$ where $A$ is the
multiplicative fading coefficient.
%%%SVMar31: make the point of noncoherent case

\begin{comment}
% DG041019
Using similar arguments as in the above, nothing prevents us from
generalizing the continuous-time results in Section \ref{s:ct} to a
much broader family of models:
\end{comment}
Using similar arguments as in the above, nothing prevents us from
generalizing Theorem \ref{th:i} to a much broader family of models:
\begin{equation}
  \intd Y_t = \sqrt{\snr} \, Z_t \,\intd t + \intd W_t,
  \label{e:ht}
\end{equation}
where $\rp{Z}$ is a random process jointly distributed with $X$, and
$\rp{W}$ is a Wiener process independent of $X$ and $\rp{Z}$.
\begin{theorem}  \label{th:ppc}
  As long as the input $\rp{Z}$ to the channel \eref{e:ht} has finite
  average power,
  \begin{equation}
    \begin{split}
      \pd{\snr} I & \left(X;Y_0^T\right)
      = \oneon{2T} \int_0^T
      \expect{ \left( Z_t - \expcnd{Z_t}{Y_0^T;\snr} \right)^2 } \\
      & -
      \expect{ \left( Z_t - \expcnd{Z_t}{Y_0^T,X;\snr} \right)^2 } \,\intd t.
    \end{split}
    \label{e:ppc}
  \end{equation}
\end{theorem}
In case $Z_t=g_t(X)$, where $g_t(\cdot)$ is an arbitrary deterministic
one-to-one time-varying mapping, Theorems \ref{th:i}-\ref{th:e} hold
verbatim except that the finite-power requirement now applies to
$g_t(X)$, and the MMSEs in this case refer to the minimum errors in
estimating $g_t(X)$.

\subsection{Gaussian Channels With Feedback}
\label{s:fb}

Duncan's Theorem (Theorem \ref{th:dun}) can be generalized to the
continuous-time additive white Gaussian noise channel with feedback
\cite{KadZak71IT}:
\begin{equation}
  \intd Y_t = \sqrt{\snr} \, Z\left(t,Y_0^t,X\right) \,\intd t
  + \intd W_t, \quad t\in[0,T],
  \label{e:fbc}
\end{equation}
where $X$ is any random message (including a random process indexed by
$t$) and the channel input $\rp{Z}$ is dependent on the message and
past output only.  
% DG041019: Like in Duncan's Theorem, 
The input-output mutual information of this channel with feedback can
be expressed as the time-average of the optimal filtering
mean-square error: %%%%%%%%%%%%%%%%SVMar30
\begin{theorem}[Kadota, Zakai and Ziv \cite{KadZak71IT}]
  \label{th:kzz}
  If the power of the input $\rp{Z}$ to the channel \eref{e:fbc} is
  finite, then
  \begin{equation}
    \begin{split}
    I\left(X;Y_0^T\right) = & \frac{\snr}{2} \int_0^T
    \Exp \Bigl\{ \Bigl( Z\left(t,Y_0^t,X\right) \\
        & \; - \expcnd{Z\left(t,Y_0^t,X\right)}{Y_0^t;\snr} \Bigr)^2 \Bigr\} \,\intd t.
    \end{split}
    \label{e:kzz}
  \end{equation}
\end{theorem}
Theorem \ref{th:kzz} is proved by showing that Duncan's proof of
Theorem \ref{th:dun} remains essentially intact as long as the channel
input at any time is independent of the future noise process
\cite{KadZak71IT}.  A new proof can be conceived by considering the
time-incremental channel, for which \eref{e:ity} holds literally.
Naturally, the counterpart of the discrete-time result (Theorem
\ref{th:in}) in the presence of feedback is
also feasible.  %%%%%%%%%%%%%%%%%SVMar30

One is tempted to also generalize the relationship between the mutual
information and smoothing error (Theorem \ref{th:i}) to channels with
feedback.  Unfortunately, it is not possible to construct a meaningful
SNR-incremental channel like \eref{e:icc} in this case, since changing
the SNR affects not only the amount of Gaussian noise, but also the
statistics of the feedback, and consequently the transmitted signal
itself.  We give two examples to show that in general the derivative
of the mutual information with respect to the SNR has no direct
connection to the noncausal MMSE, and, in particular,
\begin{equation} \label{e:fbn}
  \begin{split}
  \pd{\snr} I\left(X;Y_0^T\right) \neq & \half \int_0^T
  \Exp \Bigl\{ \bigl( Z\left(t,Y_0^t,X\right) \\
  -& \expcnd{Z\left(t,Y_0^t,X\right)}{Y_0^T;\snr} \bigr)^2 \Bigr\} \,\intd t.
  \end{split}
\end{equation}

% DG041019: Shlomo's suggestion seems not good since we scale the
% signal instead of the noise--calculating the quadratic variations
% of the received process in an arbitrarily short period of time
% doesn't tell the SNR unless the scaling is on the noise.
Having access to feedback allows the transmitter to determine the SNR
as accurate as desired by transmitting known signals and observing the
realization of the output for long enough.\footnote{The same technique
  applies to discrete-time channels with feedback.  If instead the
  received signal is in the form of $\intd Y_t= Z_t(t,Y_0^t,X) \intd t
  + \left(1/\sqrt{\snr}\right) \intd W_t$, then the SNR can also be
  determined by computing the quadratic variation of $Y_t$ during an
  arbitrarily small interval.}  Once the SNR is known, one can choose
a pathological signaling:
\begin{comment}
Having access to feedback allows the transmitter to determine the SNR
as accurate as it wishes by transmitting known signals and observing
the realization of the output during any time interval.\footnote{The
  SNR can be estimated to arbitrary precision in discrete-time
  channels with feedback, provided that the observation interval is
  long enough.}  Once the SNR is known, one can choose a pathological
signaling, e.g.,
\end{comment}
\begin{equation}
   Z(t, Y_0^t, X) = X / \sqrt{\snr}.
\end{equation}
Clearly the output of channel \eref{e:fbc} remains the same regardless
of the SNR.  Hence the mutual information has zero derivative, while
the MMSE is nonzero.  In fact, one can choose to encode the SNR in the
channel input in such a way that the derivative of the mutual
information is arbitrary (e.g., negative).
\begin{comment}
\begin{equation}
   Z(t, Y_0^t, X) =
   \begin{cases}
     X & \quad \text{if}\;\; \snr<\gamma, \\
     0 & \quad \text{otherwise},
     \end{cases}
\end{equation}
where $\gamma>0$ is an arbitrary threshold.  Clearly, the mutual
information is nonzero for $\snr<\gamma$ but vanishes for
$snr\geq \gamma$.  One cannot hope its derivative to be positive (or
even exists at all).  Hence \eref{e:fbn}.  
\end{comment}

%%% SVMar31 %%%
The same conclusion can be drawn from an alternative viewpoint by
noting that feedback can help to achieve essentially symbol error-free
communication at channel capacity by using a signaling specially
tailored for the SNR, e.g., capacity-achieving error-control codes.
More interesting is the variable-duration modulation scheme of Turin
\cite{Turin65IT} for the infinite-bandwidth continuous-time Gaussian
channel, where the capacity-achieving input is an explicit
deterministic function of the message and the feedback.  From this
scheme, we can derive a suboptimal noncausal estimator of the channel
input by appending the encoder at the output of the decoder.  Since
arbitrarily low block error rate can be achieved by the coding scheme
of \cite{Turin65IT} and the channel input has bounded power, the
smoothing MMSE achieved by the suboptimal noncausal estimator can be
made as small as desired.  On the other hand, achieving channel
capacity requires that the mutual information be nonnegligible.

%%% Shlomo: good ECC????

Note that a fundamental proviso for our mutual information-MMSE
relationship is that the input distribution not be allowed to depend
on SNR.  However, in general, feedback removes such restrictions.

\begin{comment}
Apparently, feedback allows the transmission to adapt to the SNR so
that it does not make sense in general to investigate the change of
mutual information due to change in the SNR.
\end{comment}

%%% DGMar31.  A new subsection. %%%
%%% It seems nicer to bring all vector generalization together. %%%
\subsection{Generalization to Vector Models}

Just as Theorem \ref{th:di} obtained under a scalar model has its
counterpart (Theorem \ref{th:dv}) under a vector model, all the
results in Sections \ref{s:ct} and \ref{s:dc} can be generalized to
vector models, under either discrete- or the continuous-time setting.
For example, the vector continuous-time model takes the form of
\begin{equation}
  \intd \Y_t = \sqrt{\snr} \, \X_t \,\intd t + \intd \W_t,
  \label{e:vt}
\end{equation}
where $\rp{\W}$ is an $m$-dimensional Wiener process, and $\rp{\X}$
and $\rp{\Y}$ are $m$-dimensional random processes.  Theorem
\ref{th:i} holds literally, while the mutual information rate,
estimation errors, and power are now defined with respect to the
vector signals and their Euclidean norms.  In fact, Duncan's Theorem
was originally given in vector form \cite{Duncan70JAM}.  It should be
noted that the incremental-channel devices are directly applicable to
the vector models.

In view of the above generalizations, the discrete- and
continuous-time results in Sections \ref{s:pi} and \ref{s:fb} also
extend straightforwardly to vector models.

% DG041019: Needless to say, 
Furthermore, colored additive Gaussian noise can be treated by first
filtering the observation to whiten the noise and recover the
canonical model of the form \eref{e:yz}.

\subsection{Complex-valued Channels}

The results in the discrete-time regime (Theorems
\ref{th:di}--\ref{th:dd} and Corollaries \ref{cr:cv}--\ref{cr:dn})
hold verbatim for complex-valued channel and signaling if the noise
samples are i.i.d.\ circularly symmetric complex Gaussian, whose real
and imaginary components have unit variance.  In particular, the
factor of 1/2 in \eref{e:ie}, \eref{e:vie}, \eref{e:dn} and
\eref{e:in} remains intact.  However, with the more common definition
of $\snr$ in complex-valued channels where the complex noise has real
and imaginary components with variance 1/2 each, the factor of 1/2 in
the formulas disappears.

The above principle holds also under continuous-time models as long as
the complex-valued Wiener process is appropriately defined.  This is
straightforward by noting that in general complex-valued models can be
regarded as two independent uses of the real-valued ones (with
possibly correlated inputs in the two uses).

\begin{comment}
Following the arguments in \cite{KadZak71IT}, Theorem \ref{th:i} can
also be generalized to channels with feedback:
\begin{theorem}
  If the input process $\rp{Z}$ to the channel \eref{e:fbc} has finite
  power, then
  \begin{equation}
    \pd{\snr} I\left(X;Y_0^T\right) = \half \int_0^T
    \expect{ \left( Z\left(t,Y_0^t,X\right)
      - \expcnd{Z\left(t,Y_0^t,X\right)}{Y_0^T;\snr} \right)^2 } \,\intd t.
  \end{equation}
\end{theorem}
Note that
\begin{equation}
  I\left( X;Y_0^T \right) = I\left( Z_0^T;Y_0^T \right)
\end{equation}
due to the fact that $Z_0^T$ is a function of $X$ and that $X$
interacts with $Y_0^T$ through $Z_0^T$.  The above generalizations
hinge on the fact that the proofs of Theorems \ref{th:i} and
\ref{th:dun} remain essentially intact as long as the input process to
the channel at any time is independent of the future realization of
the noise.

Similar arguments allow generalizations in the discrete-time model.
Furthermore, we have also found a result encompassing
a more general model encompassing as special cases both
those of Sections \ref{s:pi} and \ref{s:fb}.  %%SV22Mar
\end{comment}

\section{New Representation of Information Measures}
\label{s:im}

The relationship between mutual information and MMSE enables other
information measures such as entropy and divergence to be expressed as
a function of MMSE as well.

Consider a discrete random variable $X$.  Assume $N\sim \mathcal{N}
(0,1)$ independent of the input throughout this section.  The mutual
information between $X$ and its observation through a Gaussian channel
converges to the entropy of $X$ as the SNR of the channel goes to
infinity.
\begin{lemma} \label{lm:hi}
  For every discrete real-valued random variable $X$,
  % and independent $N\sim \mathcal{N} (0,1)$,
  \begin{equation} \label{e:hi}
    H(X) = \liminfty{\snr} I\left( X; \sqrt{\snr}\,X+N \right).
  \end{equation}
\end{lemma}
\begin{proof}
See Appendix \ref{a:hi}.
\end{proof}
Note that if $H(X)$ is infinity then the mutual information in
\eref{e:hi} also increases without bound as $\snr\rightarrow\infty$.
Moreover, the result holds if $X$ is subject to an arbitrary
one-to-one mapping $g(\cdot)$ before going through the channel.  In
view of \eref{e:pg} and \eref{e:hi}, the following theorem is
immediate.
\begin{theorem} \label{th:H}
  For any discrete random variable $X$ taking values in $\mathcal{A}$,
  the entropy of $X$ is given by (in nats)
  \begin{comment}
  \begin{equation} \label{e:Hx}
      H(X) = \half \int_0^\infty \Exp \Bigl\{ \bigl( g(X)
      - \expcnd{g(X)}{\sqrt{\snr}\, g(X)+N} \bigr)^2 \Bigr\} \intd \snr
  \end{equation}
  \begin{equation} \label{e:Hx}
    \begin{split}
      H(X) = \half & \int_0^\infty \Exp \Bigl\{ \bigl( g(X) \\
      & - \expcnd{g(X)}{\sqrt{\snr}\, g(X)+N} \bigr)^2 \Bigr\} \,\intd \snr
%    H(X) = \half \int_0^\infty \expect{ \left( g(X) - \expcnd{g(X)}{
%      \sqrt{\snr}\, g(X)+N} \right)^2 } \,\intd \snr
    \end{split}
  \end{equation}
  \end{comment}
  \begin{equation} \label{e:Hx}
    \begin{split}
      H&(X) \\
      =& \half  \int_0^\infty \Exp \Bigl\{ \bigl( g(X)
       - \expcnd{g(X)}{\sqrt{\snr}\, g(X)+N} \bigr)^2 \Bigr\} \intd \snr
    \end{split}
  \end{equation}
  for any one-to-one mapping $g:\; \mathcal{A} \rightarrow \mathbb{R}$.
\end{theorem}

It is interesting to note that the integral on the right hand side of
\eref{e:Hx} is not dependent on the choice of $g(\cdot)$, which is not
evident from estimation-theoretic properties alone.

\begin{comment}
Other than for discrete random variables, the entropy is not defined
and the input-output mutual information is in general unbounded as SNR
increases.  One may consider the divergence between the input
distribution and a Gaussian distribution with the same mean and
variance.
\end{comment}

The ``non-Gaussianness'' of a random variable (divergence between its
distribution and a Gaussian distribution with the same mean and
variance) and, thus, the differential entropy can also be written in
terms of MMSE. To that end, we need the following auxiliary result.

\begin{lemma} \label{lm:de}
  Let $X$ be any real-valued random variable and $X'$ be Gaussian with
  the same mean and variance as $X$, i.e., $X'\sim \mathcal{N} \left(
    \Exp X, \sigma_X^2 \right)$.  Let $Y$ and $Y'$ be the output of
  the channel \eref{e:ch} with $X$ and $X'$ as the input respectively.
  Then
  \begin{equation} \label{e:de}
        \divergence{ P_X }{ P_{X'} }
        = \liminfty{\snr} \divergence{ P_Y }{ P_{Y'} }.
  \end{equation}
\end{lemma}
% DG041018: use {proof}
\begin{proof}
% The lemma can be proved using 
  By monotone convergence and the fact that data processing reduces
  divergence.
\end{proof}
Note that in case the divergence between $P_X$ and $P_{X'}$ is
infinity, the divergence between $P_Y$ and $P_{Y'}$ also increases
without bound.  Since
\begin{equation}
  \divergence{ P_Y }{ P_{Y'} } = I(X';Y') - I(X;Y),
\end{equation}
the following result is straightforward by Theorem \ref{th:di}.
\begin{theorem} \label{th:dh}
  For every random variable $X$ with $\sigma_X^2<\infty$, its
  non-Gaussianness is given by
  \begin{eqnarray} \label{e:dh}
    &&\nind D_X = \divergence{ P_X }{ \mathcal{N}( \Exp X, \sigma_X^2 ) } \\
    &=& \nsp{2} \half \int_0^\infty \nsp{2} \frac{\sigma_X^2}{1+\snr\sigma_X^2}
    - \mmse\left(X|\sqrt{\snr}X\nsp{.5}+\nsp{.5}N\right) \nsp{.5} \intd \snr.
  \end{eqnarray}
  \begin{comment}
  \begin{eqnarray} \label{e:dh}
    D_X &=& \divergence{ P_X }{ \mathcal{N}( \Exp X, \sigma_X^2 ) } \\
    &=& \half \int_0^\infty \frac{\sigma_X^2}{1+\snr\,\sigma_X^2} \nn \\
    && \qquad- \mmse\left(X|\sqrt{\snr}\,X+N\right) \, \intd \snr.
  \end{eqnarray}
  \end{comment}
\end{theorem}
Note that the integrand in \eref{e:dh} is always positive since for
the same variance, Gaussian inputs maximize the MMSE.  Also, Theorem
\ref{th:dh} holds even if the divergence is infinity, for example in
the case that $X$ is a discrete random variable.  In light of Theorem
\ref{th:dh}, the differential entropy of $X$ can be expressed as:
\begin{eqnarray}
  \nind h(X)
  \nsp{2}&=&\nsp{2} \half \log\left( 2 \pi e\, \sigma_X^2\right) - D_X \\
  \nsp{2}&=&\nsp{2} \half \log\left( 2 \pi e\, \sigma_X^2\right) \nn \\
  \nsp{2}&&\nsp{2} - \half \int_0^\infty \nsp{2}
  \frac{\sigma_X^2}{1+\snr\sigma_X^2}
    - \mmse\left(X|\sqrt{\snr}X\nsp{.5}+\nsp{.5}N\right) \nsp{.5} \intd \snr.
    \label{e:hm}
\end{eqnarray}
\begin{comment}
\begin{eqnarray}
  h(X)
  &=& \half \log\left( 2 \pi e\, \sigma_X^2\right) - D_X \\
  &=& \half \log\left( 2 \pi e\, \sigma_X^2\right) - 
      \half \int_0^\infty \frac{\sigma_X^2}{1+\snr\,\sigma_X^2} \nn \\
      && \quad - \mmse\left(X|\sqrt{\snr}\,X+N\right) \, \intd \snr.
      \label{e:hm}
\end{eqnarray}
Let
\begin{equation}
  D_X = - \half \int_0^\infty \frac{\sigma_X^2}{1+\snr\,\sigma_X^2}
      - \mmse\left(X|\sqrt{\snr}\,X+N\right) \, \intd \snr,
\end{equation}
which denotes the ``non-Gaussianness'' of $P_X$ measured in MMSE.
\end{comment}
According to (\ref{e:dh}), $\gamma_X = e^{-D_X}$ is a parameter that
measures the difficulty of estimating $X$ when observed in Gaussian
noise across the full range of SNRs. Note that $0 \leq \gamma_X \leq
1$ with the upper bound attained when $X$ is Gaussian, and the lower
bound attained when $X$ is discrete.  Adding independent random
variables results in a random variable that is harder to estimate in
the sense of the following inequality:
%24nov %A by-product of \eref{e:hm} is a new version of Shannon's entropy
%24nov %power inequality expressed in the MMSE instead of differential
%24nov %entropy:
\begin{equation}
\label{e:epi}
\alpha \gamma_{X_1}^2 + (1 - \alpha) \gamma_{X_2}^2 \leq   \gamma_{X_1+X_2}^2
\end{equation}
where $X_1$ and $X_2$ are independent random variables and $\alpha$ is
the ratio of the variance of $X_1$ to the sum of the variances of
$X_1$ and $X_2$.  Of course, (\ref{e:epi}) is nothing but Shannon's
entropy power inequality \cite{Shanno48BSTJ}.  It would be interesting
to see if (\ref{e:epi}) can be proven from estimation-theoretic
principles.

\begin{comment}
A by-product of \eref{e:hm} is a new version of Shannon's entropy
power inequality expressed in the MMSE instead of differential
entropy:
\begin{theorem}[Shannon \cite{Shanno48BSTJ}]
  For any independent random variables $X$ and $Y$,
  \begin{equation}
    \sigma^2_X e^{-D_X} + \sigma^2_Y e^{-D_Y}
    \leq \sigma^2_{X+Y} e^{-D_{X+Y}},
                                % \leq (\sigma^2_X + \sigma^2_Y) e^{D_{X+Y}},
  \end{equation}
  where the non-Gaussianness $D_{(\cdot)}$ is defined in \eref{e:dh}.
\end{theorem}
% and $\sigma^2_{(\cdot)}$ denotes the variance of a random variable.
\end{comment}

Another observation is that Theorem \ref{th:pp} provides a new means
of representing the mutual information between an arbitrary random
variable $X$ and a real-valued random variable $Z$:
\begin{equation} \label{e:wmi}
  \begin{split}
  I ( X ; Z ) = &
  \half \int_0^\infty
  \Exp \Bigl\{
    \left(\expcnd{Z}{\sqrt{\snr} \, Z + N,X} \right)^2 \\
    & -
    \left(\expcnd{Z}{\sqrt{\snr} \, Z + N} \right)^2
  \Bigr\} \, \intd \snr.
  \end{split}
\end{equation}
An arbitrary discrete valued $Z$ can be handled as in Theorem
\ref{th:H} by means of an adequate one-to-one mapping.

The above results can be generalized to continuous-time models and
vector channels.  It is remarkable that the entropy, differential
entropy, divergence and mutual information in fairly general settings
admit expressions in pure estimation-theoretic quantities.  It remains
to be seen whether such representations lead to new insights and
applications.

\section{Conclusion}
\label{s:con}

This paper reveals that the input-output mutual information and the
(noncausal) MMSE in estimating the input given the output determine
each other by a simple formula under both discrete- and
continuous-time, scalar and vector Gaussian channel models.  A
consequence of this relationship is the coupling of the MMSEs
achievable by smoothing and filtering with arbitrary signals corrupted
by Gaussian noise.  Moreover, new expressions in terms of MMSE are
found for information measures such as entropy and divergence.

The idea of incremental channels is the underlying basis for the most
streamlined proof of the main results and for their interpretation.
The white Gaussian nature of the noise is key to this approach: 1) The
sum of independent Gaussian variates is Gaussian; and 2) the Wiener
process has independent increments.  In fact, the relationship between
the mutual information and the noncausal estimation error holds in
even more general settings of Gaussian channels.  In a follow-up to
this work, Zakai has recently extended formula \eref{e:pie} to the
abstract Wiener space \cite{Zakai04pre}, which generalizes the
classical $m$-dimensional Wiener process.

The incremental-channel technique in this paper is relevant for an
entire family of channels the noise of which has independent
increments, i.e., that is characterized by L\'evy processes
\cite{Bertoi96}.  A particular interesting case, which is reported in
\cite{GuoSha04ITW}, is the Poisson channel, where the corresponding
mutual information-estimation error relationship involves an error
measure quite different from mean-square error.

\begin{comment}
In fact, the incremental-channel devices introduced in this paper are
applicable to the entire family of processes with independent
increments, known as L\'evy processes \cite{Bertoi96}.  Results on
non-Gaussian channels, primarily Poisson channels, will be reported
separately.
\end{comment}

Applications of the relationships revealed in this paper are abundant.
In addition to the application in \cite{GuoVerITsub} to multiuser
channels, \cite{MeaUrb04ITW} shows applications to key results in EXIT
charts for the analysis of sparse-graph codes. Other applications as
well as counterparts to non-Gaussian channels will be published in the
near future.  In all, the relations shown in this paper illuminate
intimate connections between information theory and estimation theory.

\begin{comment}
The fact that the mutual information and the (noncausal) MMSE
determine each other also provides a new means to calculate or bound
one quantity using the other.
\end{comment}

% \appendix
\appendices

\section{Verification of \eref{e:ie}: Binary Input}
\label{a:dim}

\begin{proof}
  From \eref{e:eb} and \eref{e:ib}, it can be checked that
  \begin{eqnarray}
    && \nind 2 \pd{\snr} I(\snr) - \mmse(\snr) \nn \\
    &=& 1 - \int^\infty_{-\infty} \oneon{\sqrt{2\pi}}
    e^{-\half y^2} \left(1-\frac{y}{\sqrt{\snr}}\right) \nn \\
    && \quad \times \tanh\left(\snr-\sqrt{\snr}\,y\right)
    \,\intd y \label{e:dimy} \\
    &=& 1 - \oneon{\sqrt{\snr}} \int^\infty_{-\infty} \oneon{\sqrt{2\pi}}
    e^{-\half (z-\sqrt{\snr})^2} \nn \\
    && \quad \times z \tanh\left(\sqrt{\snr}\, z\right) \,\intd z,
    \label{e:dimz}
  \end{eqnarray}
  where from \eref{e:dimy} to \eref{e:dimz} $\sqrt{\snr}-y$ is
  replaced by $z$.  The integral in \eref{e:dimz} can be regarded as
  the expectation of $Z\tanh \left( \sqrt{\snr}\,Z \right)$ where
  $Z\sim \mathcal{N}(\sqrt{\snr},1)$.  The expectation remains the
  same if $Z$ is replaced by $Z'\sim \mathcal{N}(-\sqrt{\snr},1)$ due
  to symmetry.  Hence the integral can be rewritten by averaging over
  the two cases as:
  \begin{eqnarray}
    && \nind \half \int^\infty_{-\infty} \left[
      e^{-\half(z-\sqrt{\snr})^2} + 
      e^{-\half(z+\sqrt{\snr})^2} \right] %\nn \\
     z \tanh \left( \sqrt{\snr}\, z \right) \,
    \frac{\intd z}{\sqrt{2\pi}}
     \nn \\
%    &=& \int^\infty_{-\infty} \oneon{\sqrt{2\pi}}
%      e^{-\frac{z^2+\snr}{2}} z \sinh z \,\intd z \\
      &=& \half \int^\infty_{-\infty} 
      \left[ e^{-\half(z-\sqrt{\snr})^2} -
      e^{-\half(z+\sqrt{\snr})^2} \right] \, z \,
    \frac{\intd z}{\sqrt{2\pi}}  \\
    % e^{-\frac{(z+\sqrt{\snr})^2}{2}} \right] \, 
    &=& \half ( \Exp Z - \Exp Z' ) \\
    &=& \sqrt{\snr}.  \label{e:zzp}
  \end{eqnarray}
  Therefore, \eref{e:dimz} vanishes by \eref{e:zzp}, and \eref{e:ie}
  holds.
\end{proof}

\section{Proof of Lemma \ref{lm:id}}
\label{a:id}

\begin{proof}
  By \eref{e:svf}, the mutual information admits the following
  decomposition:
  \begin{equation}
    I(Y;Z) = \cnddiv{P_{Y|Z}}{P_{Y'}}{P_Z} - \divergence{P_Y}{P_{Y'}},
    \label{e:isv}
  \end{equation}
  where $Y'\sim \mathcal{N} \left(\Exp Y, \sigma_Y^2 \right)$.  Let the
  variance of $Z$ be denoted by $v$.
  \begin{comment}
  The probability density function
  associated with $Y'$ is
  The probability
  density function associated with $Y'$ is
  \begin{equation}
    p_{Y'}(y) = \oneon{\sqrt{2\pi(\delta v+1)}} \expb{
      -\frac{(y-\Exp Y)^2}{2(\delta v+1)} }.
    \label{e:pyp}
  \end{equation}
  \end{comment}
  The first term on the right hand side of \eref{e:isv} is equal to a
  divergence between two Gaussian distributions, which is found as
  \begin{equation}
    \half \log(1+\delta v) = \frac{\delta v}{2} + o(\delta)
  \end{equation}
  by using the general formula:
  \begin{equation}
    \begin{split}
      & \divergence{ \mathcal{N}\left(m_1,\sigma_1^2\right) }{
        \mathcal{N}\left(m_0,\sigma_0^2\right) } \\
      & \quad = \half \log \frac{\sigma_0^2}{\sigma_1^2} + \half
      \left( \frac{(m_1-m_0)^2}{\sigma_0^2}
        + \frac{\sigma_1^2}{\sigma_0^2} - 1 \right) \log e.
    \end{split}
  \end{equation}
  It suffices then to show that
  \begin{equation}
    \divergence{ P_Y }{ P_{Y'} }
    = \expect{ \log \frac{ p_Y(Y) }{ p_{Y'}(Y) } }
    = o(\delta),  \label{e:dyp}
  \end{equation}
  which is straightforward to check by plugging in the density
  functions:
  \begin{comment}
  The unconditional output distribution can be expressed as
  \begin{equation} \label{e:pyy}
    p_Y(y) = \oneon{\sqrt{2\pi}} \expect{
      \exph{ \left( y-\sqrt{\delta}\, Z \right)^2 } }.
  \end{equation}
  By \eref{e:pyp} and \eref{e:pyy},
  \end{comment}
  \begin{eqnarray}
    && \nind \log \frac{ p_Y(y) }{ p_{Y'}(y) } \nn\\
    &=& \log \left[ \oneon{\sqrt{2\pi}} \expect{
      \exph{ \left( y-\sqrt{\delta}\, Z \right)^2 } }  \right] \nn \\
  && \; - \log \left[
    \oneon{\sqrt{2\pi(\delta v+1)}} \expb{
      -\frac{(y-\Exp Y)^2}{2(\delta v+1)} } \right] \\
    &=& \log \expect{ \expb{
        \frac{(y-\sqrt{\delta}\,\Exp Z)^2}{2(\delta v+1)}
        - \half (y-\sqrt{\delta}\,Z)^2 } } \nn \\
    && \; + \half \log(1+\delta v) \\
    &=& \log \Exp \biggl\{ 1 + \sqrt{\delta} \, y (Z-\Exp Z)
      + \frac{\delta}{2}\bigl( y^2 (Z-\Exp Z)^2 - vy^2 \nn \\
      && \; - Z^2 + (\Exp Z)^2 \bigr) + o(\delta) \biggr\}
      + \half \log(1+\delta v) \\
    &=& \log\left( 1 - \frac{\delta v}{2} \right) + \half \log(1+\delta v)
    + o(\delta) \label{e:eoe} \\
    &=& o(\delta).
    \label{e:od}
  \end{eqnarray}
  The limit and the expectation can be exchanged to obtain
  \eref{e:eoe} as long as $\Exp Z^2 <\infty$ due to the Lebesgue
  Convergence Theorem.
  \begin{comment}
    Therefore, the second
  divergence on the right hand side of \eref{e:svf} is $o(\delta)$.
  Lemma \ref{lm:id} is immediate:
  \begin{equation}
    I(Y;Z) = \frac{\delta v}{2} + o(\delta).
  \end{equation}
  \end{comment}
\end{proof}
It is interesting to note that the proof relies on the fact that the
divergence between the output distributions of a Gaussian channel
under different input distributions is sublinear in the SNR when the
noise dominates.

\section{A Fourth Proof of Theorem \ref{th:di}}
\label{a:di}

\begin{proof}
  For simplicity, it is assumed that the order of expectation and
  derivative can be exchanged freely.  A rigorous proof is relegated
  to Appendix \ref{a:dv} where every such assumption is validated in
  the more general vector model.

  \begin{comment}
  By definition \eref{e:ixy}
  and noting that the input-output conditional probability density
  function \eref{e:pyx} is Gaussian, one has
  \begin{equation} \label{e:ip0}
    I(\snr) = -\half \log(2\pi e) - \int \pyi{0} \log \pyi{0} \,\intd y,
  \end{equation}
  \end{comment}
  
  Let $\pyi{i}$ be defined as in \eref{e:pyi}.  It can be checked that
  for all $i$,
  \begin{eqnarray}
    && \nind \pd{\snr} \pyi{i} \nn \\
    &=& \oneon{2\sqrt{\snr}} \, y \, \pyi{i+1} - \half \, \pyi{i+2}
    \label{e:smy} \\
    &=& - \oneon{2\sqrt{\snr}} \pd{y} \pyi{i+1}.
    \label{e:sy}
  \end{eqnarray}
  The derivative of the mutual information, expressed as \eref{e:ip0},
  can be obtained as
  \begin{eqnarray}
    && \nind \pd{\snr} I(\snr) \nn \\
    \nsp{1}&=&\nsp{1} - \int \left[ \log \pyi{0} + 1 \right] \pd{\snr} \pyi{0} \intd y \\
    \nsp{1}&=&\nsp{1} \oneon{2\sqrt{\snr}} \int \log \pyi{0} \, \pd{y} \pyi{1} \intd y \\
    \nsp{1}&=&\nsp{1} -\oneon{2\sqrt{\snr}} \int \frac{
      \pyi{1} }{ \pyi{0} } \, \pd{y} \pyi{0} \intd y \label{e:ibp} \\
%    &=& -\oneon{2\sqrt{\snr}} \int \frac{ \pyi{1} }{ \pyi{0} } \,
%    \left[ \sqrt{\snr}\, \pyi{1} - y\, \pyi{0} \right] \intd y \\
    \nsp{1}&=&\nsp{1} \oneon{2\sqrt{\snr}} \int \pyi{1} \,
    \left[ y - \sqrt{\snr} \,\frac{\pyi{1}}{\pyi{0}} \right] \nsp{1}\intd y,
%    &=& \oneon{2\sqrt{\snr}} \int \frac{ \pyi{1} }{ \pyi{0} } \,
%    \left[ y - \sqrt{\snr} \,\frac{\pyi{1}}{\pyi{0}} \right] \nn \\
%    && \quad \pyi{0} \intd y,
    \label{e:frp}
  \end{eqnarray}
  where \eref{e:ibp} follows by integrating by parts.  Noting that the
  fraction in \eref{e:frp} is exactly the conditional mean estimate
  (cf.~\eref{e:x2}),
  \begin{eqnarray}
%    && \nind \pd{\snr} I(\snr) \nn \\
    \pd{\snr} I(\snr)
    \nsp{1}&=&\nsp{1} \oneon{2\sqrt{\snr}} \Exp \Bigl\{ \expcnd{X}{Y;\snr} \nn \\
    && \; \times
      \left[ Y - \sqrt{\snr} \,\expcnd{X}{Y;\snr} \right] \Bigr\} \\
    \nsp{1}&=&\nsp{1} \half\, \expect{ \frac{XY}{\sqrt{\snr}} - \left( \expcnd{X}{Y;\snr} \right)^2 } \\
    \nsp{1}&=&\nsp{1} \half\, \expect{ \left( X - \expcnd{X}{Y;\snr} \right)^2 } \\
    \nsp{1}&=&\nsp{1} \half\, \mmse(\snr).
  \end{eqnarray}
\end{proof}
%Using the above techniques, it is not difficult to find the derivative
%of the conditional mean estimate $\hX(y;\snr)$ \eref{e:p10} with
%respect to the signal-to-noise ratio.  In fact, one can find any
%derivative of the mutual information in this way.

\section{Proof of Theorem \ref{th:dv}}
\label{a:dv}

% DG041019: removed \H since it suffices to treat \H\X as the input
\begin{proof}
  It suffices to prove the theorem assuming $\H=\I$ since one can
  always regard $\H\X$ as the input.  The vector channel \eref{e:vch}
  has a Gaussian conditional density \eref{e:vpyx}.  The unconditional
  density of the channel output is given by \eref{e:vpy}, which is
  strictly positive for all $\y$.  The mutual information can be
  written as (cf.\ \eref{e:ip0})
  \begin{comment}
  \begin{eqnarray}
    I(\snr)
    &=& \mathsf{D}( p_{\Y|\X;\snr} || p_{\Y;\snr} | p_\X ) \nn \\
    &=& \expect{ \log \vpYX } - \expect{ \log \vpY } \\
    &=& -\frac{N}{2} \log(2\pi e) - \int \vpy \, \log \vpy \, \intd \y.
  \end{eqnarray}
  \end{comment}
  \begin{equation}
    \begin{split}
      I(\snr) = & -\frac{L}{2} \log(2\pi e) \\
      & \quad - \int \vpy \, \log \vpy \, \intd \y.
    \end{split}
  \end{equation}
  Hence,
  \begin{eqnarray}
    \pd{\snr} I(\snr)
    \nsp{1.5} &=& \nsp{1.5} - \int \left[ \log \vpy + 1 \right] \nn \\
    && \nsp{1} \times \pd{\snr} \vpy\, \intd \y \label{e:vds} \\
    \nsp{1.5} &=& \nsp{1.5} - \int \left[ \log \vpy + 1 \right] \nn \\
    && \nsp{1} \times \expect{ \pd{\snr} \vpyX } \intd \y,  \label{e:vii}
  \end{eqnarray}
  where the derivative penetrates the integral in \eref{e:vds} by the
  Lebesgue Convergence Theorem, and the order of taking the derivative
  and expectation in \eref{e:vii} can be exchanged by Lemma
  \ref{lm:ds}, which is shown below in this Appendix.  It can be
  checked that (cf.\ \eref{e:smy} and \eref{e:sy})
  \begin{eqnarray}
    %    && \pd{\snr} \vpyx \nn \\
    && \nind \pd{\snr} \vpyx \nn \\
    &=& \oneon{2\sqrt{\snr}} \, \xT \left(\y
    -\sqrt{\snr}\,\x\right) \, \vpyx \\
    &=& - \oneon{2\sqrt{\snr}} \, \xT \nabla \vpyx.
    \label{e:dsy}
  \end{eqnarray}
  %  where the derivative with respect to a vector is defined as $\pd{\y}
  %  =\tran{\left [\ppd{ y_1 },\dots,\ppd{ y_L }\right] }$.
  Using \eref{e:dsy}, the right hand side of \eref{e:vii} can be
  written as
  \begin{equation}
    \begin{split}
%      & \pd{\snr} \mathsf{D}( p_{\Y|\X;\snr} || p_{\Y;\snr} | p_\X ) \\
%      =&
    \oneon{2 \sqrt{\snr}} \Exp \biggl\{ \XT \int [ & \log \vpy + 1 ] \\
    & \times \nabla \vpyX \, \intd \y \biggr\}.
    \end{split}
    \label{e:bp}
  \end{equation}
  The integral in \eref{e:bp} can be carried out by parts to obtain
  \begin{equation}
    - \int \vpyX \, \nabla \left[ \log \vpy + 1 \right] \, \intd \y,
  \end{equation}
  since for all $\x$, as $\|\y\| \rightarrow \infty$,
  \begin{equation}
    \vpyx\, \left[ \log \vpy + 1 \right] \rightarrow 0.
  \end{equation}
  Hence, the expectation in \eref{e:bp} can be further evaluated as
  \begin{equation}    \label{e:exy}
    \begin{split}
      %- \oneon{2 \sqrt{\snr}} \int \expect{ \XT \, \frac{
      - \int \expect{ \XT \, \frac{
          \vpyX }{ \vpy } } \nabla \vpy \, \intd \y
    \end{split}
  \end{equation}
  where we have changed the order of the expectation with respect to
  $\X$ and the integral (i.e., expectation with respect to $\Y$).  By
  \eref{e:dsy} and Lemma \ref{lm:dy} (shown below in this Appendix),
  \eref{e:exy} can be written as
  \begin{equation}
    \begin{comment}
    \oneon{2\sqrt{\snr}} \int \expcnd{ \XT }{\Y=\y;\snr} \,
    \expect{ \left( \y - \sqrt{\snr}\, \X \right) \, \vpyX } \, \intd \y.
    \end{comment}
    \begin{split}
    & \int \expcnd{ \XT }{\Y=\y;\snr} \\
    & \quad \times
    \expect{ \left( \y - \sqrt{\snr}\, \X \right) \, \vpyX } \, \intd \y.
    \end{split}
    \label{e:dpy}
  \end{equation}
  Therefore, \eref{e:vii} can be rewritten as
  \begin{eqnarray}
      && \nind \pd{\snr} I(\snr) \nn \\
%    \pd{\snr} I(\snr)
    \nsp{2}&=&\nsp{2} \oneon{2\sqrt{\snr}} \int \expcnd{ \XT }{\Y=\y;\snr}
        \nn \\
    &&  \nsp{2} \times
    \Exp\bigl\{\y-\sqrt{\snr} \X \bigl| \Y=\y;\snr\bigl\} \vpy \intd \y \\
    \nsp{2}&=&\nsp{2} \expect{ \Exp\bigl\{ \XT | \Y;\snr \bigl\}
      \Exp \biggl\{ \frac{\Y}{2 \sqrt{\snr}} - \half \X \Bigl| \Y;\snr \biggr\} } \\
%    \nsp{2}&=&\nsp{2} \oneon{2 \sqrt{\snr}} \expect{ \expcnd{ \XT }{\Y;\snr}
%      \expcnd{ \Y - \sqrt{\snr}\, \X }{\Y;\snr} } \\
%    \nsp{2}&=&\nsp{2} \oneon{2 \sqrt{\snr}} \expect{ \XT\Y } - \half \expect{
%    \left\| \expcnd{\X}{\Y;\snr} \right\|^2 } \\
%    \nsp{2}&=&\nsp{2} \half \expect{ \|\X\|^2 } - \half \expect{
%    \left\| \expcnd{\X}{\Y;\snr} \right\|^2 } \\
    \nsp{2}&=&\nsp{2} \expect{ \half \|\X\|^2 - \half
    \left\| \expcnd{\X}{\Y;\snr} \right\|^2 } \\
    \nsp{2}&=&\nsp{2} \half \expect{ \left\| \X - \expcnd{\X}{\Y;\snr} \right\|^2 }.
    \label{e:die}
  \end{eqnarray}
  Hence the proof of Theorem \ref{th:dv}.
\end{proof}

The following two lemmas were needed to justify the exchange of
derivatives and expectation with respect to $P_\X$ in the above proof.
\begin{lemma}  \label{lm:ds}
  If\; $\Exp\|\X\|^2 <\infty$, then
  \begin{equation}
    \begin{split}
      \pd{\snr} & \expect{ \vpyX } \\
      & = \expect{ \pd{\snr} \vpyX }.
    \end{split}
  \end{equation}
\end{lemma}
\begin{proof}
  Let
  \begin{equation}
    \begin{split}
    f_\delta(\x,\y,\snr) = & \oneon{\delta} \bigl[
      p_{\Y|\X;\snr}(\y|\X;\snr+\delta) \\
      & \qquad - \vpyX \bigr]
    \end{split}
  \end{equation}
  and
  \begin{equation}
    f(\x,\y,\snr) = \pd{\snr} \vpyx.
  \end{equation}
  Then, $\forall\,\x,\y,\snr$, $f_\delta(\x,\y,\snr) \rightarrow
  f(\x,\y,\snr)$ as $\delta \rightarrow 0$.
%  \begin{equation}
 %   \limzero{\delta} f_\delta(\x,\y,\snr) = f(\x,\y,\snr).
  %\end{equation}
  Lemma \ref{lm:ds} is equivalent to
  \begin{equation}    \label{e:lct}
    \limzero{\delta} \int f_\delta(\x,\y,\snr) P_\X(\intd \x)
    = \int f(\x,\y,\snr) P_\X(\intd \x).
  \end{equation}
  Suppose we can show that for every $\delta,\x,\y$ and $\snr$,
  \begin{equation}
    | f_\delta(\x,\y,\snr) | < \|\x\|^2 + \oneon{\sqrt{\snr}} |\yT\x|.
    \label{e:fl}
  \end{equation}
  % DG041019 rewording
  Then \eref{e:lct} holds by the Lebesgue Convergence Theorem since
  the right hand side of \eref{e:fl} is integrable with respect to
  $P_\X$ by the assumption in the lemma.  Note that
  \begin{equation}
    \begin{comment}
      f_\delta(\x,\y,\snr) = (2\pi)^{-\frac{L}{2}} \oneon{\delta}
      \, \left\{  \exph{ \|\y-\sqrt{\snr+\delta}\,\x\|^2} 
      - \exph{ \|\y-\sqrt{\snr}\,\x\|^2} \right\}.
    \end{comment}
    \begin{split}
      f_\delta(\x,\y,\snr) =& (2\pi)^{-\frac{L}{2}} \oneon{\delta}
      \, \left(  \exph{ \|\y-\sqrt{\snr+\delta}\,\x\|^2} \right.\\
    &\quad \left. - \exph{ \|\y-\sqrt{\snr}\,\x\|^2} \right).
    \end{split}
  \end{equation}
  If
  \begin{equation}
    \oneon{\delta} \leq \|\x\|^2 + \oneon{\sqrt{\snr}} |\yT\x|,
  \end{equation}
  then \eref{e:fl} holds trivially.  Otherwise,
  \begin{eqnarray}
    && \nind | f_\delta(\x,\y,\snr) | \nn \\
%    | f_\delta(\x,\y,\snr) |
    \nsp{2}&<&\nsp{2} \oneon{\delta} \biggl| \exp \biggl[ \half\|\y-\sqrt{\snr}\,\x\|^2 \nn\\
      && \qquad \qquad
      -\half \|\y-\sqrt{\snr+\delta}\,\x\|^2 \biggr] -1 \biggr| \\
    \nsp{2}&<&\nsp{2} \oneon{2\delta} \left[ \exp\left|
      \delta \|\x\|^2 - ( \sqrt{\snr+\delta} \!-\! \sqrt{\snr} )
      \yT\x \right| -1 \right] \\
    \nsp{2}&<&\nsp{2} \oneon{2\delta} \left[ \expb{\delta \,
      \left( \|\x\|^2 + \oneon{\sqrt{\snr}} |\yT\x| \right) } - 1 \right].
  \end{eqnarray}
  % DG041019 rewording
  The inequality \eref{e:fl} holds for all $\x,\y,\snr$ due to the
  fact that
  \begin{equation} \label{e:et}
    e^t - 1 < 2 t, \quad \forall\, 0\leq t<1.
  \end{equation}
\end{proof}

\begin{lemma}  \label{lm:dy}
  If\; $\Exp\X$ exists, then for $i=1,\dots,L$,
  \begin{equation} \label{e:dy}
    \begin{split}
      \ppd{y_i} & \expect{ \vpYX } \\
      & \quad = \expect{ \ppd{y_i} \vpYX }.
    \end{split}
  \end{equation}
\end{lemma}
\begin{proof}
  The proof is similar to that for Lemma \ref{lm:ds}.    Let
%  \begin{equation}
%    g(\x,\y,\snr) = \ppd{y_i} \vpyx
%  \end{equation} 
% and
  \begin{equation}
    \begin{split}
    g_\delta(\x,\y,\snr) = & \oneon{\delta} \bigl[
      p_{\Y|\X;\snr}(\y+\delta\,\e_i|\X;\snr) \\
      & \qquad - \vpyX \bigr]
    \end{split}
%    g_\delta(\x,\y,\snr) = \oneon{\delta} \left[
%      p_{\Y|\X;\snr}(\y+\delta\,\e_i|\X;\snr) - \vpyX \right]
  \end{equation}
  where $\e_i$ is a vector with all zero except on the $\ith i$ entry,
  which is 1.  Then, $\forall \x,\y,\snr$,
  \begin{equation}
    \limzero{\delta} g_\delta(\x,\y,\snr) = \ppd{y_i} \vpyx.
    % \limzero{\delta} g_\delta(\x,\y,\snr) = g(\x,\y,\snr).
  \end{equation}
  % DG041019: commented
  \begin{comment}
  Lemma \ref{lm:dy} is equivalent to
  \begin{equation}
    \limzero{\delta} \int g_\delta(\x,\y,\snr) P_\X(\intd \x)
    = \int g(\x,\y,\snr) P_\X(\intd \x).
    \label{e:lcg}
  \end{equation}
  \end{comment}
  We show that
  \begin{equation}
    | g_\delta(\x,\y,\snr) | < |y_i| + 1 + \sqrt{\snr}\, |x_i|,
    \label{e:gl}
  \end{equation}
  so that \eref{e:dy} holds by the Lebesgue Convergence Theorem (cf.\ 
  \eref{e:lct}).  Note that
  \begin{equation}
    \begin{comment}
      g_\delta(\x,\y,\snr) = (2\pi)^{-\frac{L}{2}} \oneon{\delta}
      \, \biggl(
      \exph{ \|\y+\delta\,\e_i-\sqrt{\snr}\,\x\|^2}
    - \exph{ \|\y-\sqrt{\snr}\,\x\|^2} \biggr).
    \end{comment}
    \begin{split}
      g_\delta(\x,\y,\snr) =& (2\pi)^{-\frac{L}{2}} \oneon{\delta}
      \, \biggl( \exph{ \|\y+\delta\,\e_i-\sqrt{\snr}\,\x\|^2} \\
      &\quad - \exph{ \|\y-\sqrt{\snr}\,\x\|^2} \biggr).
    \end{split}
  \end{equation}
  If
  \begin{equation}
    \oneon{\delta} \leq |y_i| + 1 + \oneon{\sqrt{\snr}} |x_i|,
  \end{equation}
  then \eref{e:gl} holds trivially.  Otherwise,
  \begin{eqnarray}
    && \nind | g_\delta(\x,\y,\snr) | \nn \\
    % | g_\delta(\x,\y,\snr) |
    &<& \oneon{2\delta} \biggl( \exp \biggl|
    \half\|\y-\sqrt{\snr}\,\x\|^2 \nn\\
      && \quad -\half \|\y+\delta\,\e_i-\sqrt{\snr}\,\x\|^2 \biggr|
    -1 \biggr) \\
    &=& \oneon{2\delta} \left( \exp \left| \frac{\delta}{2} \,
    \left( 2y_i + \delta - 2 \sqrt{\snr}\, x_i \right) \right|-1\right),
%    &<& \oneon{2\delta} \left( \expb{\delta \,
%      \left( |y_i| + 1 + \sqrt{\snr}\, |x_i| \right) } - 1 \right) \\
%    &<& |y_i| + 1 + \sqrt{\snr}\, |x_i|.
  \end{eqnarray}
  and \eref{e:gl} holds by \eref{e:et}.
  % since the   right hand side of \eref{e:gl} is integrable with respect to $P_\X$ by assumption.  
\end{proof}

\begin{comment}
 \end{comment}

  \begin{comment}
  Note that
  \begin{equation}
    \pd{\snr} \pyi{i} = \oneon{2\sqrt{\snr}} y \pyi{i+1} - \half \pyi{i+2},
  \end{equation}
  and
  \begin{equation}
    \pd{y} \pyi{i} = \sqrt{\snr}\cdot \pyi{i+1} - y \pyi{i}.
  \end{equation}
  Clearly,
  \begin{equation}
    \pd{\snr} \pyi{i} = - \oneon{2\sqrt{\snr}} \pd{y} \pyi{i+1}.
    \label{e:sy}
  \end{equation}
  \end{comment}

\section{Verification of \eref{e:e}: Random Telegraph Input}
\label{a:rt}

Let $\xi= -\frac{2\nu}{\snr}$ and define
\begin{equation}
  f(i,j) =
  \int_1^\infty u^\frac{i}{2} (u-1)^\frac{j}{2} e^{\xi u} \,\intd u.
\end{equation}
It can be checked that
\begin{eqnarray}
  f(i,j) &=& f(i+2,j) - f(i,j+2), \label{e:pfd} \\
  \pd{\xi} f(i,j) &=& f(i+2,j), \label{e:pfx} \\
  -\xi \, f(i,j) &=& \frac{i}{2} f(i-2,j) + \frac{j}{2} f(i,j-2),
  \label{e:pft}
\end{eqnarray}
where verifying \eref{e:pft} entails integration by parts.  Then
\eref{e:rtc} can be rewritten as
\begin{equation}
  \cmmse(\snr) = f(-1,-1) / f(1,-1)
\end{equation}
and hence
\begin{equation}
  \begin{split}
  & \pd{\snr} [ \snr \cdot \cmmse(\snr) ]
  = \bigl[ f(-1,-1)f(1,-1) \\
  & \quad - \xi f^2(1,-1) + \xi f(-1,-1)f(3,-1) \bigr] / f^2(1,-1).
  \end{split}
\end{equation}
With the change of variables $t=(1-x^2)^{-1}$ and $u=(1-y^2)^{-1}$,
\eref{e:rtn} can also be rewritten:
\begin{equation} \label{e:di}
  \begin{split}
    & \mmse(\snr) = f^{-2}(1,-1) \\
    & \int_1^\infty \int_1^\infty
    \frac{e^{(t+u)\xi}}{t+u-1} t^\half u^\half (t-1)^{-\half} (u-1)^{-\half}
    \, \intd t\intd u.
  % (t+u-1)^{-1} t^\half u^\half (t-1)^{-\half} (u-1)^{-\half}
  \end{split}
%  \mmse(\snr) = f^{-2}(1,-1) \int_1^\infty \int_1^\infty
%  (t+u-1)^{-1} t^\half u^\half (t-1)^{-\half} (u-1)^{-\half}
%  \expb{(t+u)\xi} \, \intd t\intd u.
\end{equation}
The denominator in \eref{e:di} prevents the double integral from being
separated.  This can be circumvented by taking derivative with respect
to $\xi$.  Noting that
\begin{equation}
e^\xi \, \pd{\xi} \left[ e^{-\xi} f^2(1,-1) \, \mmse(\snr) \right] = f^2(1,-1),
\end{equation}
the identity \eref{e:e} is equivalent to
\begin{equation}
  \begin{split}
    e^\xi \, \pd{\xi} \bigl[ e^{-\xi} &
      \bigl( f(-1,-1)f(1,-1) - \xi f^2(1,-1) \\
        & + \xi f(-1,-1)f(3,-1) \bigr) \bigr] = f^2(1,-1)  \label{e:fff}
  \end{split}
\end{equation}
since both sides of \eref{e:e} tend to 0 as $\xi\rightarrow -\infty$.
With the help of \eref{e:pfd}--\eref{e:pft}, verifying \eref{e:fff} is
a matter of algebra.

\section{Proof of Lemma \ref{lm:dx}}
\label{a:dx}

% {\bf SV- This proof should also be made more elegant in view of the
% more elegant proof of Lemma 1.}

Lemma \ref{lm:dx} can be regarded as a consequence of Duncan's
Theorem.  The mutual information can be expressed as a time-integral
of the causal MMSE:
\begin{equation}
  I\left(Z^T_0;Y^T_0\right) = \frac{\delta}{2} \int^T_0 \Exp
  \left( Z_t - \expcnd{Z_t}{Y_0^t;\delta} \right)^2 \intd t,
  \label{e:id}
\end{equation}
As the SNR $\delta\rightarrow0$, the observation $Y_0^T$ becomes
inconsequential in estimating the input signal.  Indeed, the causal
MMSE estimate converges to the unconditional mean in mean-square
sense:
\begin{equation}
  \expcnd{Z_t}{Y_0^t;\delta} \rightarrow \Exp Z_t.
  \label{e:hx}
\end{equation}
Putting \eref{e:id} and \eref{e:hx} together proves Lemma \ref{lm:dx}.

% DG041019: rewording
In parallel with the proof of Lemma \ref{lm:id}, another reasoning of
Lemma \ref{lm:dx} from first principles without invoking Duncan's
Theorem is presented in the following.  In fact, Lemma \ref{lm:dx} is
established first in this way so that a more intuitive proof of
Duncan's Theorem is given in Section \ref{s:ti} using the idea of
time-incremental channels.

% DG: Make this proof RIGOROUS!!!!!

\begin{proof}[Lemma \ref{lm:dx}]
  By definition~\eref{e:iT}, the mutual information is the expectation
  of the logarithm of the Radon-Nikodym derivative~\eref{e:rnd}, which
  can be obtained by the chain rule as
  \begin{equation}
    \Phi = \frac{ \intd \mu_{YZ} }{ \intd \mu_Y \intd \mu_Z }
    = \frac{\intd \mu_{YZ}}{\intd \mu_{WZ}} \,
    \left( \frac{\intd \mu_Y}{\intd \mu_W} \right)^{-1}.
    \label{e:dch}
  \end{equation}

  First assume that $\rp{Z}$ is a bounded uniformly stepwise process,
%  i.e., there exists a finite subdivision of $[0,T]$, $0=t_1 <\dots
%  <t_n <t_{n+1} =T$, and a finite constant $M$ such that
  i.e., there exists a finite subdivision of $[0,T]$, $0=t_0 <t_1
  <\dots <t_n=T$, and a finite constant $M$ such that
  \begin{equation}
    Z_t(\omega) = Z_{t_i}(\omega), \quad t\in[t_i,t_{i+1}), \;
    i=0,\dots,n-1,
  \end{equation}
  and $Z_t(\omega)<M$, $\forall\, t\in[0,T]$.  Let
  $\Z=[Z_{t_0},\dots,Z_{t_n}]$, $\Y=[ Y_{t_0},\dots, Y_{t_n}]$, and
  $\W=[W_{t_0},\dots,W_{t_n}]$ be $(n+1)$-dimensional vectors formed
  by the samples of the random processes.  Then, the input-output
  conditional density is Gaussian:
  \begin{equation}
    \begin{split}
      p_{\Y|\Z}(&\y|\z) = \prod^{n-1}_{i=0} \oneon{\sqrt{2\pi(t_{i+1}-t_i)}} \\
      & \times \expb{ - \frac{ \left( y_{i+1} - y_i - \sqrt{\delta}\, z_i
            (t_{i+1}-t_i) \right)^2 }{2(t_{i+1}-t_i)} }.
    \end{split}
  \end{equation}
  Easily,
  \begin{eqnarray}
    && \nind \frac{p_{\Y\Z}(\b,\z)}{p_{\W\Z}(\b,\z)}
    = \frac{ p_{\Y|\Z}(\b|\z) }{ p_{\W}(\b) } \\
    \nsp{4}&=&\nsp{2} \exp \Biggl[ \sqrt{\delta}
    \sum^{n-1}_{i=0} z_i(b_{i+1}\!-\!b_i)
    -\frac{\delta}{2} \sum^{n-1}_{i=0} z_i^2 (t_{i+1}\!-\!t_i) \Biggr].
    \label{e:yb}
  \end{eqnarray}
  \begin{comment}
  \begin{eqnarray}
    \frac{p_{\Y\Z}(\b,\z)}{p_{\W\Z}(\b,\z)}
    &=& \frac{ p_{\Y|\Z}(\b|\z) }{ p_{\W}(\b) } \\
    &=& \expb{ \sqrt{\delta} \sum^{n-1}_{i=0} z_i(b_{i+1}\!\!-\!\!b_i)
      \nsp{1}-\nsp{1} \frac{\delta}{2} \sum^{n-1}_{i=0}
      z_i^2 (t_{i+1}\!\!-\!\!t_i) }.
    \label{e:yb}
  \end{eqnarray}
  Consider a set of continuous functions in $[0,T]$ of the form:
  \begin{equation}
    A = \left\{ f(t): \; f(t_i)\in[a_i,b_i], i=1,\dots,n \right\}.
  \end{equation}
  \end{comment}
  Thus the Radon-Nikodym derivative can be established as
  \begin{equation}
%    \phi_t = \expb{ \stdlt \int_0^t Z_s \,\intd W_s
    \frac{\intd \mu_{YZ}}{\intd \mu_{WZ}} = \expb{ \sqrt{\delta}
      \int_0^T Z_t \,\intd W_t - \frac{\delta}{2} \int_0^T Z_t^2 \,\intd t }
    \label{e:dyzb}
  \end{equation}
  using the finite-dimensional likelihood ratios~\eref{e:yb}.  It is
  clear that $\mu_{YZ} \ll \mu_{WZ}$.
  
  For the case of a general finite-power process (not necessarily
  bounded) $\rp{Z}$, a sequence of bounded uniformly stepwise
  processes which converge to the $\rp{Z}$ in $L^2(\intd t\intd P)$
  can be obtained.  The Radon-Nikodym derivative \eref{e:dyzb} of the
  sequence of processes also converges.  Absolute continuity is
  preserved.  Therefore, \eref{e:dyzb} holds for all such processes
  $\rp{Z}$.\footnote{A shortcut to the proof of \eref{e:dyzb} is by
    the Girsanov Theorem \cite{Oksend03}.}

  The derivative \eref{e:dyzb} can be rewritten as
  \begin{equation}    \label{e:yzbz}
    \begin{split}
    & \frac{\intd \mu_{YZ}}{\intd \mu_{WZ}}
    = 1 + \sqrt{\delta} \int_0^T Z_t \,\intd W_t \\
    & \qquad + \frac{\delta}{2} \left[ \biggl(\int_0^T Z_t\,\intd W_t \biggr)^2
      - \int_0^T Z_t^2 \,\intd t\right] + o(\delta).
    \end{split}
  \end{equation}
  By the independence of the processes $\rp{W}$ and $\rp{Z}$, the
  measure $\mu_{WZ}=\mu_W\mu_Z$.  Thus integrating on the measure
  $\mu_Z$ gives
  \begin{equation}    \label{e:dyb}
    \begin{split}
      & \frac{\intd \mu_Y}{\intd \mu_W}
      = 1 + \sqrt{\delta} \int_0^T \Exp Z_t \,\intd W_t \\
      & \; + \frac{\delta}{2}
    \left[ \Exp_{\mu_Z} \biggl(\int_0^T Z_t\,\intd W_t \biggr)^2
      - \int_0^T \Exp Z_t^2 \,\intd t\right] + o(\delta).
    \end{split}
  \end{equation}
  Using \eref{e:yzbz}, \eref{e:dyb} and the chain rule \eref{e:dch},
  the Radon-Nikodym derivative $\Phi$ exists and is given by
  \begin{eqnarray}
%    && \frac{\intd \mu_{YZ}}{\intd \mu_{WZ}} \,
%    \left( \frac{\intd \mu_Y}{\intd \mu_W} \right)^{-1} \nn \\
    \Phi
    \nsp{1}&=&\nsp{1} 1 + \sqrt{\delta} \int_0^T Z_t - \Exp Z_t \,\intd W_t
    + \frac{\delta}{2} \,
    \Biggl[ \biggl( \int_0^T Z_t \,\intd W_t \biggr)^2 \nn \\
    && - \int_0^T Z_t^2 \,\intd t
      - 2 \int_0^T \Exp Z_t \,\intd W_t
      \int_0^T Z_t-\Exp Z_t \,\intd W_t \nn \\
      && - \Exp_{\mu_Z} \biggl( \int_0^T Z_t\,\intd W_t \biggr)^2
          + \int_0^T \Exp Z_t^2 \,\intd t \Biggr] + o(\delta) \\
    \nsp{1}&=&\nsp{1} 1 + \sqrt{\delta} \int_0^T Z_t - \Exp Z_t \,\intd W_t \nn \\
    &&  + \frac{\delta}{2} \, \Biggl[ \biggl( \int_0^T Z_t - \Exp Z_t
      \,\intd W_t \biggr)^2 \nn \\
      && - \Exp_{\mu_Z} \biggl( \int_0^T Z_t
      - \Exp Z_t \,\intd W_t \biggr)^2 \nn \\
    && - \int_0^T Z_t^2
      - \Exp Z_t^2 \,\intd t \Biggr] + o(\delta).
    \label{e:phi}
  \end{eqnarray}
  Note that the mutual information is an expectation with respect to
  the measure $\mu_{YZ}$.  It can be written as
  \begin{equation}
    I\left(Z_0^T;Y_0^T\right) = \int \log \Phi' \,\intd \mu_{YZ}
  \end{equation}
  where $\Phi'$ is obtained from $\Phi$ \eref{e:phi} by substituting
  all occurrences of $\intd W_t$ by $\intd Y_t = \sqrt{\delta}\, Z_t +
  \intd W_t$:
  \begin{eqnarray}
    \Phi'
    \nsp{1}&=&\nsp{1} 1 + \sqrt{\delta} \int_0^T Z_t - \Exp Z_t \,\intd Y_t \nn \\
    && + \frac{\delta}{2} \, \Biggl[ \biggl( \int_0^T Z_t - \Exp Z_t
      \,\intd Y_t \biggr)^2 \nn \\
      && - \Exp_{\mu_Z} \biggl( \int_0^T Z_t
      - \Exp Z_t \,\intd Y_t \biggr)^2 \nn \\
    && - \int_0^T Z_t^2
      - \Exp Z_t^2 \,\intd t \Biggr] + o(\delta) \\
    \nsp{1}&=&\nsp{1} 1 + \sqrt{\delta} \int_0^T Z_t - \Exp Z_t \,\intd W_t \nn \\
    && + \frac{\delta}{2} \, \Biggl[ \biggl( \int_0^T Z_t - \Exp Z_t
      \,\intd W_t \biggr)^2 \nn \\
    && - \Exp_{\mu_Z} \biggl( \int_0^T Z_t
      - \Exp Z_t \,\intd W_t \biggr)^2
    + \int_0^T (Z_t - \Exp Z_t)^2 \,\intd t \nn \\
%      && \quad + \int_0^T (Z_t - \Exp Z_t)^2 \,\intd t \nn \\
      && + \int_0^T \Exp (Z_t - \Exp Z_t)^2 \,\intd t \Biggr] + o(\delta) \\
    \nsp{1}&=&\nsp{1} 1 + \sqrt{\delta} \int_0^T \tilde{Z}_t \,\intd W_t \nn \\
    && + \frac{\delta}{2} \, \Biggl[ \biggl( \int_0^T \tilde{Z}_t
      \,\intd W_t \biggr)^2 - \Exp_{\mu_Z} \biggl( \int_0^T \tilde{Z}_t
      \,\intd W_t \biggr)^2  \nn \\
    && + \int_0^T \tilde{Z}^2_t \,\intd t
      + \int_0^T \Exp \tilde{Z}^2_t \,\intd t \Biggr] + o(\delta)
  \end{eqnarray}
  where $\tilde{Z}_t =Z_t -\Exp Z_t$.  Hence
  \begin{equation}
    \begin{split}
      \log \Phi'
      = & \sqrt{\delta} \int_0^T \tilde{Z}_t \,\intd W_t
    + \frac{\delta}{2} \, \Biggl[ - \Exp_{\mu_Z} \biggl( \int_0^T
      \tilde{Z}_t \,\intd W_t \biggr)^2 \\
    & + \int_0^T \tilde{Z}_t^2 \,\intd t
      + \int_0^T \Exp \tilde{Z}_t^2 \,\intd t \Biggr] + o(\delta).
    \end{split}
  \end{equation}
  Therefore, the mutual information is
  \begin{eqnarray}
    && \nind \Exp \log \Phi' \nn \\
    \nsp{1}&=&\nsp{1} \frac{\delta}{2} \, \left[ 
      2 \int_0^T \!\Exp \tilde{Z}_t^2 \intd t
      - \Exp \biggl( \int_0^T \!\tilde{Z}_t \intd W_t \biggr)^2
    \right] + o(\delta) \\
%    \nsp{1}&=&\nsp{1} \frac{\delta}{2} \, \left[ - \Exp \biggl( \int_0^T
%      \tilde{Z}_t \,\intd W_t \biggr)^2
%      + 2 \int_0^T \Exp \tilde{Z}_t^2 \,\intd t \right] + o(\delta) \\
    \nsp{1}&=&\nsp{1} \frac{\delta}{2} \, \left[
      2 \int_0^T \Exp \tilde{Z}_t^2 \,\intd t 
      - \int_0^T \Exp \tilde{Z}_t^2 \,\intd t
    \right] + o(\delta) \\
%    \nsp{1}&=&\nsp{1} \frac{\delta}{2} \, \left[ - \int_0^T
%      \Exp \tilde{Z}_t^2 \,\intd t
%      + 2 \int_0^T \Exp \tilde{Z}_t^2 \,\intd t \right] + o(\delta) \\
    \nsp{1}&=&\nsp{1} \frac{\delta}{2} \int_0^T \Exp \tilde{Z}_t^2 \,\intd t + o(\delta),
  \end{eqnarray}
  and the lemma is proved.
\end{proof}

\section{Proof of Lemma \ref{lm:hi}}
\label{a:hi}

\begin{proof}
  Let $Y=\sqrt{\snr}\, g(X)+N$.  Since
  \begin{equation}
    0 \leq H(X) - I(X;Y) = H(X|Y),
  \end{equation}
  it suffices to show that the uncertainty about $X$ given $Y$
  vanishes as $\snr\rightarrow\infty$:
  \begin{equation} \label{e:hxy}
    \liminfty{\snr} H(X|Y) = 0.
  \end{equation}
  
  Assume first that $X$ takes a finite number ($m<\infty$) of distinct
  values.  Given $Y$, let $\hX$ be the decision for $X$ that achieves
  the minimum probability of error, which is denoted by $p$.  Then
  \begin{eqnarray} \label{e:fn}
    H(X|Y) \leq H(X|\hX) \leq p \,\log(m-1) + H_2(p),
  \end{eqnarray}
  where $H_2(\cdot)$ stands for the binary entropy function, and
  the second inequality is due to Fano \cite{CovTho91}.  Since
  $p\rightarrow0$ as $\snr \rightarrow \infty$, the right hand side of
  \eref{e:fn} vanishes and \eref{e:hxy} is proved.

  In case $X$ takes a countable number of values and that
  $H(X)<\infty$, for every natural number $m$, let $U_m$ be an
  indicator which takes the value of 1 if $X$ takes one of the $m$
  most likely values and 0 otherwise.  Let $\hX_m$ be the function of
  $Y$ which minimizes $\Prob \left\{ X\neq \hX_m | U_m=1 \right\}$.
  Then for every $m$,
  \begin{eqnarray}
    && \nind H(X|Y) \nn \\
    &\leq& H(X|\hX_m) \\
        &=& H(X,U_m|\hX_m) \\
        &=& H(X|\hX_m,U_m) + H(U_m|\hX_m) \\
        &\leq& \Prob\{U_m=1\} H(X|\hX_m,U_m=1) \nn \\
        && + \Prob\{U_m=0\} H(X|\hX_m,U_m=0)
                + H(U_m) \\
        &\leq& \Prob\{U_m=1\} H(X|\hX_m,U_m=1) \nn \\
        && + \Prob\{U_m=0\} H(X)
                + H_2(\Prob\{U_m=0\}). \label{e:hxu}
  \end{eqnarray}
  The conditional probability of error $\Prob\left\{ X\neq\hX_m |
    U_m=1 \right\}$ vanishes as $\snr \rightarrow \infty$ and so does
  $H(X|\hX_m,U_m=1)$ by Fano's inequality.  Therefore, for every $m$,
  \begin{equation} \label{e:lhxy}
    \liminfty{\snr} H(X|Y) \leq \Prob\{U_m=0\} H(X)
                + H_2(\Prob\{U_m=0\}).
  \end{equation}
  The limit in \eref{e:lhxy} must be 0 since $\Prob \{U_m=0\}
  \rightarrow0$ as $m\rightarrow \infty$.  Thus \eref{e:hxy} is also
  proved in this case.

  In case $H(X)=\infty$, $H(X|U_m=1)\rightarrow\infty$ as
  $m\rightarrow \infty$.  For every $m$, the mutual information
  (expressed in the form of a divergence) converges:
  \begin{equation}
    \liminfty{\snr} \cnddiv{P_{Y|X,U_m=1}}{P_{Y|U_m=1}}{P_{X|U_m=1}}
    = H(X|U_m=1).
  \end{equation}
  Therefore, the mutual information increases without bound as $\snr
  \rightarrow \infty$ by also noticing
  \begin{eqnarray}
    && \nind I(X;Y) \nn \\
    \nsp{2}&\geq&\nsp{2} I(X;Y|U_m) \\
    \nsp{2}&\geq&\nsp{2} \Prob\{U_m=1\}
    \cnddiv{P_{Y|X,U_m=1}}{P_{Y|U_m=1}}{P_{X|U_m=1}}.
  \end{eqnarray}
  We have thus proved \eref{e:hi} in all cases.
\end{proof}

\section*{Acknowledgment}

We gratefully acknowledge discussions with Professors Haya Kaspi,
Tsachy Weissman, Moshe Zakai, and Ofer Zeitouni.

%\bibliographystyle{IEEEtran}
%\bibliography{def,nonumber,dguo,book,math,comms,cdma}

\end{document}